\documentstyle[epsf,righttag,citesort,makeidx]{amsbook}
\numberwithin{equation}{chapter} %AMS-LaTeX command
\makeindex

\def\pb{\not\hspace{-0.7ex}p\hspace{0.7ex}}
\def\kb{\not\hspace{-0.7ex}k\hspace{0.7ex}}

\newcommand{\beq}{\begin{equation}}
\newcommand{\eeq}{\end{equation}}
\newcommand{\bea}{\begin{eqnarray}}
\newcommand{\eea}{\end{eqnarray}}
\newcommand{\bes}{\begin{split}}
\newcommand{\ees}{\end{split}}
\newcommand{\as}{\alpha_s}
\newcommand{\D}{\displaystyle}

\newcommand{\eq}[1]{eq.~(\ref{#1})}
\newcommand{\Eq}[1]{Eq.~(\ref{#1})}
\newcommand{\rfn}[1]{(\ref{#1})}

\def\mafigura#1#2#3#4{
  \begin{figure}[htbp]
    \begin{center}
      \epsfxsize=#1
      \leavevmode
      \epsffile{#2}
    \end{center}
    \caption{#3}
    \label{#4}
  \end{figure} }
\begin{document}
%\documentstyle[epsf,times,amstex]{article}
%\begin{document}

%\hoffset -.4cm
%\textheight = 522pt
%\textwidth  = 348pt
%\baselineskip 25pt

%\thispagestyle{empty}
%
%\vspace*{\stretch{1}}
%\begin{center}
%{\sc \huge  Quark mass effects in QCD jets}
%\end{center}
%\vspace*{\stretch{2}}

\pagebreak
\thispagestyle{empty}
\vspace*{\stretch{1}}
\begin{flushright}
{\sc \huge  Quark mass effects in QCD jets}
\rule{\linewidth}{1mm}
\end{flushright}
\vspace*{\stretch{2}}
\begin{center}
{\sc \large Tesis Doctoral 1996}
\end{center}

\pagebreak
\thispagestyle{empty}
\vspace*{\stretch{1}}

\pagebreak
\thispagestyle{empty}

\begin{center}

{\sc \huge  Quark mass effects in QCD jets}

\vspace*{\stretch{1}}

\epsfverbosetrue
\begin{figure}[hbtp]
   \begin{center}
   \epsfxsize=7cm
   \leavevmode
   \hspace{1cm}
   \epsffile{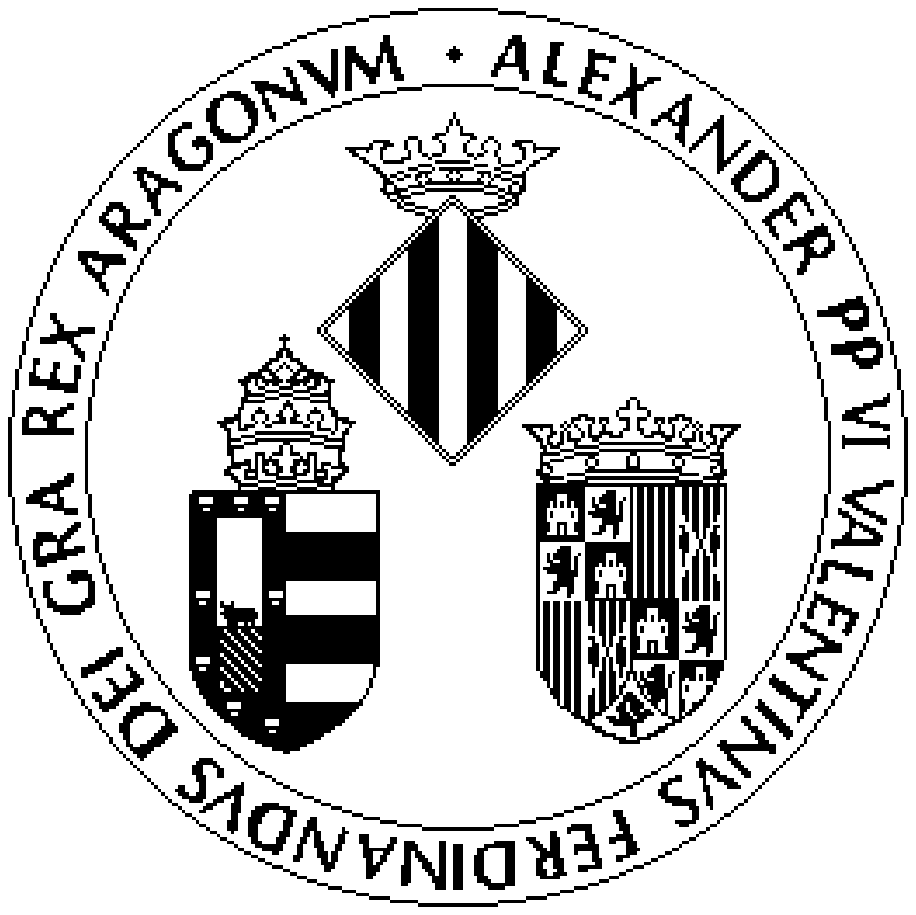}
%   \epsffile{escudo.ps}
   \end{center}
   \vskip 1cm
   \begin{center}
      {\sc \Large Departament de F\'{i}sica Te\`{o}rica}
      \vskip .5cm
      {\sc \Large Universitat de Val\`encia}
      \vskip .5cm
      {\sc  \large -- Estudi General --}
   \end{center}
\end{figure}

\end{center}

\vspace*{\stretch{1}}

\begin{center}
{\sc \large Tesis Doctoral} \vskip .1cm

{\sc \large presentada por} \vskip .5cm

{\sc \large Germ\'{a}n Vicente Rodrigo Garc\'{i}a} \vskip .1cm

{\sc \large 28 de Octubre de 1996}

{\sc \large isbn 84-370-2989-9}

\end{center}

\pagebreak

\thispagestyle{empty}

%\vspace*{\stretch{2}}
%$\bigcirc$\hspace{-1.7ex}{\sf c}\hspace{1.7ex} Germ\'an Rodrigo, 1996.

\vspace*{\stretch{1}}

Portada: {\it ``Anciana''}, Jos\'e de Ribera, Aguafuerte.

Contraportada: de la Suite Vollard {\it ``Hombre descubriendo una mujer''},
Pablo Picasso, Punta seca.

%\end{document}

\thispagestyle{empty}

\vspace*{\stretch{1}}

     N' {\it Arcadi Santamaria i Luna}, Professor Titular del
Departament de F\'{\i}sica Te\`{o}rica de la Universitat de
Val\`{e}ncia, \vskip .5cm

     {\bf Certifica:} \vskip .5cm

     Que la present Mem\`{o}ria, 
{\sc Quark Mass Effects in QCD Jets},
ha estat realitzada sota la seua direcci\'{o} al Departament
de F\'{\i}sica Te\`{o}rica de la Universitat de Val\`{e}ncia
per En {\it Germ\'{a}n Vicente Rodrigo i Garc\'{\i}a},
i constitueix la seua Tesi per a optar al grau de 
Doctor en F\'{\i}sica. \vskip .3cm

     I per a que aix\'{\i} conste, en acompliment de la 
legislaci\'{o} vigent, presenta en la Universitat de Val\`{e}ncia
la referida Tesi Doctoral, i signa el present certificat, a
\vskip .5cm

\begin{flushright}
     Burjassot, 12 de Setembre de 1996 \\  \vskip 6cm

     Arcadi Santamaria i Luna
\end{flushright}

\vspace*{\stretch{1}}

\pagebreak

\thispagestyle{empty}

\vspace*{\stretch{1}}

\thispagestyle{empty}

\vspace*{\stretch{1}}

\begin{flushright}
{\large \sl a Chelo}
\end{flushright}

\vspace*{\stretch{2}}

\pagebreak

\thispagestyle{empty}

\vspace*{\stretch{1}}

\pagenumbering{roman}
\chapter*{Pre\'{a}mbulo}

     La presente memoria esta dedicada al estudio de 
los efectos inducidos por la masa de los quarks en 
observables relacionados con la desintegraci\'on
a tres jets del bos\'on de gauge $Z$ y su aplicaci\'on
experimental en LEP.

Para la mayor parte de los observables estudiados en 
LEP el efecto de la masa de los quarks puede despreciarse
puesto que \'esta siempre aparece como 
el cociente $m_q^2/m_Z^2$, donde $m_q$ es la masa del quark 
y $m_Z$ es la masa del bos\'on $Z$, lo cual supone correcciones
muy peque\~nas incluso para el quark m\'as pesado producido  
en LEP, el quark $b$, muy por debajo de la precisi\'on 
experimental actualmente accesible.

No obstante, aunque este \'ultimo argumento sea cierto para secciones 
eficaces totales no ocurre as\'{\i} cuando estudiamos observables
que, aparte de la masa del quark, dependan de variables
adicionales como por ejemplo secciones eficaces a n-jets.
En tal caso, puesto que ponemos en juego una nueva escala 
de energ\'{\i}as $E_c=m_Z \sqrt{y_c}$, donde $y_c$ es el par\'ametro 
que define la multiplicidad del jet,
aparecen contribuciones del tipo $m_q^2/E_c^2=(m_q/m_z)^2/y_c$
que, para valores de $y_c$ lo suficientemente peque\~nos, 
podr\'{\i}an incrementar considerablemente  
el efecto de la masa de los quarks y permitir 
su estudio en LEP. As\'{\i} mismo, el efecto de la masa de los quarks
podr\'{\i}a verse favorecido por logaritmos de la 
masa, $m_q^2 \log(m_q^2/m_Z^2)$, provenientes de la
integraci\'on sobre espacio f\'asico.

Dos son las razones que motivan nuestro estudio.
En primer lugar, el error sistem\'atico m\'as importante en la
medida de la constante de acoplamiento fuerte $\as^b(m_Z)$
a partir del cociente entre las anchuras de desintegraci\'on 
del bos\'on $Z$ a dos y tres jets
en la producci\'on de $b \bar{b}$ en LEP~\cite{juano}
procede de la incertidumbre debida al desconocimiento de
los efectos de la masa del quark $b$.
Una mejor comprensi\'on de estos efectos contribuir\'{\i}a 
obstensiblemente a una mejor medida de la constante de
acoplamiento fuerte $\as^b(m_Z)$.

En segundo lugar, asumiendo universalidad de sabor para 
las interacciones fuertes, dicho estudio podr\'{\i}a 
permitir por primera vez una medida experimental de la
masa del quark $b$ a partir de los datos de LEP.
Debido a que los quarks no aparecen en la naturaleza como
part\'{\i}culas libres el estudio de su masa presenta 
serias dificultades te\'oricas. De hecho, la masa de los
quarks debe considerarse como una constante de 
acoplamiento m\'as.
La masa de los quarks pesados, como el $b$ y el $c$, 
puede extraerse a bajas energ\'{\i}as a partir del espectro
del bottomium y del charmonium con Reglas de Suma de QCD y 
c\'alculos en el ret\'{\i}culo.
Una medida de la masa del quark $b$ a altas energ\'{\i}as,
como por ejemplo en LEP, presentar\'{\i}a la ventaja de
permitir una determinaci\'on de la masa del quark $b$ 
a una escala de energ\'{\i}as muy por encima de su
umbral de producci\'on, a diferencia de lo que ocurre 
en los dos m\'etodos anteriormente descritos.
Es m\'as, dicha medida permitir\'{\i}a por primera vez comprobar 
como la masa del quark $b$ evoluciona 
seg\'un predice el Grupo de Renormalizaci\'on 
desde escalas del order de la masa del quark misma, $\mu=m_b$,
hasta altas energ\'{\i}as, $\mu=m_Z$, del mismo modo 
que fue posible hacerlo para la constante de acoplamiento 
fuerte $\as(\mu)$ y constituir\'{\i}a una nueva confirmaci\'on 
de QCD como teor\'{\i}a para describir las interacciones fuertes.

En el Cap\'{\i}tulo 1 comenzamos revisando las distintas definiciones
de masa, repasamos cuales son las Ecuaciones del Grupo de 
Renormalizaci\'on en QCD, analizamos como conectar 
los par\'ametros de una teor\'{\i}a con los de su teor\'{\i}a
efectiva a bajas energ\'{\i}as mediante las adecuadas Condiciones
de Conexi\'on y realizamos una peque\~na recopilaci\'on de las
determinaciones m\'as recientes que de las masas de todos los quarks 
han sido realizadas a partir de Reglas de Suma de QCD,
c\'alculos en el ret\'{\i}culo y Teor\'{\i}a de Perturbaciones Quirales.
Finalmente, evolucionamos todas las masas mediante las 
Ecuaciones del Grupo de Renormalizaci\'on hasta la escala 
de energ\'{\i}as de la masa del bos\'on $Z$
poniendo especial incapi\'e en la masa del quark $b$ cuyos efectos 
pretendemos estudiar en profundidad en el resto de esta tesis.

En el Cap\'{\i}tulo 2 justificamos por qu\'e es posible medir 
la masa del quark $b$ en LEP a partir de observables a tres jets,
comparamos el comportamiento de la anchura de desintegraci\'on 
inclusiva del bos\'on $Z$ expresada en funci\'on 
de las distintas definiciones de masa, 
definimos cuales son los algoritmos de reconstrucci\'on de jets, 
en particular los cuatro sobre los cuales basaremos nuestro 
an\'alisis: EM, JADE, E y DURHAM.
Finalmente, analizamos a primer order en la constante de 
acoplamiento fuerte algunos observables a tres jets y 
proporcionamos funciones sencillas que parametrizan su comportamiento 
en funci\'on de la masa del quark y del par\'ametro que 
define la multiplicidad del jet. 

Puesto que el primer order no permite distinguir entre las 
posibles definiciones de masa y puesto que para el quark 
$b$ la diferencia entre ellas es num\'ericamente muy 
importante, en los Cap\'{\i}tulos 3, 4 y 5 nos centramos 
en el an\'alisis del orden siguiente.
En el Cap\'{\i}tulo 3 presentamos y clasificamos las 
distintas amplitudes de transici\'on que debemos calcular 
a este orden. La principal dificultad de dicho c\'alculo 
radica en la aparici\'on de divergencias infrarrojas
debido a la presencia de part\'{\i}culas sin masa como 
son los gluones. En el Cap\'{\i}tulo 4 analizamos 
el comportamiento infrarrojo de estas amplitudes de
transici\'on, la integramos anal\'{\i}ticamente 
en la regi\'on de espacio f\'asico que contiene las 
divergencias y finalmente mostramos como dichas divergencias
se cancelan cuando sumamos las contribuciones procedentes 
de las correcciones virtuales y reales. 
En el Cap\'{\i}tulo 5 presentamos los resultados de la
integraci\'on num\'erica de las partes finitas
as\'{\i} como ajustes sencillos de estos resultados 
para facilitar su manejo,
exploramos a segundo orden los observables a tres 
jets que hab\'{\i}amos estudiado en el Cap\'{\i}tulo 2
y discutimos su comportamiento en funci\'on de la 
escala.
Finalmente, en los ap\'endices, recopilamos algunas 
de las funciones necesarias para el c\'alculo que 
hemos realizado: integrales a un loop, 
reducci\'on de Passarino-Veltman, el
problema de $\gamma_5$ en Regularizaci\'on Dimensional,
espacio f\'asico en $D$-dimensiones, etc.

El resto de esta tesis esta escrita en ingl\'es para cumplir 
con la normativa vigente sobre el {\bf Doctorado Europeo}.

\vspace*{\stretch{1}}

\pagebreak

\vspace*{\stretch{1}}

\chapter*{Preamble}

     This thesis is devoted to the study of the 
effects induced by the quark mass in some 
three-jet observables related to the decay 
of the $Z$ gauge boson and its experimental 
application in LEP.

Quark masses can be neglected for many observables 
at LEP because usually they appear as the ratio
$m_q^2/m_Z^2$, where $m_q$ is the quark mass and 
$m_Z$ is the $Z$-boson mass. 
Even for the heaviest quark produced at LEP, the 
$b$-quark, these corrections are very small 
and remain bellow the LEP experimental precision.

While this argument is correct for total cross sections,
it is not completely true for quantities that depend on 
other variables. In particular n-jet cross sections.
In this case we introduce a new scale $E_c=m_Z \sqrt{y_c}$,
where $y_c$ is the jet-resolution parameter that defines 
the jet multiplicity, and for small values of $y_c$ there
could be contributions like $m_q^2/E_c^2=(m_q/m_z)^2/y_c$
which could enhance the quark mass effect considerably 
allowing its study at LEP.
Furthermore, the quark mass effect could be favoured by 
logarithms of the mass, $m_q^2 \log(m_q^2/m_Z^2)$, coming
from phase space integration.

Our motivation is twofold.
First, it has been shown~\cite{juano} that the biggest
systematic error in the measurement of $\as^b(m_Z)$
($\as$ obtained from $b\bar{b}$-production at LEP from 
the ratio of three to two jets) comes from the uncertainties in 
the estimate of the quark mass effects.
A better understanding of such effects will contribute to 
a better determination of the strong coupling constant 
$\as^b(m_Z)$.

Second, by assuming flavour universality of the strong 
interaction we study the possibility of measuring the 
bottom quark mass, $m_b$, from LEP data.
A precise theoretical framework is needed for the 
study of the quark mass effects in physical 
observables because quarks are not free particles. 
In fact, quark masses have to be treated 
more like coupling constants.
The heavy quark masses, like the $b$ and the $c$-quark masses, 
can be extracted at low energies from the bottomia and the
charmonia spectrum from QCD Sum Rules and Lattice 
calculations. 
Nevertheless, a possible measurement of the bottom quark 
mass at high energies, like in LEP, would present the 
advantage of determining the bottom quark mass 
far from threshold in contrast
to the other methods described above.
Furthermore, such a measurement would allow to test for the 
first time how the bottom quark mass evolve 
following the Renormalization Group prediction from 
scales of the order of the quark mass itself,  $\mu=m_b$,
to high energy scales, $\mu=m_Z$, in the same way it 
was possible with the strong coupling constant
and would provide a new test of QCD as a good theory 
describing the strong interaction.

In Chapter 1 we start by reviewing the different
theoretical mass definitions,
we solve the QCD Renormalization Group Equations,
we analyze how to connect the parameters of a theory with
the parameters of its low energy effective theory through 
the adequate Matching Conditions, in particular how to pass
a heavy quark threshold and we review the most recent 
determinations of all the quark masses form QCD Sum Rules, 
Lattice, and Chiral Perturbation Theory.
Finally we run all the quark masses to the $Z$-boson mass 
scale and we focus our study in the bottom 
quark mass which we will study in the rest of this thesis.

In Chapter 2 we justify the possibility of extracting the 
bottom quark mass in LEP from some three-jet observables, 
we compare the behaviour of the $Z$-boson inclusive decay width 
expressed in terms of the different quark mass definitions,
we define what is a jet clustering algorithm, in particular 
the four on which we will work: EM, JADE, E and DURHAM.
Finally, we analyze at first order in the strong 
coupling constant some three-jet observables
and we give simple functions parametrizing their 
behaviour in terms of the quark mass and the jet 
resolution parameter. 

Since the first order does not allow to distinguish 
which mass definition should we use in the theoretical 
expressions and since for the bottom quark mass 
the difference among the possible mass definitions is 
quite important, in Chapters 3, 4 and 5 we focus
in the analysis at the next order.
In Chapter 3 we present and classify the transition 
amplitudes we should compute at second order 
in the strong coupling constant.
The main difficulty of such calculation is the 
appearance, in addition to renormalized UV 
divergences, of infrared (IR) singularities since we are
dealing with massless particles like gluons.
In chapter 4 we analyze the infrared behaviour of 
our transition amplitudes, we integrate them analytically 
in the phase space region containing the infrared 
singularities and finally we show how the 
infrared divergences cancel when we sum up 
the contributions coming from the real and the
one-loop virtual corrections.
In Chapter 5 we present the results of the numerical
integration over the finite parts and we give simple fits 
to these results in order to facilitate their handling,
we explore at second order the three-jet observables
we have studied in Chapter 2 and we discuss their
scale-dependent behaviour.
Finally, in the appendices, we collect some of the 
functions needed for the calculation we have performed: 
one-loop integrals, Passarino-Veltman reduction,
the $\gamma_5$ problem in Dimensional Regularization, 
phase space in $D$-dimensions, etc. 

This thesis fulfils the {\bf European Ph.D.} conditions.

\vspace*{\stretch{1}}

%\pagebreak

%\thispagestyle{empty}

%\vspace*{\stretch{1}}

\tableofcontents
\pagenumbering{arabic}
\setcounter{page}{1}
\chapter{The quark masses}

     \nocite{UIMP}

     \label{UIMPchapter}

     GUT and SUSY theories predict some relations among the fermion 
masses (or more properly among 
Yukawa couplings\index{Yukawa coupling}) at the
unification\index{Unification} scale, 
$M_{GUT} \sim  10^{16} (GeV)$. 
For instance, in $SU(5)$ we have the usual lepton-bottom quarks 
unification, $h_b=h_{\tau}$, $h_s=h_{\mu}$, $h_d=h_e$, or the 
modified Georgi-Jarlskog relation, $h_b=h_{\tau}$, $h_s=h_{\mu}/3$,
$h_d=3 h_e$, while for $SO(10)$ typically we get unification of 
the third family, $h_t=h_b=h_{\tau}=h_{\nu_{\tau}}$.
This together with the RGE's provide a powerful tool
for predicting quark masses at low energies. 
 
     This chapter is not devoted to Unification.
Rather, we try to make a short review of 
the most recent determinations of the quark masses and to 
calculate the running until $m_Z$, the mass of the $Z$-boson. Why $m_Z$?.
For model building purposes it is a good idea to have a reference
scale and extensions of the Standard Model\index{Standard Model}
appear above $m_Z$.
Furthermore, the strong coupling constant 
$\as$ is really strong below $m_Z$ and then special care
has to be taken on the matching in passing a heavy quark 
threshold\index{Threshold} and on the running.

      To define what is the mass of a quark
is not an easy task because quarks are not free particles.
For leptons it is clear that the physical 
mass is the pole of the propagator.
Quark masses, however, have to be treated more like coupling
constants.
We can find in the literature several quark mass definitions:
the Euclidean mass,\index{Quark mass!Euclidean}
$M_E(p^2=-M^2)$, defined as the mass
renormalized at the Euclidean point $p^2=-M^2$. It is gauge 
dependent but softly dependent on $\Lambda_{QCD}$.
Nevertheless, it is not used anymore in the most recent works.
The perturbative pole 
mass~\cite{Tarrach81}\index{Quark mass!perturbative pole}
and the running mass\index{Quark mass!running}
are the two most commonly used quark mass definitions.
The ``perturbative'' pole mass, $M(p^2=M^2)$,
is defined as the pole of the renormalized quark propagator in
a strictly perturbative sense.
It is gauge invariant and scheme independent. However, 
it suffers from renormalon ambiguities.\index{Renormalon}
The running mass, $\bar{m}(\mu)$, the renormalized mass
in the $\overline{MS}$ scheme\index{Renormalization!scheme!$\overline{MS}$}
or its corresponding Yukawa coupling related to it through the
vev of the Higgs\index{Higgs}, $\bar{m}(\mu)=v(\mu) \bar{h}(\mu)$,
is well defined and since it is a true short distance parameter
it has become the preferred mass definition in the last years.

%%%%%%%%%%%%%%%%%%%%%%%%%%%%%%%%%
\begin{table}[t]
\begin{center}

\caption{Recent determinations of the light quark masses from 
second order $\chi$PT, QCD Sum Rules and lattice.
\label{light0}}

\vspace{.3cm}

\begin{tabular}{|c|c|c|l|} \hline 
Gasser     & $\chi$PT & $O(p^4)$ 
           & $\frac{\D m_d-m_u}{\D m_s-\hat{m}}
              \frac{\D 2\hat{m}}{\D m_s+\hat{m}}
             = 2.35 \times 10^{-3}$\\
Leutwyler  & &
           & $m_s/\hat{m} = 25.7 \pm 2.6$ \\ \hline
Donoghue   & $\chi$PT & $O(p^4)$ 
           & $\frac{\D m_d-m_u}{\D m_s-\hat{m}}
              \frac{\D 2\hat{m}}{\D m_s+\hat{m}}
             = 2.11 \times 10^{-3}$\\
 et al.    & &
           & $m_s/\hat{m} = 31. $ \\ \hline
\end{tabular}

\vspace{.3cm}

\begin{tabular}{|c|c|c|l|c|} \hline 
Bijnens    & FESR & NNLO 
           & $(\bar{m}_u+\bar{m}_d)(1GeV)$
           & \\
Prades     & Laplace SR& 
           & \qquad $= 12.0 \pm 2.5$  
           & $\alpha_s(m_Z) = 0.117(5)$ \\
de Rafael  & (pseudo)& 
           & 
           & \\ \hline

Ioffe      & Isospin viol.& NNLO 
           & $(\bar{m}_d-\bar{m}_u)(1GeV)$
           & $\Lambda = 150.$ \\ 
et al.     & in QCD SR & 
           & \qquad $= 3.0 \pm 1.0$ 
           & \\ \hline
\end{tabular}

\vspace{.3cm}

\begin{tabular}{|c|c|c|c|c|} \hline 
Narison    & $\tau$-like SR& NNLO 
           & $\bar{m}_s(1GeV) = 197(29)$ 
           & $\alpha_s(m_Z) = 0.118(6)$ \\ 
           & & NLO
           & $\bar{m}_s(1GeV) = 222(22)$ 
           & \\ \hline

Jamin      & QSSR & NNLO 
           & $\bar{m}_s(1GeV) = 189(32)$ 
           & $\alpha_s(m_Z) = 0.118(6)$ \\ 
M\"unz     & (scalar) & 
           & 
           & \\ \hline

Chetyrkin  & QCD SR & NNLO 
           & $\bar{m}_s(1GeV) = 171(15)$ 
           & $\alpha_s(m_Z) = 0.117(5)$ \\ 
et al.     & & 
           & 
           & \\ \hline \hline

Allton     & quenched  & NLO 
                 & $\bar{m}_s(2GeV) = 128(18)$ 
                 & $\Lambda^5 = 240. \pm 90.$ \\ 
et al.     & Lattice & 
           & 
           & \\ \hline
\end{tabular}
\end{center}

\end{table}

%%%%%%%%%%%%%%%%%%%%%%%%%%%%%%%%%%%

     For the light quarks, up, down and strange, chiral 
perturbation theory~\cite{GL1,GL2,GL3,DON1,DON2}
\index{Chiral Perturbation Theory}
provides a powerful tool for determining renormalization group invariant
quark mass ratios. The absolute values, usually the running
mass at $1(GeV)$, can be extracted from different 
QCD Sum Rules~\cite{BI,IO1,IO2,NAs,JM,CH} \index{Sum Rules}
or for the strange quark mass from
lattice~\cite{AL}\index{Lattice}.
For the heavy quarks, bottom and charm, we can deal either with
QCD Sum Rules~\cite{NAb,DOM1,DOM2,DOM3,TY,NE1,NE2} or 
lattice calculations~\cite{CRI,VG1,DA1,DA2,KH}.
The bulk of this thesis is devoted to explore the 
possibility of extracting the bottom quark mass from 
jet physics at LEP~\cite{RO95,FU,QCD96,FUQCD96}.
For the top quark we have the 
recent measurements from CDF and D\O\ at\index{CDF}\index{D\O\} 
FERMILAB~\cite{top1,top2,sandra,top3,top4}
\index{FERMILAB}
that we will identify with the pole mass.

%%%%%%%%%%%%%%%%%%%%%%%%%%%%%%%%%%%%%%%%%%%%%%%%%%%%%%%

\begin{table}[hbtp]
\begin{center}

\caption{Recent determinations of the heavy quark masses from 
QCD Sum Rules, lattice and FERMILAB.
\label{heavy0}}

\vspace{.3cm}

\begin{tabular}{|c|c|c|c|c|} \hline 
Narison    & QSSR & NLO 
           & $\bar{m}_b(M_b) = 4.23(4)$ 
           & $\alpha_s(m_Z) = 0.118(6)$ \\ 
           & $\Psi$, $\Upsilon$ &
           & $\bar{m}_c(M_c) = 1.23^{+(4)}_{-(5)}$ 
           & \\  
           & &  
           & $M_b = 4.62(2)$ 
           &\\
& & 
           & $M_c = 1.42(3)$ 
           &\\ \hline

Narison    & non-rel & NLO 
           & $M_b^{NR} = 4.69^{+(3)}_{-(2)}$ 
           & $\alpha_s(m_Z) = 0.118(6)$ \\ 
           & Laplc.SR & 
           & $M_c^{NR} = 1.45^{+(5)}_{-(4)}$ 
           & \\  \hline

Dominguez  & rel,non-rel & LO 
           & $M_b = 4.70(7)$ 
           & $\Lambda^4 = 200-300$ \\ 
et al.     & Laplc.SR & $1/m_q^2$
           & $M_c = 1.46(7)$ 
           & $\Lambda^5 = 100-200$\\
           &  $J/\Psi$, $\Upsilon$ &  
           & & \\  \hline

Titard     & $q\bar{q}$ & 
           & $\bar{m}_b(\bar{m}_b) = 4.397^{+(18)}_{-(33)}$ 
           & $\alpha_s(m_Z) = 0.117(5)$ \\ 
Yndur\'ain & potential & 
           & $\bar{m}_c(\bar{m}_c) = 1.306^{+(22)}_{-(35)}$ 
           & \\ \hline

Neubert    & QCD SR & NLO
           & $M_b = 4.71(7)$ 
           & \\ 
           & & $1/m_q$
           & $M_c = 1.30(12)$
           & \\ \hline
\end{tabular}

\vspace{.3cm}

\begin{tabular}{|c|c|c|c|} \hline 
Crisafulli & Lattice
           & $\bar{m}_b(\bar{m}_b) = 4.17(6)$ 
           & \\ 
et al.     & in B-meson 
           & 
           & \\  \hline

Gim\'enez  & Lattice
           & $\bar{m}_b(\bar{m}_b) = 4.15(20)$ 
           & \\ 
et al.     & in B-meson 
           & 
           & \\  \hline

Davies     & NRQCD + leading
           & $M_b = 5.0(2)$
           & \\ 
et al.     & rel and Lattice 
           & $\bar{m}_b(M_b) = 4.0(1)$ 
           & $\alpha^{(5)}_{\overline{MS}} = 0.115(2)$ \\
           & spacing, $b\bar{b}$ 
           & & \\  \hline

El-Khadra  & Fermilab action 
           & $M_c = 1.5(2)$ 
           & \\ 
Mertens    & in quenched Lat
           & 
           & \\ \hline
\end{tabular}

\vspace{.3cm}

\begin{tabular}{|c|l|c|} \hline
  CDF        & $M_t = 176.8 \pm 4.4(stat) \pm 4.8(sys)$ & $mean$ \\ \hline
  D\O\       & $M_t = 169. \pm 8.(stat) \pm 8. (sys)$
& $M_t = 175. \pm 6.$ \\ \hline   
\end{tabular}

\end{center}

\end{table}

%%%%%%%%%%%%%%%%%%%%%%%%%%%%%%%%%%%%%%%%%

     We have summarized in tables \ref{light0} and \ref{heavy0} all of
these recent quark mass determinations. Of course the final result 
depends on the strong gauge coupling constant used in the analysis,
for this reason we quote it too. In the running we will take 
the world average~\cite{BethkeQCD96} strong coupling constant value
$\as^{(5)}(m_Z) = 0.118 \pm 0.006$\index{alphas@$\as(m_Z)$!world average} 
for masses obtained from QCD Sum Rules 
but for lattice masses we will run with the lattice~\cite{DA1} result
$\as^{(5)}(m_Z) = 0.115 \pm 0.002$. These values are
consistent with almost all the references. For those that differ an 
update is needed but this is beyond the goals of this review.
For instance, S.~Narison~\cite{NAb} makes two different 
determinations for the bottom and the charm quark masses.
In the first, and for the first time, he gets directly the running 
mass avoiding then the renormalon ambiguity associated with the 
pole mass. The second one, from non-relativistic Laplace Sum Rules,
is in fact an update of the work of
Dominguez et al.~\cite{DOM1,DOM2,DOM3}.

     The $O(\as^2)$ strong correction to the relation between the 
perturbative pole mass, $M$, and the running mass, $\bar{m}(\mu)$,
was calculated in~\cite{GB}
\beq
\frac{M}{\bar{m}(M)} = 1 + \frac{4}{3} \frac{\as(M)}{\pi} 
+ K \left( \frac{\as(M)}{\pi} \right)^2 + O(\alpha_s^3(M)), 
\label{fuerte}
\end{equation}
where $K_t \simeq 10.95$ for the top quark, 
$K_b \simeq 12.4$ for the bottom and 
$K_c \simeq 13.3$ for the charm. 
As pointed out by S.~Narison~\cite{NAb} \eq{fuerte} is consistent
with three loops running but for two loops running we can drop 
the $O(\as^2)$ term. Recently, the electroweak correction to the
relation between the perturbative pole mass and the Yukawa coupling
has been calculated~\cite{HE1,HE2}. However, this correction is 
small, for instance for the top quark it is less than $0.5 \%$ in the $SM$
for a mass of the Higgs lower than $600 (GeV)$ and at most 
$3. \%$ for $M_H \simeq 1 (TeV)$, and for consistency one has to 
include it only if two loop electroweak running is done.

     Instead of expressing the solution of the QCD renormalization
group equations\index{Renormalization!Group Equations} 
for the strong gauge coupling constant and the quark
masses in terms of $\Lambda_{QCD}$\index{lambda@$\Lambda_{QCD}$}
we perform an expansion in the strong coupling 
constant at one loop \cite{RO93}. At three loops we get
\bea
   \as(\mu) &=& \as^{(1)}(\mu) \left( 1 + c_1(\mu) \as^{(1)}(\mu)
                + c_2(\mu) (\as^{(1)}(\mu))^2 \right), \\
   \bar{m}(\mu) &=& \bar{m}^{(1)}(\mu) \left( 1 + d_1(\mu) \as^{(1)}(\mu)
                + d_2(\mu) (\as^{(1)}(\mu))^2 \right), \label{massrun} 
\eea 
where $\as^{(1)}(\mu)$ and $\bar{m}^{(1)}(\mu)$ are the one loop 
solutions 

\beq
   \as^{(1)}(\mu) = \frac{\as(\mu_0)}{1+\as(\mu_0) \beta_0 t}, \qquad
   \bar{m}^{(1)}(\mu) = \bar{m}(\mu_0) K(\mu)^{-2\gamma_0/\beta_0},
\end{equation}
with $t = 1/(4\pi) \log \mu^2/\mu_0^2$, $K(\mu)$ the ratio 
$K(\mu) = \as(\mu_0)/\as^{(1)}(\mu)$, and 

\bea
& &c_1(\mu) = - b_1 \log K(\mu), \nonumber \\
& &c_2(\mu) = b_1^2 \log K(\mu) \left[\log K(\mu) - 1 \right] 
               - (b_1^2-b_2) \left[ 1 - K(\mu) \right], \nonumber \\
& &d_1(\mu) = - \frac{2 \gamma_0}{\beta_0} \left[
             (b_1-g_1)\left[1-K(\mu)\right] + b_1 \log K(\mu) \right], \\
& &d_2(\mu) = \frac{\gamma_0}{\beta_0^2} \biggl\{
             [ \beta_0(b_2-b_1^2) + 2\gamma_0(b_1-g_1)^2 ] 
               \left[1-K(\mu)\right]^2  \nonumber\\
           &+& \beta_0 (g_2-b_1 g_1) \left[1-K^2(\mu)\right] \nonumber \\
           &+& \biggl[ 4\gamma_0 b_1 (b_1-g_1) \left[1-K(\mu)\right] 
             - 2 \beta_0 b_1 g_1 + b_1^2(\beta_0+2\gamma_0) \log K(\mu) 
             \biggr] \log K(\mu) \biggr\}, \nonumber
\eea
where
\bea
   b_1 &=& \frac{\beta_1}{4\pi \beta_0}, \qquad
   b_2 = \frac{\beta_2}{(4\pi)^2 \beta_0}, \qquad
   g_1 = \frac{\gamma_1}{\pi \gamma_0}, \qquad
   g_2 = \frac{\gamma_2}{\pi^2 \gamma_0}, 
\eea
are the ratios of the well known beta and gamma functions in
the $\overline{MS}$ scheme\index{Beta function}\index{Gamma function}
\index{Renormalization!scheme!$\overline{MS}$}
\bea
     \beta_0&=&11 - \frac{2}{3} N_F, \qquad    \gamma_0=2, \nonumber \\
     \beta_1&=&102 - \frac{38}{3} N_F, \qquad
     \beta_2 = \frac{1}{2} \left( 2857 - \frac{5033}{9} N_F 
                 + \frac{325}{27} N_F^2 \right), \nonumber \\
     \gamma_1&=&\frac{101}{12} - \frac{5}{18} N_F, \qquad 
     \gamma_2 = \frac{1249}{32} - \frac{277 + 180 \zeta(3)}{108} N_F
                 - \frac{35}{648} N_F^2, 
\label{beta}
\eea
and $\zeta(3) = 1.2020569 \dots$ is the 
Riemann zeta-function\index{Riemann zeta-function}. 
Our initial condition for the strong coupling constant will be 
$\as(m_Z)$. Then we will run $\as$ from $m_Z$ to lower scales,
i.e. for instance $\mu_0 = m_Z$ or the upper threshold.\index{Threshold}
On the other side, we will run the masses from low to higher
scales. We need the inverted version of \rfn{massrun} 
\beq
    \bar{m}(\mu_0) = \bar{m}(\mu) K(\mu)^{2\gamma_0/\beta_0}
\left( 1 - d_1(\mu) \as^{(1)}(\mu)
         + ( d_1^2(\mu)-d_2(\mu) ) (\as^{(1)}(\mu))^2 \right).
\end{equation}

     The beta and gamma functions depend on the 
number of flavours $N_F$. Therefore, we have to decide whether
we have five or four flavours.
The trick~\cite{BE1,BE2}~\footnote{See also~\cite{RO93}.
The mass independent constant coefficient of 
the $O(\as^3)$ matching condition for the strong coupling 
constant has been recently corrected by~\cite{LA}.
Nevertheless, the numerical effect of this correction is very
small.} is to built below the heavy quark threshold\index{Threshold}
an effective theory\index{Effective Theory} where the heavy 
quark has been integrated out. Imposing agreement of both 
theories, the full and the effective one, at low energies 
they wrote $\mu$ dependent 
matching conditions\index{Matching conditions} that 
express the parameters of the effective theory, with $N-1$
quark flavours, as a perturbative expansion in terms of the 
parameters of the full theory with $N$ flavours

%%%%%%%%%%%%%%%%%%%%%%%%%%%%%%%%%%%%%%%%%%%

\begin{table}[hbtp]
\begin{center}
\caption{Running at the NLO and NNLO of the top quark mass to $m_Z$,  
$\as^{(5)}(m_Z) = 0.118 \pm 0.006$, $\as^{(6)}(m_Z) = 0.117 \pm 0.006$.
\label{heavytop}}
\begin{tabular}{|l|ccc|} \hline 
     & $\quad \bar{m}_t(M_t) (GeV)       \quad$ 
     & $\quad \bar{m}_t(\bar{m}_t) (GeV) \quad$ 
     & $\quad \bar{m}_t(m_Z) (GeV)       \quad$   \\ \hline
NLO  & $167.\pm6.$ & $168.\pm6.$ & $176.\pm6.$ \\
NNLO & $165.\pm6.$ & $166.\pm6.$ & $174.\pm7.$ \\ \hline
\end{tabular}
\end{center}
\end{table}

%%%%%%%%%%%%%%%%%%%%%%%%%%%%%%%%%%%%%%%%%%%

\begin{table}[hbtp]
\begin{center}
\caption{Running at the NLO of the bottom quark mass to $m_Z$ 
and running masses at the running mass scale needed for thresholds.
For masses extracted from QCD SR 
$\as^{(5)}(m_Z) = 0.118 \pm 0.006$, for lattice
$\as^{(5)}(m_Z) = 0.115 \pm 0.002$
\label{heavyb}}
\begin{tabular}{|c|cc|} \hline 
           & $\qquad \bar{m}_b(\bar{m}_b) (GeV) \qquad$ 
           & $\qquad \bar{m}_b(m_Z)       (GeV) \qquad$ \\ \hline

Narison    & $4.29 \pm 0.04$ & $2.97 \pm 0.13$ \\  
        
Narison    & $4.35 \pm 0.05$ & $3.03 \pm 0.13$ \\  

Titard / Yndur\'ain 
           & $4.397^{+0.018}_{-0.033}$ & $3.07 \pm 0.11$ \\  

Neubert    & $4.37 \pm 0.09$ & $3.04 \pm 0.17$ \\  \hline 

Crisafulli et al.
           & $4.17 \pm 0.06$ & $2.93 \pm 0.08$ \\

Gim\'enez  et al. 
           & $4.15 \pm 0.20$ & $2.91 \pm 0.19$ \\

Davies et al. 
           & $4.13 \pm 0.11$ & $2.89 \pm 0.12$ \\ \hline 

    $mean$ & $4.33 \pm 0.06$ & $3.00 \pm 0.12$ \\ \hline 
\end{tabular}
\end{center}
\end{table}

%%%%%%%%%%%%%%%%%%%%%%%%%%%%%%%%%%%%%%%%%%%

\begin{table}[hbtp]
\begin{center}
\caption{Running at the NLO of the charm quark mass to $m_Z$ 
and running masses at the running mass scale needed for thresholds.
For masses extracted from QCD SR 
$\as^{(5)}(m_Z) = 0.118 \pm 0.006$, for lattice
$\as^{(5)}(m_Z) = 0.115 \pm 0.002$
\label{heavyc}}
\begin{tabular}{|c|cc|} \hline 
           & $\qquad \bar{m}_c(\bar{m}_c) (GeV) \qquad$ 
           & $\qquad \bar{m}_c(m_Z) (GeV)       \qquad$ \\ \hline

Narison    & $1.28 \pm 0.04$ & $0.52 \pm 0.09$ \\ 
        
Narison    & $1.31 \pm 0.06$ & $0.54 \pm 0.10$ \\ 

Titard / Yndur\'ain 
           & $1.306^{+0.022}_{-0.035}$ & $0.52 \pm 0.08$ \\ 

Neubert    & $1.17 \pm 0.12$ & $0.45 \pm 0.14$ \\ \hline 

El-Khadra et al. 
           & $1.36 \pm 0.19$ & $0.61 \pm 0.15$ \\ \hline  

    $mean$ & $1.30 \pm 0.08$ & $0.52 \pm 0.10$ \\ \hline 
\end{tabular}
\end{center}
\end{table}

%%%%%%%%%%%%%%%%%%%%%%%%%%%%%%%%%%%%%%%%%%%

%%%%%%%%%%%%%%%%%%%%%%%%%%%%%%%%%%%%%%%%%%%

\begin{table}[hbtp]
\begin{center}
\caption{Running of the light quark masses to $m_Z$. For masses
extracted from QCD SR  $\as^{(5)}(m_Z) = 0.118 \pm 0.006$,
for lattice$^*$ $\as^{(5)}(m_Z) = 0.115 \pm 0.002$. First box is NLO running,
second and third boxes are NNLO running.
\label{light}}
\begin{tabular}{|c|cc|} \hline 
\qquad \qquad \qquad \qquad \qquad & 
$\qquad \bar{m}_s(1GeV) (MeV) \qquad$ &
$\qquad \bar{m}_s(m_Z)  (MeV) \qquad$ \\ \hline
Narison       & $222. \pm 22.$ & $105. \pm 28.$ \\  
Allton et al.$^*$ & $156. \pm 17.$ & $ 78. \pm 15.$ \\  \hline
\end{tabular}

\vspace{.5cm}

\begin{tabular}{|c|cc|} \hline 
\qquad \qquad \qquad \qquad \qquad & 
$\qquad \bar{m}_s(1GeV) (MeV) \qquad$ &
$\qquad \bar{m}_s(m_Z)  (MeV) \qquad$ \\ \hline
Narison          & $197. \pm 29.$ & $88. \pm 31.$ \\  
Jamin / M\"unz   & $189. \pm 32.$ & $85. \pm 32.$ \\ 
Chetyrkin et al. & $171. \pm 15.$ & $75. \pm 23.$ \\ \hline 
$mean$           & $186. \pm 30.$ & $83. \pm 30.$ \\ \hline 
\end{tabular}

\vspace{.5cm}

\begin{tabular}{|cc|} \hline 
Bijnens et al. & Ioffe et al. \\  \hline
$(\bar{m}_u+\bar{m}_d)(1GeV) = (12.0 \pm 2.5) MeV$ & 
$(\bar{m}_d-\bar{m}_u)(1GeV) = ( 3.  \pm 1. ) MeV$ \\
$(\bar{m}_u+\bar{m}_d)(m_Z)  = ( 5.4 \pm 2.2) MeV$ &
$(\bar{m}_d-\bar{m}_u)(m_Z)  = ( 1.4 \pm 0.7) MeV$ \\  \hline
\end{tabular}
\end{center}
\end{table}

%%%%%%%%%%%%%%%%%%%%%%%%%%%%%%%%%%%%%%%%%%%

\bea 
   \as^{N-1}(\mu) &=& \as^N(\mu) \left[ 1 
        + \frac{x}{6} \frac{\as^N(\mu)}{\pi}
        + \frac{1}{12} 
\left(\frac{x^2}{3} + \frac{11x}{2} + \frac{11}{6} \right) 
\left(\frac{\as^N(\mu)}{\pi}\right)^2 \right], \nonumber \\
   \bar{m}^N_l(\mu) &=& \bar{m}^{N-1}_l(\mu) \left[ 1 
        - \frac{1}{12} \left( x^2 + \frac{5x}{3} + \frac{89}{36} \right) 
\left(\frac{\as^{N-1}(\mu)}{\pi}\right)^2 \right],
\label{continuous} 
\eea
with $x = \log \bar{m}^2(\mu)/\mu^2$,
where $\bar{m}(\mu)$ is the heavy quark mass which decouples at the 
energy scale $\mu$ and $\bar{m}_l(\mu)$ are the lighter quark masses.
These matching conditions make the strong coupling
constant and the light quark masses discontinuous at  
thresholds.\index{Threshold} 
However, taking the matching in this way we ensure,
as pointed out explicitly in \cite{RO93}, that the final result
is independent of the particular matching point we choose for 
passing the threshold. As it is independent, the easiest way to implement
a heavy quark decoupling\index{Decoupling}
is to take the threshold scale
as the running mass at the running mass scale, i.e., 
$\mu_{th} = \bar{m}(\bar{m})$ or equivalently $x=0$,
then the discontinuity appears only at two loops matching.

     We have summarized in tables \ref{heavytop}, 
\ref{heavyb}, \ref{heavyc} and \ref{light}
the results for the quark masses running until $m_Z$. By NLO we mean 
connection between the perturbative pole mass and the running mass
dropping the $O(\as^2)$ term, running to two loops and matching at
one loop, i.e., strong gauge coupling and masses continuous at
$\mu_{th} = \bar{m}(\bar{m})$. Three loop running and matching
as expressed in \eq{continuous} with $x=0$ correspond to NNLO.
For consistency with the original works we perform the 
bottom and the charm quark mass running just at NLO. For the 
light quarks the running is consistent to NNLO using the threshold masses,
$\bar{m}(\bar{m})$, of the bottom and charm quarks determined at NLO.
\index{Threshold}

We propagate the errors in the running in such a way we
maximize them. The relative uncertainty in the strong coupling
constant decreases in the running from low to 
high energies as the ratio of the strong coupling constants
at both scales. On the contrary, in the way we propagate 
the errors, induced by the strong coupling constant error,
the quark mass relative uncertainty increases following
\beq
\varepsilon_r (\bar{m}(m_Z)) \simeq 
\varepsilon_r (\bar{m}(\mu)) 
+ \frac{2 \gamma_0}{\beta_0}
\left(\frac{\as(\mu)}{\as(m_Z)}-1 \right)
\varepsilon_r (\as(m_Z)).
\end{equation}
The absolute quark mass propagated error value at high energies
depends on the balance between the quark mass at low 
energies and the strong coupling constant errors.
Case the quark mass error contribution dominates the relative 
uncertainty remains almost fixed and since the running mass decreases 
at high energies its absolute error decreases too.
Case the strong coupling constant error contribution 
is the biggest the absolute quark mass uncertainty increases.
In this last situation,
a possible evaluation of the bottom quark mass
at the $m_Z$ scale would be considered even competitive to
low energy QCD Sum Rules and lattice calculations
with smaller errors.

We have depicted in figure~\ref{running}.
the running of the bottom quark mass.
we run $\bar{m}_b(\bar{m}_b)=4.39(GeV)$ with 
$\as (m_Z)=0.112$ and $\bar{m}_b(\bar{m}_b)=4.27(GeV)$ with 
$\as (m_Z)=0.124$, the extreme quark mass and strong coupling
constant values for $\bar{m}_b(\bar{m}_b)=(4.33\pm 0.06) GeV$ 
and $\as (m_Z)=0.118 \pm 0.006$, and we take the difference as
the propagated error. 
In this case, since the strong coupling constant uncertainty is 
the biggest, the absolute bottom quark mass uncertainty 
increases in the running to $m_Z$.

%%%%%%%%%%%%%%%
\mafigura{6.5cm}{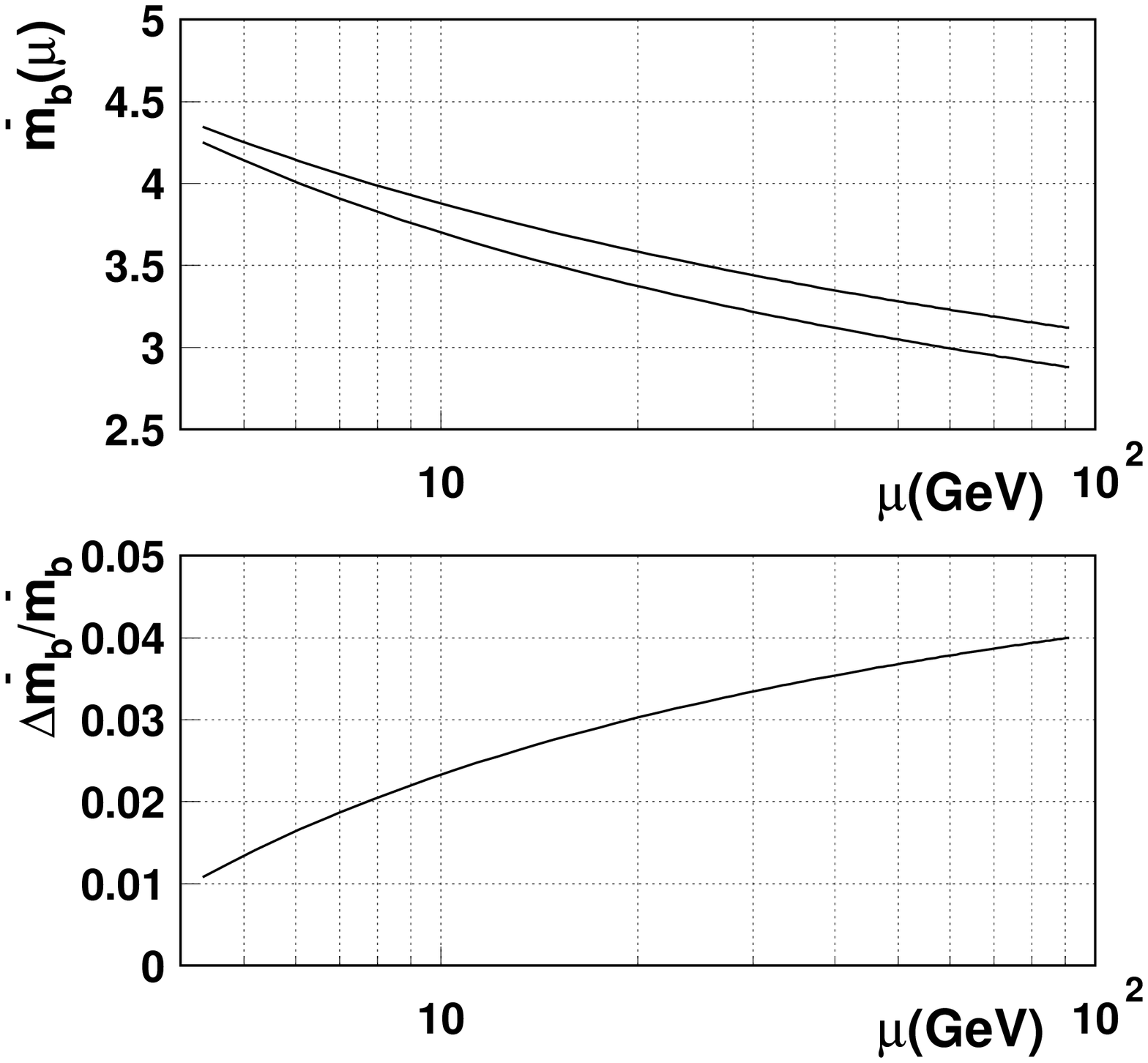}
{Running of the bottom quark mass from low energies to the
$m_Z$ scale.
Upper line is the run of $\bar{m}_b(\bar{m}_b)=4.39(GeV)$ with 
$\as (m_Z)=0.112$. Bottom line is the run of  
$\bar{m}_b(\bar{m}_b)=4.27(GeV)$ with $\as (m_Z)=0.124$.
Second picture is the difference of both, our estimate for
the propagated error.}
{running}
%%%%%%%%%%%%%%%
 
     It is informative to notice that the running mass of the
top quark\index{Top quark} is shifted about $7(GeV)$ down
from its perturbative pole mass. This shift is of the order
of its experimental error. Therefore it is important
to clarify which mass is measured at the CDF and D\O\ experiments.
\index{CDF}\index{D\O\} 
We have decoupled the top quark at $m_Z$ otherwise it makes no sense
to run the top down. This fact shifts down slightly the strong coupling
constant in $m_Z$, from $\as^{(5)}(m_Z) = 0.118 \pm 0.006$ we get 
$\as^{(6)}(m_Z) = 0.117 \pm 0.006$ but has no effect on the masses 
because the errors screen the difference between the theory with 
5 and 6 flavours. Curiously the running of the top until $m_Z$ cancels
the difference between the running and the pole mass,
$\bar{m}_t(m_Z)\sim M_t$.

     One has to be very careful in comparing the running of the masses 
obtained from QCD Sum Rules and those obtained from lattice 
because we took different values for the strong coupling constant 
at $m_Z$. Furthermore, we have to remember that the error in the
running is dominated by the error in the strong coupling constant.
However it is impressive to notice the good agreement  
of the results obtained in lattice~\cite{DA2,CRI}
with the running of the masses from QCD Sum Rules.
Nevertheless, we have to keep in mind that  
a recent lattice evaluation~\cite{VG1,VG2} 
has enlarged the initial estimated error on the bottom quark 
mass up to $200 MeV$, $\bar{m}_b(\bar{m}_b) = (4.15 \pm 0.20) GeV$,
due to unknown higher orders in the perturbative matching of the 
Heavy Quark Effective Theory (HQET)\index{HQET}
to the full theory.
 
     We can now play the game of combining the light quark masses 
of table~\ref{light} with the ratios obtained from $\chi$PT. The mean 
value of the strange quark mass together with the Bijnens et al.~\cite{BI} 
result gives 
\beq 
   \frac{2 \bar{m}_s}{\bar{m}_u+\bar{m}_d} = 33. \pm 12.,
\end{equation}
in agreement with the $\chi$PT result. Being conservative we 
can also get for the up and down quarks 
$\bar{m}_u(1GeV) = (3. \pm 2.) MeV$ and $\bar{m}_d(1GeV) = (9. \pm 2.) MeV$  
that translate into $\bar{m}_u(m_Z) = (1.5 \pm 1.2) MeV$ and 
$\bar{m}_d(m_Z) = (4.1 \pm 1.7) MeV$.

     To summarize, the running of the quark masses to the $m_Z$ energy 
scale gives a running top mass that is around its perturbative pole 
mass, $\bar{m}_t(m_Z) = (176. \pm 6.)GeV$, for the bottom and 
the charm quarks we get $\bar{m}_b(m_Z) = (3.00 \pm 0.12) GeV$ and
$\bar{m}_c(m_Z) = (0.52 \pm 0.10) GeV$ respectively, while for the
strange quark we have a result affected by a big error
$\bar{m}_s(m_Z) = (83. \pm 30.)MeV$. The same happens for the 
up and down quarks, we get $\bar{m}_u(m_Z) = (1.5 \pm 1.2) MeV$ and 
$\bar{m}_d(m_Z) = (4.1 \pm 1.7) MeV$.

%\documentstyle[12pt,times,cite,epsf]{article}
%       LOCAL DEFINITIONS
%\newcommand{\elltbar}{\overline{\widetilde{\ell}\, }}
%\newcommand{\com}[1]{{\it \bf #1}\\}
%\newcommand{\mysection}[1]{\setcounter{equation}{0}\section{#1}}
%\renewcommand{\theequation}{\thesection.\arabic{equation}}
%\newcommand{\mathbold}[1]{\mbox{\rm\bf #1}}
\newcommand{\mrm}[1]{\mbox{\rm #1}}
\newcommand{\half}{{1\over 2}}
\newcommand{\bla}{\hspace{1cm}}
\newcommand{\db}{\hspace{-0.2ex}\not\hspace{-0.7ex}D\hspace{0.1ex}}
\newcommand{\sla}[1]{\hspace{-0.1ex}\not\hspace{-0.5ex} #1\hspace{0.1ex}}
\newcommand{\delte}{\Delta_\epsilon}
\newcommand{\gh}{\frac{g}{4c_W}}
\newcommand{\zbb}{Z \rightarrow b\bar{b}}
\newcommand{\zbbg}{Z \rightarrow b\bar{b}g}
\newcommand{\ap}{\frac{\alpha_s}{\pi}}
\newcommand{\cs}{\frac{4\alpha_s}{3\pi}}
\newcommand{\yc}{y_{c}}
\newcommand{\rb}{r_b}
\newcommand{\za}{z_\alpha}
\newcommand{\zb}{z_\beta}
\newcommand{\zg}{z_\gamma}
\newcommand{\be}{\beta}
\newcommand{\el}[1]{^{(#1)}}
\newcommand{\abs}[1]{\left| #1\right|}
\renewcommand{\Re}[1]{\mathop{\mrm{Re}}\left\{ #1 \right\}}
%\newcommand{\vev}[1]{\left\langle #1\right\rangle}
%\newcommand{\bra}[1]{\left\langle #1\right|}
%\newcommand{\ket}[1]{\left| #1\right\rangle}
%\newcommand{\lrover}[1]{
%      \raisebox{1.3ex}{\rlap{$\leftrightarrow$}} \raisebox{ 0ex}{$#1$}}
%%%%%%%%%%%%%%%%%%%%%%%%%%%
%\hoffset-1in
%\voffset-1in
%\oddsidemargin25mm
%\evensidemargin25mm
%\marginparwidth25mm
%\marginparsep 0pt
%\topmargin 2cm
%\headheight 0pt
%\headsep 0pt
%\footheight 12pt
%\footskip 30pt
%\textwidth 16cm
%\textheight 25cm
%%%%%%%%%%%%%%%
\def\mafigura#1#2#3#4{
  \begin{figure}[hbtp]
    \begin{center}
      \epsfxsize=#1
      \leavevmode
      \epsffile{#2}
    \end{center}
    \caption{#3}
    \label{#4}
  \end{figure} }
%%%%%%%%%%%%%%%

\chapter{Three-jet observables at leading order}

%\section{Introduction}

As we mentioned in the previous chapter
in the Standard Model\index{Standard Model} 
of electroweak interactions
all fermion masses are free parameters, 
and their origin, although linked to the 
spontaneous symmetry breaking\index{Spontaneous symmetry breaking}
mechanism, remains secret.
Masses of charged leptons are well measured experimentally and 
neutrino masses, if they exist, are also bounded. 
In the case of quarks the situation is more complicated because free
quarks are not observed in nature. 
Therefore, one can only get some indirect information on the values
of the quark masses. For light quarks
($m_q < 1$~GeV, the scale at which QCD interactions become
strong), that is, for $u$-,$d$- and $s$-quarks, one can 
define the quark masses as the parameters of the Lagrangian that break
explicitly the chiral symmetry\index{Chiral symmetry}
of the massless QCD Lagrangian. Then,
these masses can be extracted from a careful analysis of meson
spectra and meson decay constants. For heavy quarks ($c$- and $b$-quarks)   
one can obtain the quark masses from the known spectra of the hadronic
bound states by using, e.g., QCD sum rules\index{Sum Rules}
or lattice\index{Lattice} calculations. However,
since the strong gauge coupling constant
is still large at the scale of heavy
quark masses, these calculations are plagued by uncertainties and
nonperturbative effects.\index{Non-perturbative effects}

It would be very interesting to have some experimental
information on the quark masses obtained at much larger scales
where a perturbative quark mass definition can be used and,
presumably, non-perturbative effects are negligible. 
The measurements at LEP\index{LEP} will combine
this requirement with very high experimental statistics.

The effects of quark masses can be neglected for many
observables in LEP studies, 
as usually quark masses appear in the ratio
$m_q^2/m_Z^2$. For the bottom quark, the heaviest
quark produced at LEP, and taking 
a $b$-quark mass of about 5~GeV this ratio is $0.003$, even
if the coefficient in front is 10 we get a correction of about 3\%. 
Effects of this order are measurable at LEP, however, as we will see 
later, in many cases the actual mass that should be used in the
calculations is the {\it running} mass of the $b$-quark computed at
the $m_Z$ scale: $\bar{m}_b(m_Z) \approx 3$~GeV rendering the effect
below the LEP precision for most of the observables. 

While this argument
is correct for total cross sections for production of $b$-quarks it
is not completely true for quantities that depend on other variables.
In particular it is not true for jet cross sections which depend
on a new variable, $y_c$ 
(the jet-resolution parameter\index{Jet!resolution parameter}
that defines the jet multiplicity) and
which introduces a new scale in the analysis, $E_c=m_Z\sqrt{y_c}$. Then,
for small values of $y_c$ there could be contributions coming like
$m_b^2/E_c^2 = (m_b/m_Z)^2 / y_c$ which could enhance the mass effect
considerably. In addition mass effects could also be enhanced by 
logarithms of the mass. For instance, the ratio of the phase space
for two massive quarks and a gluon to the phase 
space for three massless
particles is $1+8 (m_q/m_Z)^2 \log(m_q/m_Z)$. 
This represents a 7\% effect 
for $m_q=5$~GeV and a 3\% effect for $m_q=3$~GeV. 

The high precision achieved at LEP\index{LEP}
makes these effects relevant.
In fact, they have to be taken into account in the
test of the flavour independence of
$\as(m_Z)$~\cite{l3,delphi,opal,aleph,chrin}. 
In particular it has been shown~\cite{juano} that the biggest systematic
error in the measurement of 
$\alpha_s^b(m_Z)$\index{alphasb@$\alpha_s^b(m_Z)$} 
($\alpha_s$ obtained 
from $b\bar{b}$-production at LEP from the ratio of three to two jets) 
comes from the uncertainties in the estimate of the quark mass effects.
This in turn means that mass effects have already been seen. 
Now one can reverse the question
and ask about the possibility of measuring the mass of the bottom
quark, $m_b$, at LEP by assuming the 
flavour universality\index{Flavour universality}
of the strong interactions.

Such a measurement
will also allow to check the running of 
$\bar{m}_b(\mu)$ from $\mu=m_b$ to $\mu=m_Z$ as has been done before
for $\alpha_s(\mu)$. In addition
$\bar{m}_b(m_Z)$ is the crucial input parameter
in the analysis of the unification\index{Unification} of 
Yukawa couplings\index{Yukawa coupling}
predicted by many grand unified theories
and which has attracted much attention in the last years~\cite{guts}.

The importance of quark mass effects in $Z$-boson decays
has already been discussed in the literature~\cite{rev1,rev2}.
The complete order $\as$ results for
the inclusive decay rate of $Z\rightarrow b\bar{b}+b\bar{b} g+\cdots$
\index{Inclusive decay rate}
can be found\footnote{The order $\as$ corrections
to the vector part, including
the complete mass dependences, were already known from
QED\index{QED} calculations~\cite{schwinger}.}
in~\cite{complas1,complas2,complas3}.
The total cross section and forward-backward asymmetry for
$e^+e^-  \rightarrow \gamma^*, Z \rightarrow
q \bar{q}$ with one-loop QCD corrections
were calculated in~\cite{zerwas}. Analytical results with cuts
for these quantities
were obtained in~\cite{bardin}.
The leading quark mass effects
for the inclusive $Z$-width are known
to order $\as^3$ for the vector part~\cite{kuhn1}
and to order $\as^2$ for the axial-vector part~\cite{kuhn2}. 
Quark mass effects for three-jet final states in the process
$e^+e^-$ annihilation into $q\bar{q}g$ were 
considered in~\cite{gjets1,gjets2,gjets3}
for the photonic channel and extended
later to the $Z$ channel in~\cite{zjets1,zjets2}.
Polarization effects have been studied for instance  
in~\cite{Tung1,Tung2,Kuhnpol} for massive quarks.
In~\cite{grun} the possibility of the measurement of the 
quark mass from the angular distribution of heavy-quark jets
in  $e^+e^-$-annihilation was discussed.
Recently~\cite{Ballestrero92,Ballestrero94}
calculations of the three-jet event rates,
including mass effects, were done for the most popular jet
clustering algorithms using the Monte Carlo approach.\index{Monte Carlo}

We will discuss the possibility of measuring 
the $b$-quark mass at LEP, in particular,
we will study bottom quark mass effects in $Z$ decays into 
two and three jets.

\nocite{jade}

\section{The inclusive decay rate $\zbb$}\index{Inclusive decay rate}

\label{inclusivesection}

To calculate at order $\as$ the total decay rate of the
$Z$-boson into massive quarks one has
to sum up the virtual one-loop gluonic corrections to 
$Z \rightarrow b \bar{b}$ with the real 
gluon bremsstrahlung,\index{Gluon!bremsstrahlung}
see figure~\ref{feynman}.
In addition to renormalized UV divergences\index{UV divergences}
\footnote{Note that conserved currents or partially conserved
currents as the vector and axial currents do not get renormalized. 
Therefore, all UV divergences cancel when one sums properly self-energy
and vertex diagrams. The remaining poles in $\epsilon$ correspond to
IR divergences. 
One can see this by separating carefully the poles 
corresponding to UV divergences from the poles corresponding to IR 
divergences.},
IR singularities, either collinear or soft,
\index{IR singularities}
\index{IR singularities!collinear}
\index{IR singularities!soft}
appear because of the presence of massless
particles like gluons. Therefore, some regularization
method for the IR divergences is needed.
Bloch-Nordsiek~\cite{IR1} and Kinoshita-Lee-Nauenberg~\cite{IR2,IR3}
\index{Bloch-Nordsiek theorem}
\index{Kinoshita-Lee-Nauenberg theorem}
theorems assure IR divergences cancel for inclusive cross sections.
Technically this means, if we use 
Dimensional Regularization\index{Dimensional Regularization}
to regularize the IR divergences~\cite{dimreg1,dimreg2,dimreg3}
of the loop diagrams we 
should express the phase space for the tree-level diagrams
in arbitrary dimension $D=4-2\epsilon$.
At order $\as$ and for massive quarks all IR divergences
appear as simple poles $1/\epsilon$.
The IR singularities cancel when we integrate over the
full phase space.

%%%%%%%%%%%%%%%
\mafigura{10cm}{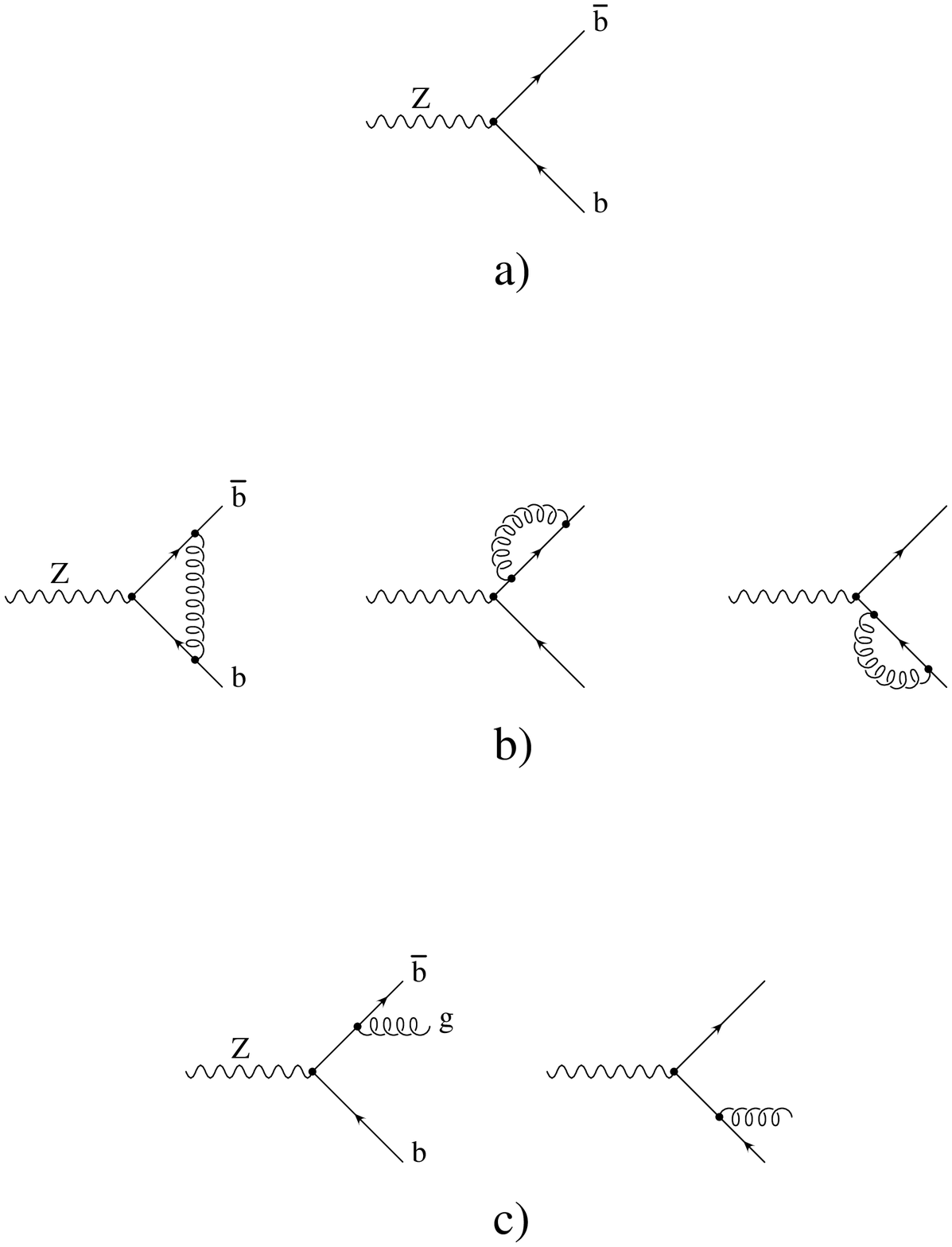}
{Feynman diagrams contributing to the decay rates
$Z\rightarrow b\bar{b}$, $Z\rightarrow b\bar{b} g$
 at order $\as$.}
{feynman}
%%%%%%%%%%%%%%%

At the end, we obtain the well-known result~\cite{djuadi}
\begin{multline}
\label{inclusiveL}
\Gamma_b =
m_Z \frac{g^2}{c_W^2 64 \pi} \left[
g_V^2  \left(
1+\frac{\as}{\pi}(1+12 r_b)
\right)  \right. \\ \left.
+ g_A^2 \left(
1-6 r_b + \frac{\as}{\pi}\left(1-6 r_b (2\log r_b+1) \right)
\right) \right]~.
\end{multline}
where $r_b=m_b^2/m_Z^2$ and $g_V$ ($g_A$)
are the vector (axial-vector) neutral current
\index{Neutral current!vector coupling} 
\index{Neutral current!axial-vector coupling} 
couplings of the quarks in the Standard Model. 
At tree level and for the $b$-quark we have
\begin{equation}
g_V=-1+\frac{4}{3} s^2_W~, \bla
g_A= 1~.
\end{equation}
We denote by $c_W$ and $s_W$ the cosine and the sine of
the weak mixing angle.
Here and below we will conventionally use 
$\as=\as(m_Z)$\index{alphas@$\as(m_Z)$} to
designate the value of the running strong
coupling constant at the $m_Z$-scale.

It is interesting to note the presence
of the large logarithm, $\log(m_b^2/m_Z^2)$, proportional to 
the quark mass in the axial part of the QCD corrected
width, \eq{inclusiveL}. The mass that appears in all above
calculations should be interpreted as 
the perturbative {\it pole} mass\index{Quark mass!perturbative pole}
of the quark. But in principle the
expression \rfn{inclusiveL} could also be
written in terms of the so-called
{\it running} quark mass\index{Quark mass!running}
at the $m_Z$ scale by using 
\begin{equation}
\label{poltorunning}
m_b^2 = \bar{m}_b^2(m_Z) \left[1+2\frac{\as}{\pi}
\left(\frac{4}{3}-\log \frac{m_b^2}{m_Z^2} \right)\right]~.
\end{equation}
Then, we see that all large logarithms are absorbed 
in the running of the quark mass from the $m_b$ scale
to the $m_Z$ scale~\cite{kuhn1} and we have
\begin{multline}
\label{inclusiveL2}
\Gamma_b =
m_Z \frac{g^2}{c_W^2 64\pi} \left[
g_V^2  \left(
1+\frac{\as}{\pi}(1+12\bar{r}_b)\right)
\right. \\ \left.
+ g_A^2  \left(
1-6\bar{r}_b +
\frac{\as}{\pi}(1-22\bar{r}_b)
\right) \right]~,
\end{multline}
where $\bar{r}_b = \bar{m}_b^2(m_Z)/m_Z^2$. 

This result means that the bulk of the QCD\index{QCD}
corrections depending on the mass could be accounted for
by using tree-level expressions for the decay
width but interpreting the quark mass as the running mass.
The same point has been stressed in~\cite{Mendez90}
for the hadronic width of the charged Higgs boson.
On the other
hand, since $\bar{m}_b(m_Z) \approx 3$~GeV is much smaller than the
pole mass, $m_b \approx 5$~GeV, it is clear that the quark mass
corrections are much smaller than expected from the na\"{\i}ve use
of the tree-level result with  $m_b \approx 5$~GeV, which would give mass 
corrections at the 1.8\% level while in fact, once QCD corrections are
taken into account, the mass corrections are only at the 0.7\% level. 

In fact, this remarkable nice feature of the running mass 
holds to all orders in perturbation
theory~\cite{Chetyrkin82,Chetyrkin94}, i.e.,
all the potentially dangerous terms of the type $m^2 \log m^2/s$
can be absorbed in the 
$\overline{MS}$ scheme.\index{Renormalization!scheme!$\overline{MS}$}

\section{Jet clustering algorithms}

\index{Jet!clustering algorithms}

According to our current understanding of the strong interactions,
coloured partons\index{Parton}, produced  in hard processes,
are hadronized and, at experiment, one only observes colourless particles.
It is known empirically that, in
high energy collision, final particles group in several
clusters by forming energetic jets, 
which are related to the primordial partons. 
Thus, in order to compare theoretical predictions 
with  experiments, it is necessary
to define precisely what is a jet in both, parton level calculations
and experimental measurements.

At order $\as$, the decay
widths of $Z$ into both two and three partons are IR divergent.
The two-parton decay rate is divergent
due to the massless gluons running in the loops.
The $Z$-boson decay width into three-partons has an IR divergence
because massless gluons could be radiated with zero energy.
The sum, however, is IR finite.
Then it is clear that at the parton-level one can define
an IR finite {\it two-jet decay rate}, 
by summing the two-parton decay rate
and the IR divergent part of the three-parton decay width, e.g. 
integrated over the part of the phase space which contains soft
gluon emission~\cite{sw}. The integral over the rest of the phase
space will give the {\it three-jet decay rate}.
Thus we need to introduce a 
``resolution parameter''\index{Jet!resolution parameter}
in the theoretical calculations in order to define  IR-safe observables.
Obviously, the resolution parameter, which defines the two- and 
the three-jet parts of the three-parton 
phase space should be related to the one
used in the process of building jets from real particles.

\begin{table}
\begin{center}
\caption{The jet-clustering algorithms 
\label{table1}}
\begin{tabular}{lll}
\hline
Algorithm & Resolution & Combination \\
\hline
EM & $2(p_i \cdot p_j)/s$ & $p_k= p_i+p_j$ \\
JADE & $2(E_i E_j) (1-\cos \theta_{ij})/s $ & $p_k= p_i+p_j$ \\ 
E & $(p_i+p_j)^2/s$ & $p_k = p_i+p_j$ \\
DURHAM  & $2 \min(E_i^2,E_j^2) (1-\cos \theta_{ij})/s \ \ $ & 
$p_k = p_i+p_j$\\
\hline
\end{tabular}
\end{center}    
\end{table}

In the last years the most popular definitions of jets
are based on the so-called jet clustering algorithms. 
These algorithms can be applied at the parton
level in the theoretical calculations and also to the
bunch of real particles observed at experiment.
It has been shown that, for some of the algorithms,
the passage from partons to hadrons (hadronization)\index{Hadronization}
does not change much the behaviour of the observables~\cite{Bethke92},
thus allowing to compare theoretical predictions
with experimental results. In what follows we will use the word
particles for both partons and real particles.

In the jet-clustering algorithms jets are defined as follows:
starting from  a bunch of particles
with momenta $p_i$ one computes, for example, a quantity like
\[
y_{ij}=2 \frac{E_i E_j}{s} (1-\cos \theta_{ij})
\]
for all pairs $(i,~j)$ of particles. Then one takes the minimum of
all $y_{ij}$ and if it satisfies that it is smaller than a given quantity 
$y_c$ (the resolution parameter, y-cut)  the two particles 
which define this $y_{ij}$ are regarded
as belonging to the same jet, therefore, they are recombined into a new
pseudoparticle by defining the four-momentum of the pseudoparticle according
to some rule, for example
\[
p_k = p_i + p_j~.
\]
After this first step one  has a bunch of pseudoparticles and
the algorithm can be applied again and again until all the pseudoparticles
satisfy  $y_{ij} > y_c$. 
The number of pseudoparticles found in the end
is the number of jets in the event.

Of course, with such a jet definition the number of jets found in an
event and its whole topology will depend on the value of $y_c$. 
For a given event, larger 
values of $y_c$ will result in a smaller number of jets.
In theoretical calculations one can define
cross sections or decay widths into jets as a function of
$y_c$, which are
computed at the parton level, by following exactly the same algorithm.
This procedure leads automatically to IR finite quantities 
because one excludes the regions of phase space that cause trouble.
The success of the jet-clustering algorithms is due, mainly,
to the fact that the cross sections
obtained after the hadronization\index{Hadronization}
process agree quite well with 
the cross-sections calculated at the parton level when the 
same clustering algorithm is used in both theoretical predictions
and experimental analyses. 

%%%%%%%%%%%%%%%
\mafigura{12cm}{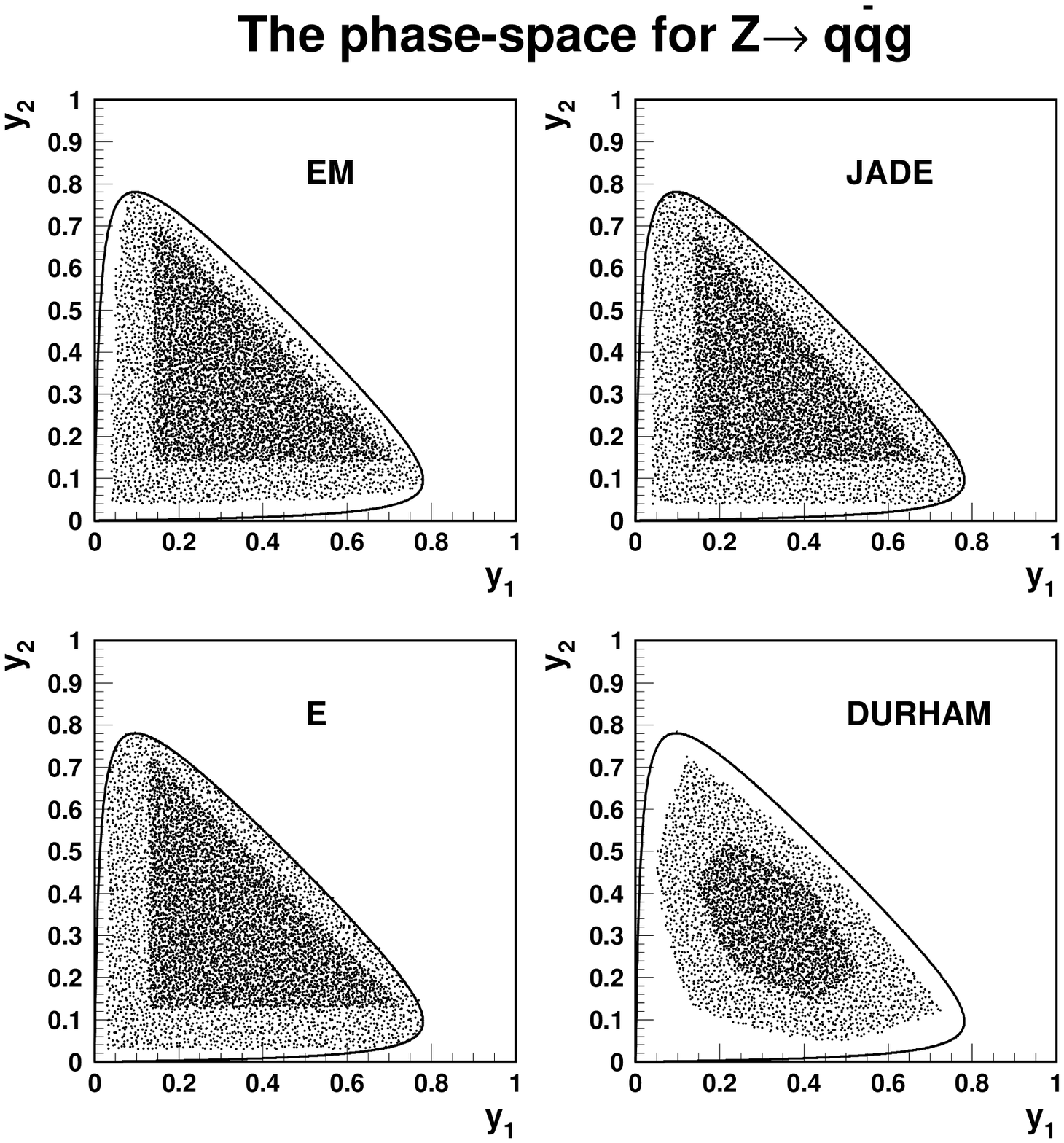}
{The phase 
space for $Z\rightarrow b(p_1) \bar{b}(p_2) g(p_3)$ in the plane
$y_1=2(p_1\cdot p_3)/s$ and $y_2=2(p_1\cdot p_3)/s$        
with cuts ($y_c=0.04$ and $y_c=0.14$) for the different algorithms. The mass
of the quark has been set to 10 GeV to enhance mass effects in the plot.}
{phase}
%%%%%%%%%%%%%%%

There are different successful jet-clustering algorithms and we refer
to~\cite{Bethke92,Kunszt89} for a detailed discussion and comparison of
these algorithms in the case of massless quarks.

In the following we will use the four jet-clustering algorithms
listed in the table~\ref{table1}, where $\sqrt{s}$ is the total
centre of mass energy.
In addition to the well-known JADE\index{JADE algorithm},
E\index{E algorithm} and DURHAM\index{DURHAM algorithm} algorithms we will
use a slight modification of the JADE scheme particularly useful for 
analytical calculations with massive quarks~\cite{RO95}.
It is defined by the two following equations
\[
y_{ij} =  \frac{2(p_i \cdot p_j)}{s}
\]
and
\[
p_k= p_i+p_j. 
\]
We will denote this algorithm as the EM\index{EM algorithm} scheme.
For massless particles and at the lowest order E, JADE and EM give
the same answers. However already at order $\as^2$ they give different
answers since after the first recombination the 
pseudoparticles are not massless anymore and the resolution functions
are different. 

For massive quarks the three algorithms, E, JADE and EM are
already different at order $\as$. The DURHAM ($K_T$) algorithm,
which has been  recently considered in order to avoid
exponentiation\index{Exponentiation}
problems present in the JADE algorithm 
\cite{durham1,durham2,Bethke92}, 
is of course completely different from the other
algorithms we use, both in the massive and the massless cases.

In figure~\ref{phase} we plotted the phase-space for two values of $y_c$ 
($y_c=0.04$ and $y_c=0.14$) for all four schemes (the solid line
defines the whole phase space for $Z \rightarrow q \bar{q} g$ with
$m_q=10$ GeV).

There is an ongoing discussion on which is the best algorithm for 
jet clustering in the case of massless quarks. The main criteria 
followed to choose them are based in two requirements:
\begin{enumerate}
\item Minimize higher order corrections.
\item Keep the equivalence between parton\index{Parton} and
hadronized cross sections. 
\end{enumerate}
To our knowledge no complete comparative study of the jet-clustering
algorithms has been done for the case of massive quarks. The properties
of the different algorithms with respect to the above criteria
can be quite different in the case of massive quarks from those in 
the massless case.
The first one because the leading terms containing
double-logarithms of y-cut  ($\log^2(y_c)$) that
appear in the massless calculation (at order $\as$) 
and somehow determine the size
of higher order corrections are softened in the case of massive 
quarks by single-logarithms of $y_c$ times a logarithm of the quark mass.
The second one because hadronization\index{Hadronization}
corrections for massive quarks could be different from the
ones for massless quarks. 

Therefore, we will not stick to any
particular algorithm but rather present results and compare them
for all the four algorithms listed in the table~\ref{table1}.

\section{Two- and three-jet event rates}

At the parton level the two-jet region in the decay
$Z \rightarrow b(p_1) \bar{b}(p_2) g(p_3)$ is given, in terms of the 
variables~\footnote{Only two are independent, in the 
EM algorithm $y_{12}=1-2 r_b-y_{13}-y_{23}$.}
$y_{13}$, $y_{23}$ and $y_{12}$, by the following conditions:
\beq
\label{cuts}
y_{13} < y_c \quad \mrm{or} \quad 
y_{23} < y_c \quad \mrm{or} \quad 
y_{12} < y_c~.
\end{equation}
This region contains the IR singularity, $y_{13}=y_{23}=0$ and the
rate obtained by the integration of the amplitude over this part
of the phase space should be added to the one-loop
corrected decay width for $\zbb$. The sum of these two 
quantities is of course IR finite and it is 
the so-called two-jet decay width at order $\as$.  
The integration over the rest of the phase space defines
the three-jet decay width at the leading order. 
It is obvious that the sum of the two-jet and three-jet
decay widths is independent of the resolution
parameter $y_c$, IR finite and given by the quantity 
$\Gamma_b = \Gamma(Z\rightarrow b\bar{b}+b\bar{b} g+\cdots)$
calculated in section~\ref{inclusivesection}. Therefore we have
\beq
\Gamma_{b}= \Gamma^b_{2j}(\yc)+
\Gamma^b_{3j}(\yc)+\cdots~.
\end{equation}
Clearly, at order $\as$, knowing 
$\Gamma_{b}$ and $\Gamma^b_{3j}(\yc)$ we can obtain
$\Gamma^b_{2j}(\yc)$ as well.

The calculation of
$\Gamma^b_{3j}(\yc)$ at order $\as$ is a tree-level calculation
and does not
have any IR problem since the soft gluon region has been excluded
from phase space. Therefore the calculation can be done in 
four dimensions without trouble.  
The final result can be written in the following form
\index{Three-jet decay rate}
\beq
\Gamma^b_{3j} = m_Z \frac{g^2}{c_W^2 64 \pi} \ap
\left( g_V^2 H\el{0}_V(\yc,\rb)+g_A^2 H\el{0}_A(\yc,\rb)\right)~,
\label{g3j}
\end{equation}
where the superscript $^{(0)}$ in the functions
$H^{(0)}_{V(A)}(y_c,r_b)$ reminds us that this is
only the lowest order result.
Obviously, the general form \ref{g3j} is independent of
what particular jet-clustering algorithm has been used. 

In the limit of zero masses, $r_b=0$, chirality\index{Chiral symmetry}
is conserved and the two functions $H^{(0)}_V(y_c,r_b)$ 
and $H^{(0)}_A(y_c,r_b)$ become identical
\beq
H^{(0)}_V(y_c,0) =  H^{(0)}_A(y_c,0) \equiv A^{(0)}(y_c)~.
\end{equation}
In this case we obtain the known result for the JADE-type 
algorithms,\index{JADE algorithm!A0@$A^{(0)}(y_c)$ function} which
is expressed in terms of the function $A^{(0)}(y_c)$
given by~\footnote{Note that with our normalization 
$A^{(0)}(y_c) = \frac{1}{2} A(y_c)$,
with $A(y_c)$ defined in \cite{Bethke92}.}.
\beq
\bes
A^{(0)}(y_c) & =  \frac{C_F}{2} \left[
- \frac{\pi^2}{3} + \frac{5}{2} - 6 y_c - \frac{9}{2} y_c^2
+(3-6 y_c) \log \left( \frac{y_c}{1-2 y_c} \right)\right. \\
& + \left. 2 \log^2 \left( \frac{y_c}{1-y_c} \right)
+ 4 \mrm{Li}_2 \left(\frac{y_c}{1-y_c}\right) \right]~.
\label{massless}
\ees
\end{equation}
The $A^{(0)}(y_c)$ function is also known analytically for 
the DURHAM algorithm \cite{durham1,durham2}.
Analytical expressions for the functions 
$H^{(0)}_V(y_c,r_b)$ and $H^{(0)}_A(y_c,r_b)$ 
are given for the EM algorithm in \cite{RO95}.

To see more clearly the size of mass effects we are going to study 
the following ratio of jet fractions\index{R3bd@$R_3^{bd}$} 
\begin{equation}
\label{r30}
R^{bd}_3 \equiv \frac{\Gamma^b_{3j}(y_c)/\Gamma^b}
{\Gamma^d_{3j}(y_c)/\Gamma^d}=
\left(c_V \frac{H^{(0)}_V(y_c,r_b)}{A^{(0)}(y_c)} +
c_A \frac{H^{(0)}_A(y_c,r_b)}{A^{(0)}(y_c)}\right)
\left(1+6 r_b c_A + O(r_b^2)\right)~,
\end{equation}
where we have defined\index{cv@$c_V$}\index{ca@$c_A$}
\[
c_V= \frac{g_V^2}{g_V^2+g_A^2}~, \quad
c_A= \frac{g_A^2}{g_V^2+g_A^2}~.
\]
In \eq{r30} we have kept only the lowest order terms in $\as$ and
$r_b$.  The last factor is due to the normalization to total rates.
This normalization is important from the experimental point of view
but also from the theoretical point of view because in these quantities
large weak corrections dependent on the 
top quark mass~\cite{mybb1,mybb2,mybb3,mybb4}\index{Top quark}
cancel. Note that, for massless quarks, the ratio
$\Gamma^d_{3j}(y_c)/\Gamma^d$ is independent on the neutral current
couplings of the quarks and, therefore,
 it is the same for up- and down-quarks
and given by the  
function $A\el{0}$. This means that we could equally use the normalization
to any other light quark or to the sum of all of them (including also
the c-quark if its mass can be neglected).

%%%%%%%%%%%%%%%
\mafigura{12cm}{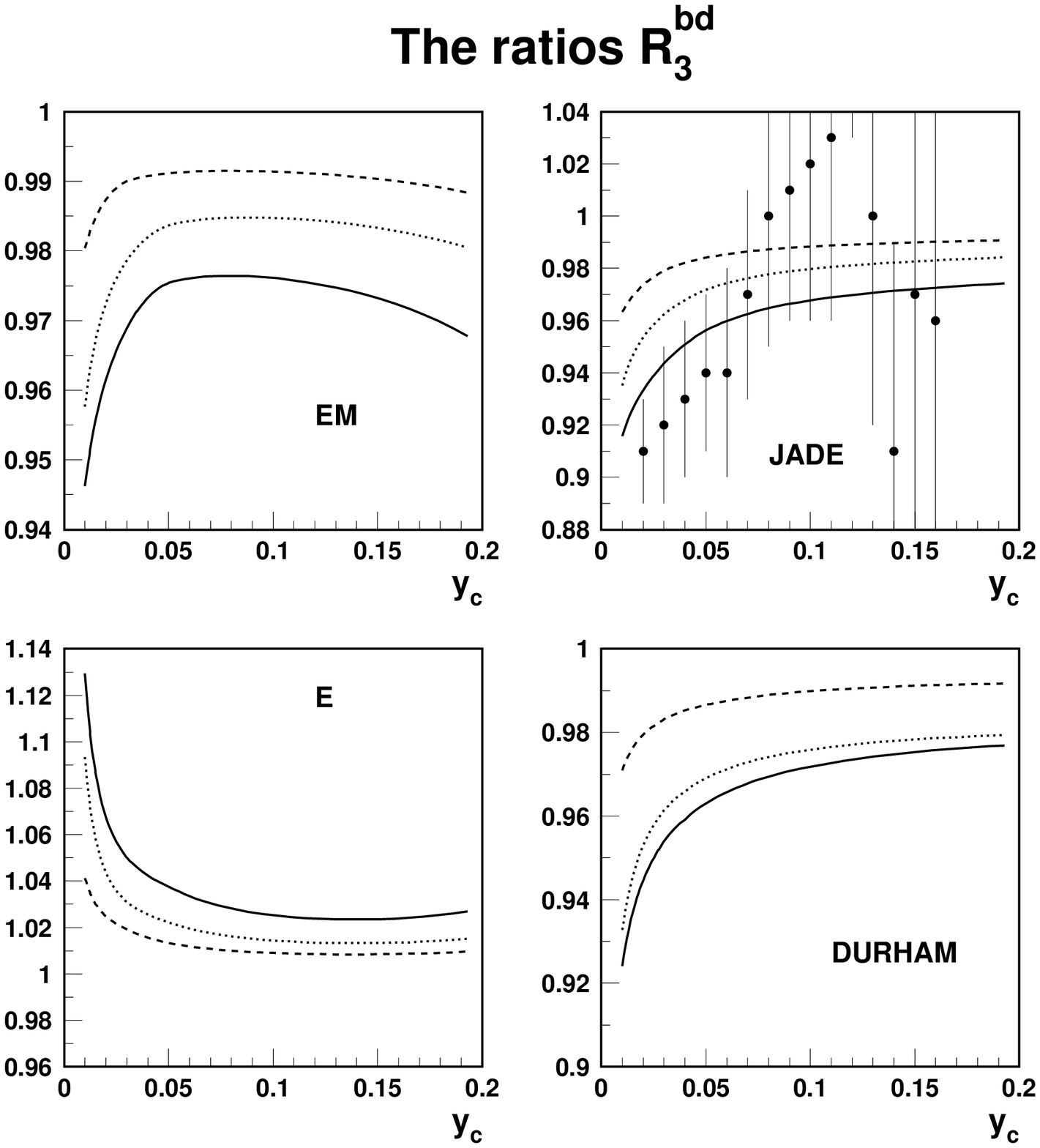}
{The ratios $R^{bd}_3$ (see \protect\eq{r30}) for the four algorithms. 
Solid lines correspond to $m_b=5$~GeV,
dashed lines correspond to $m_b=3$~GeV and dotted lines give our estimate
of higher order corrections to the $m_b=5$~GeV curve. For the JADE
algorithm we have also included the results of the 
analysis of the data collected during 1990-1991 by 
the DELPHI group~\protect\cite{delphi}.\protect\index{DELPHI}}
{r3}
%%%%%%%%%%%%%%%

\subsection{Estimate of higher order contributions}

All previous results come from a 
tree-level calculation, however, 
as commented in the introduction,
we do not know what is the
value of the mass we should use in the final results  since the
difference among the pole mass, the running mass at $\mu=m_b$ or
the running mass at $\mu= m_Z$ are next-order effects in $\as$.

In the case of the  
inclusive decay rate we have shown that one
could account (with very good precision)
for higher order corrections by using the running
mass at the $m_Z$ scale in the lowest order calculations.
Numerically the effect of running the quark mass from $m_b$ to $m_Z$
is very important.

One could also follow a similar 
approach  in the case of jet rates and try to account
for the next-order corrections by using the running quark mass at 
different scales.
We will see below that the dependence of $R^{bd}_3$ on the quark
mass is quite strong (for all clustering schemes);
using the different masses (e.g. $m_b$ or $\bar{m}_b(M_Z)$)
could amount to almost a factor 2 in the mass effect.
This suggests that
higher order corrections could be important.
Here, however, the situation is quite different, since in the 
decay rates to jets we have an additional scale given
by $y_c$, $E_c \equiv m_Z \sqrt{y_c}$, e.g. for $y_c=0.01$ we have
$E_c=9$~GeV and for $y_c=0.05$, $E_c=20$~GeV.
Perhaps one can absorb large logarithms, $\log(m_b/m_Z)$
by using the running coupling and the running mass 
at the $\mu=m_Z$ scale, 
but there will remain logarithms of the resolution parameter, 
$\log(y_c)$. For not very small $y_c$ one can expect that the 
tree-level results obtained by using 
the running mass at the $m_Z$ scale are a good approximation,
however, as we already said, 
the situation cannot be settled completely until
a next-to-leading calculation including mass effects is available.  

Another way to estimate higher order effects in 
$R^{bd}_3$ is to use the known results for the massless case 
\cite{Ellis81,ellis1,ellis2,ellis3,Bethke92,Kunszt89,Fabricius81,Kramer89}.
Including  higher order corrections the general form 
of \eq{g3j} is still valid with the change
$H_{V(A)}^{(0)}(y_c,r_b) \rightarrow H_{V(A)}(y_c,r_b)$. Now we can
expand the functions $H_{V(A)}(y_c,r_b)$ in $\as$ and factorize the
leading dependence on the quark mass
as follows
\begin{multline}
H_{V(A)}(y_c,r_b) = A^{(0)}(y_c)+\ap A^{(1)}(y_c) \\ + 
r_b \left( B_{V(A)}^{(0)}(y_c,r_b)
+ \ap B_{V(A)}^{(1)}(y_c,r_b)\right) + \cdots~.
\end{multline}
In this equation we already took into account that for massless 
quarks vector and axial contributions are 
identical\footnote{This is not
completely true at $O(\as^2)$ 
because the triangle anomaly:\index{Triangle anomaly}
there are one-loop triangle diagrams 
contributing to $Z\rightarrow b\bar{b} g$  with the top and the bottom quarks 
running in the loop. Since $m_t \not= m_b$ the anomaly cancellation is 
not complete. These diagrams contribute to the axial part even for
$m_b=0$ and  lead to a deviation from $A^{(1)}_V(y_c) = A^{(1)}_A(y_c)$
\cite{hagiwara}. 
This deviation is, however, small \cite{hagiwara} and we are
not going to consider its effect here.}

Then, we can rewrite 
the ratio $R^{bd}_3$, at order $\as$, as follows\index{R3bd@$R_3^{bd}$} 
\bea
\label{r3a}
R^{bd}_3 = \left[
1 + r_b \left\{
 c_V
\frac{B_V^{(0)}(y_c,r_b)}{A^{(0)}(y_c)} \left[
1 + \ap \left( \frac{B_V^{(1)}(y_c,r_b)}{B_V^{(0)}(y_c,r_b)}
- \frac{A^{(1)}(y_c)}{A^{(0)}(y_c)}
\right) \right] \right. \right. \nonumber \\
\left. \left. +
  c_A
\frac{B_A^{(0)}(y_c,r_b)}{A^{(0)}(y_c)} \left[
1 + \ap \left( \frac{B_A^{(1)}(y_c,r_b)}{B_A^{(0)}(y_c,r_b)}
- \frac{A^{(1)}(y_c)}{A^{(0)}(y_c)}
\right) \right]
\right\} \right] \nonumber \\
\times \left[1 + 6 \: r_b \left\{
c_A (1 + 2 \ap \log r_b ) - c_V 2 \ap \right\} \right]~.
\end{eqnarray} 

The functions $B_V^{(0)}(y_c,r_b)$ and $B_A^{(0)}(y_c,r_b)$
where calculated in~\cite{RO95}.
The lowest order function for the massless case, 
$A^{(0)}(y_c)$, is also known analytically for JADE-type
algorithms, \eq{massless} and refs.~\cite{Bethke92,Kunszt89},
and for the DURHAM algorithm \cite{durham1,durham2}.
A parametrization of the  function $A^{(1)}(\yc)$ can be found 
in~\cite{Bethke92} for the different algorithms~\footnote{With our
choice of the normalization $A^{(1)}(y_c)=B(y_c)/4$, where $B(y_c)$ 
is defined in \cite{Bethke92}.}.
As we already mentioned
this function is different for different clustering algorithms.
The only unknown functions in \eq{r3a}
are $B_V^{(1)}(y_c,r_b)$ and $B_A^{(1)}(y_c,r_b)$,
which must be obtained from a complete calculation
at order $\as^2$ including mass effects
(at least at leading order in $\rb $). 
The second part of this thesis is devoted to this 
calculation.

Nevertheless, in order to estimate
the impact of higher order corrections in our calculation 
we will assume for the moment that 
$B_{V,A}^{(1)}(y_c,r_b)/B_{V,A}^{(0)}(y_c,r_b) \ll
A^{(1)}(y_c)/A^{(0)}(y_c)$ and take 
$A^{(1)}(y_c)/A^{(0)}(y_c)$ from\footnote{For the
EM algorithm this function has not yet been computed. To make
an estimate of higher order corrections we will use in this case
the results for the E algorithm.} \cite{Bethke92,Kunszt89}.
Of course this does not need to be the case but at least it gives
an idea of the size of higher order corrections. 
We will illustrate the numerical effect of these corrections for 
$R^{bd}_3$ in the next subsection.
As we will see, the estimated effect of next-order corrections is quite
large, therefore in order to obtain the $b$-quark mass
from these ratios the calculation of the 
functions $B_{V,A}^{(1)}(y_c,r_b)$ is mandatory.

\subsection{Numerical results for $R^{bd}_3$ for different clustering
algorithms}

To complete this section we present  
the numerical results for $R^{bd}_3$ calculated with the different 
jet-clustering algorithms.
For the JADE, E and DURHAM algorithms we obtained the three-jet rate
by a numerical integration over the phase-space given by the cuts 
(see fig.~\ref{phase}).
For the EM scheme we used our analytical
results~\cite{RO95} which were also
employed to cross check  the numerical procedure.

In fig.~\ref{r3} we present the ratio $R^{bd}_3$, obtained by
using the tree-level expression, \eq{r30}, against $y_c$
for $m_b= 5$~GeV and $m_b= 3$~GeV. We also plot the results given by
\eq{r3a} (with $B_{V,A}\el{1}(\yc,\rb)/B_{V,A}\el{0}(\yc,\rb)=0$)
for $m_b=5$~GeV, which gives an estimate of higher
order corrections.
For $y_c < 0.01$ we do not
expect the perturbative calculation to be valid. 

As we see from the figure, the behaviour of $R^{bd}_3$ is quite different in
the different schemes. The mass effect has a negative sign for all schemes
except for the  E-algorithm.
For $y_c > 0.05$
the mass effects are at the 4\% level for $m_b= 5$~GeV and
at the 2\% level for $m_b= 3$~GeV (when the tree level expression is used). 
Our estimate of higher order effects, with
the inclusion of the next-order effects in $\as$
for massless quarks, shifts the curve for $m_b=5$~GeV in the direction
of the 3~GeV result and amounts to about of  20\% to 40\% of the difference
between the tree-level calculations with the two different masses.
For both E and EM schemes we used the higher order results
for the E scheme.

%For the JADE algorithm we have also plotted in fig.~\ref{r3} 
%the experimental results for $R^{bd}_3$
%obtained by the DELPHI group \cite{delphi} on the basis
%of the data collected in 1990-1991.
%The experimental errors, due to the limited statistics analyzed,
%are rather large.
%However, one can already see the effect of the quark mass.
%If the $b$-quark mass would be zero,
%one should obtain a ratio $R^{bd}_3$ constant and equal to 1.
%It is clearly seen from the figure that
%for $\yc < 0.08$ the 
%data are significantly below 1. 
%For larger values of $\yc$, the number
%of events decreases, the errors 
%become too large and the data are consistent with 1.
%When larger amount of data is analyzed and the experimental
%error is decreased, it will be very interesting to see if
%data will exhibit the
%different signs of the mass
%effect in $R^{bd}_3$ (positive for the E scheme and negative for 
%the other schemes) as
%predicted by our parton level calculations
%(see fig.~\ref{r3}).

%In spite of the fact that the effect of the quark mass in $R^{bd}_3$
%has been seen, it is too early, in our opinion, to extract now
%the value of the $b$-quark mass from the data.
%As discussed  above 
%the higher order corrections
%to $R^{bd}_3$ are presumably rather large and should be included in
%the theoretical calculations.
%However, it is clear, that once
%the essential next-to-leading order corrections will be available
%and all LEP data will be included in the analysis,
%the ratios $R^{bd}_3$ will certainly
%allow for a reasonable determination of the $b$-quark
%mass and for a check of its running from $m_b$ to $m_Z$.

\begin{table}
\begin{center}
\caption{Results of the three parameter fits of the functions
$B_{V,A}^{(0)}(y_c,r_b)/A^{(0)}(y_c)
= \sum_{n=0}^2 k_{V,A}^{(n)}$ log${}^n y_c$ 
in the range $0.01 < y_c < 0.20$
\label{table2}}
\begin{tabular}{lrrrrrr}
\hline
Algorithm   & $k_V^{(0)} \;$ & $k_V^{(1)} \;$ & $k_V^{(2)} \;$
            & $k_A^{(0)} \;$ & $k_A^{(1)} \;$ & $k_A^{(2)} \;$ \\
\hline
EM     & $-28.58$ & $-14.64$ & $-2.72$ & $-30.67$ & $-13.54$ & $-2.61$ \\
JADE   & $-13.25$ & $- 5.19$ & $-2.01$ & $-15.42$ & $ -4.13$ & $-1.90$ \\
E      & $ 25.97$ & $ 19.04$ & $ 4.68$ & $ 23.39$ & $ 19.81$ & $ 4.71$ \\
DURHAM & $-12.70$ & $- 4.76$ & $-1.69$ & $-15.48$ & $ -4.28$ & $-1.65$ \\
\hline
\end{tabular}
\end{center}
\end{table}

%%%%%%%%%%%%%%%
\mafigura{12cm}{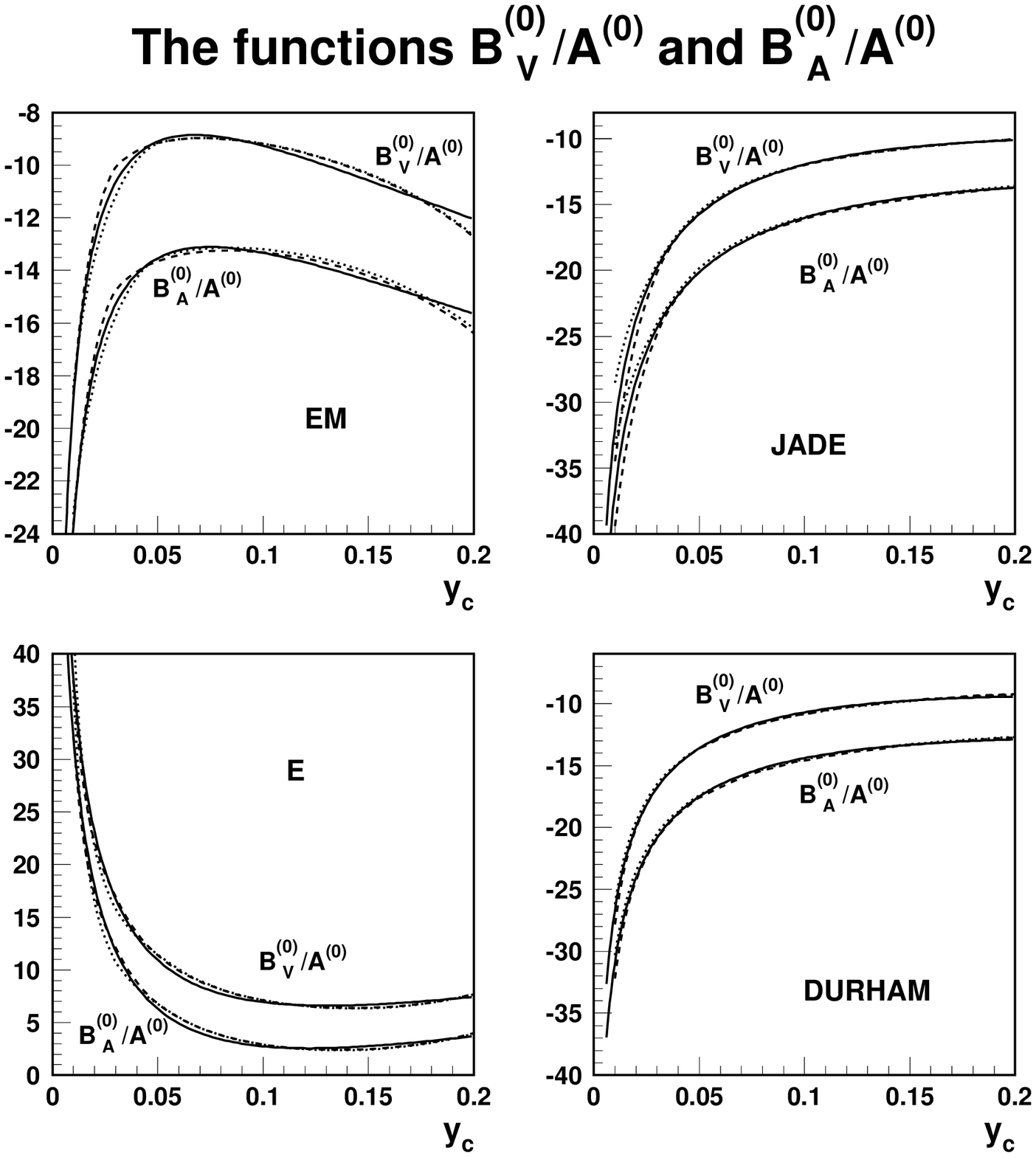}
{The functions $B_V\el{0}/A\el{0}$ and $B_A\el{0}/A\el{0}$ for the four 
algorithms. Dashed lines for $m_b=3$~GeV, dotted lines for $m_b=5$~GeV
and solid lines for our three-parameter fit, \protect\eq{parfit}.}
{bes}
%%%%%%%%%%%%%%%

To simplify the use of our results we present
simple fits in $\log y_c$ to the ratios
$B_{V,A}^{(0)}(y_c,r_b)/A^{(0)}(y_c)$, which define $R^{bd}_3$ at
lowest order, for the different clustering algorithms.
We use the following parametrization:
\beq
B_{V,A}^{(0)}(y_c,r_b)/A^{(0)}(y_c)=\sum_{n=0}^2 k_{V,A}^n \log^n y_c~.
\label{parfit}
\end{equation}
The results of the fits for the range $0.01 < y_c < 0.20$
are presented in table~\ref{table2}.

For completeness we include also in table~/ref{tableA0}
a five parameter fit to the LO massless function $A^{(0)}(y_c)$.
For the DURHAM algorithm we take the result from~\cite{Bethke92}.
For the EM, JADE and E algorithms that give the same answer 
we have fitted the analytical expression of \Eq{massless}.

\begin{table}
\begin{center}
\caption{Five parameter fits to the LO massless function
$A^{(0)}(y_c) = \sum_{n=0}^4 k^{(n)}$ log${}^n y_c$ 
in the range $0.01 < y_c < 0.20$
\label{tableA0}}
\begin{tabular}{lrrrrr}
\hline
Algorithm   & $k^{(0)} \;$ & $k^{(1)} \;$ & $k^{(2)}\;$
            & $k^{(3)} \;$ & $k^{(4)} \;$ \\
\hline
EM/JADE/E   & $1.19248$ & $2.64729$ & $1.359065$ & $-.0173022$ 
            & $-.001971025$ \\
DURHAM      & $.8625$ & $1.835$ & $1.0745$ & $.133$ & $.01255$ \\
\hline
\end{tabular}
\end{center}
\end{table}

In fig.~\ref{bes} we plot the ratios
$B_{V,A}\el{0}(\yc,\rb)/A\el{0}(\yc)$ as a function of $y_c$ for 
the different
algorithms (dashed lines for $m_b=5$~GeV, dotted lines for $m_b=3$~GeV
and solid curves for the result of our fits).
As we see from the figure the remnant mass
dependence in these ratios (in the range of masses we are interested in
and in the range of $y_c$ we have considered)
is rather small and for actual fits we used the average of the ratios 
for the two different masses. 
We see from these figures that such a simple three-parameter fit works
reasonably well for all the algorithms.

Concluding this section we would like to make the following remark.
In this chapter we discuss the
$Z$-boson decay. In LEP\index{LEP} experiments one studies the process
$e^+e^- \rightarrow (Z \gamma^*)\rightarrow b\bar{b}$ and, apart
from the resonant $Z$-exchange cross section,
there are contributions from  the pure
$\gamma$-exchange and from the 
$\gamma-Z$-interference.\index{gZ@$\gamma-Z$-interference}
The non-resonant $\gamma$-exchange contribution 
at the peak is less than 1\% for muon production and in the case
of $b$-quark production there is an additional
suppression factor $Q_b^2=1/9$.
In the vicinity of the $Z$-peak the interference
is also suppressed because it is proportional  to $Q_b (s-m_Z^2)$ ($\sqrt s$
is the $e^+e^-$ centre of mass energy). We will neglect these terms as they
give negligible contributions compared with the uncertainties in higher
order QCD corrections to the quantities we are considering.

Obviously, QED initial-state 
radiation\index{QED!initial-state radiation}
should be taken into account in the real analysis;
the cross section for $b$-pair production 
at the $Z$ resonance can be written as
\beq
\sigma_{b\bar{b}}(s)=\int \sigma^0_{b\bar{b}}(s') F(s'/s)ds'
\end{equation}
where $F(s'/s)$ is the well-known QED radiator\index{QED!radiator} 
for the total cross section~\cite{radiator} and, the Born cross section, 
neglecting pure $\gamma$
exchange contribution and the $\gamma-Z$-interference, has the form
\beq
\sigma^0_{b\bar{b}}(s)=\frac{12 \pi \Gamma_e \Gamma_b}{m_Z^2}
\frac{s}{(s-m_Z^2)^2+m_Z^2\Gamma_Z^2}
\end{equation}
with obvious notation. Note that $\Gamma_b$ in this expression
can be an inclusive width as well as some more exclusive quantity, 
which takes into account some kinematical restrictions
on the final state.

\section{Discussion and conclusions}

In this chapter we have presented a theoretical
study of quark-mass effects in the decay of the
$Z$-boson into bottom quarks at LO in the strong 
coupling constant. Furthermore, we have analyzed
some three-jet observables which are very sensitive
to the value of the quark masses.

For a slight modification of the JADE algorithm 
(the EM algorithm)\index{EM algorithm}
we were able to calculate analytically~\cite{RO95}
the three-jet decay width of the 
$Z$-boson into $b$-quarks as a function of the 
jet resolution parameter, $y_c$, and the $b$-quark mass. 
The answer is rather involved, but can be  expressed
in terms of elementary functions. Apart from the fact that
these analytical calculations are interesting by 
themselves, they can also be used to test 
Monte Carlo\index{Monte Carlo} simulations.
For the EM, JADE, E and DURHAM clustering algorithms we
have obtained the three-jet decay width by a simple two-dimensional
numerical integration. Numerical and analytical results have been
compared in the case of the EM scheme. 

We discussed quark-mass effects by considering
the quantity\index{R3bd@$R_3^{bd}$}
\[
R^{bd}_3 = \frac{\Gamma^b_{3j}(y_c)/\Gamma^b}{\Gamma^d_{3j}(y_c)/\Gamma^d}
= 1 + \frac{m_b^2}{m_Z^2} F(m_b,y_c)~
\]
which has many advantages from both the theoretical and the experimental
point of views. In particular, at lowest order, 
the function $F(m_b,y_c)$ is almost 
independent on the quark mass (for the small values of the mass in which
we are interested in) and has absolute values ranging from 10 to 35
(depending on $y_c$ and on the algorithm), where the larger values are 
obtained for $y_c$ of about $0.01$.

At the lowest order in $\as$ we do not know what is the exact
value of the quark mass that should be used in the above equation
since the difference between the different
definitions of the $b$-quark mass,
the pole mass, $m_b \approx 5$~GeV, or the running mass
at the $m_Z$-scale, $\bar{m}_b(m_Z) \approx 3$~GeV, is order $\as$.
\index{Quark mass!perturbative pole}\index{Quark mass!running}
Therefore, we have presented all results for these two values of
the mass and have interpreted the difference as an estimate of
higher order corrections. Conversely one can keep the mass fixed 
and include in $F(m_b,y_c)$ higher order corrections
already known for the massless case.
According to these estimates
the $O(\as)$ corrections can be about 40\% of the tree-level
mass effect (depending on the clustering scheme), 
although we cannot exclude even larger corrections.

By using the lowest order result we find that for moderate
values of the resolution parameter, $y_c \approx 0.05$, the mass 
effect in the ratio $R^{bd}_3$ is about $4\%$ 
if the pole mass value of the $b$-quark, $m_b \approx 5$~GeV, is used,
and the effect decreases to 2\% if $m_b=3$~GeV.
However, in order to extract a meaningful value of the $b$-quark mass
from the data it will be necessary to include next-to-leading
order~(NLO) corrections since the leading mass effect we have
calculated does not distinguish among the different definitions
of the quark mass (pole mass, running mass at the $m_b$ scale
or running mass at the $m_Z$ scale).
Next chapters are devoted to the calculation of
the NLO corrections.

Finally, if gluon jets can be identified with enough
efficiency~\cite{salva} another interesting three-jet
observable very sensitive to the bottom quark mass is
the angular distribution\index{Angular distribution}
\[
R^{bd}_\vartheta \equiv \left.
\frac{1}{\Gamma^b}\frac{d\Gamma^b_{3j}}{d\vartheta}\right/
\frac{1}{\Gamma^d}\frac{d\Gamma^d_{3j}}{d\vartheta},
\]
where $\vartheta$ is the minimum of the angles formed between
the gluon jet and the quark and antiquark jets. 
The angular distribution $R^{bd}_\vartheta$ was studied at 
LO in~\cite{RO95}. We will leave its analysis at NLO
order for a future work.

\chapter{Transition amplitudes at next-to-leading order}

    In the previous chapter we have seen that some three-jet
observables, in particular the following ratio of three-jet
decay rate fractions\index{R3bd@$R_3^{bd}$}
\beq
R_3^{bd} = \frac{\Gamma_{3j}^b (y_c)/\Gamma^b}
                {\Gamma_{3j}^d (y_c)/\Gamma^d}, 
\end{equation}
are very sensitive to the quark masses and their
study at LEP\index{LEP} can provide a very interesting 
experimental information about the bottom quark mass.
Among other applications, such study would allow to perform 
the first measurement of the bottom quark mass 
far from threshold and, what is more important,
it would provide, for the first time, a check 
of the running of quark masses from scales of 
the order of $\mu \sim 5(GeV)$ to $\mu = m_Z$
in the same way the running of the strong 
coupling constant has been checked before.

However, as we mentioned, in order to extract a meaningful
value of the bottom quark mass from LEP data it is 
necessary to include next-to-leading order (NLO)
strong corrections in $R_3^{bd}$
since the LO QCD prediction does not allow to distinguish
among the possible theoretical definitions of the quark mass
(pole mass, running mass at the $m_b$ scale or running mass 
at the $m_Z$ scale) that numerically are quite different.

At the NLO we have contributions from three- and four-parton 
final states. The three-jet cross section is obtained
by integrating both contributions in the three-jet 
phase space region defined by the jet clustering 
algorithms considered in the previous chapter: 
EM, JADE, E and DURHAM. 
This quantity is infrared finite and well defined, 
however the three- and four-parton transition 
amplitudes independently contain infrared 
singularities. Therefore, some regularization procedure 
is needed. We use 
Dimensional Regularization\index{Dimensional Regularization}
because it preserves the QCD Ward identities.\index{Ward identities} 
Since at this order we have diagrams with a
three gluon vertex\index{Gluon!three gluon vertex} 
it is not possible anymore to regulate the infrared
divergences with a gluon mass\index{Gluon!mass}
as it would be possible at the lowest order.
 
The three-parton transition amplitudes can be 
expressed in terms of a few scalar one-loop integrals.
After UV renormalization we obtain 
analytical expressions for the terms proportional 
to the infrared poles and for the finite contributions.
The finite contributions are integrated numerically 
in the three-jet region. The infrared poles 
will cancel against the four-parton contributions.

The four-parton transition amplitudes are splited 
into a soft part in the three-jet region 
and a hard contribution.
The soft terms are integrated 
analytically in arbitrary $D$ dimensions
in the region of the phase space containing 
the infrared singularities. We obtain analytical 
expressions for the infrared behaviour 
of the four-parton transition amplitudes 
and we show how these infrared contributions 
cancel exactly the infrared singularities 
of the three-parton transition amplitudes.
The hard terms are calculated in $D=4$ dimensions.
The remaining phase space integrations giving rise to
finite contributions are performed numerically.

In this chapter we consider the NLO corrections to the 
three-jet decay rate of the $Z$-boson into massive 
bottom quarks. First we present and classify the 
matrix elements of the virtual corrections to the 
process
\beq
  Z(q) \rightarrow b(p_1) + \bar{b}(p_2) + g(p_3)~,
\end{equation}
and the tree-level processes
\beq
\bes
  Z(q) & \rightarrow b(p_1) + \bar{b}(p_2) + g(p_3) + g(p_4)~,\\
  Z(q) & \rightarrow b(p_1) + \bar{b}(p_2) + q(p_3) + \bar{q}(p_4)~,\\
  Z(q) & \rightarrow b(p_1) + \bar{b}(p_2) + b(p_3) + b(p_4)~,\\
\label{todos}
\ees
\end{equation}
where $q$ stands for a light quark and the symbols in brackets 
denote the particle momenta.
In the following we will denote by $p_{ij}$ the sum 
of the momenta of particles labelled $i$ and $j$, $p_{ij}=p_i+p_j$.
In next chapter we calculate the singular IR pieces of 
these matrix elements in the three-jet region
and show how after UV renormalization IR singularities
cancel among the one-loop corrected width of  
$Z\rightarrow b\bar{b}g$ and the tree-level processes \rfn{todos}.

We use Dimensional Regularization\index{Dimensional Regularization}
to regularize both 
the UV and the IR divergences~\cite{dimreg1,dimreg2,dimreg3}
with na\"{\i}ve anticommutating\index{gamma5@$\gamma_5$} 
$\gamma_5$ prescription,
see section~\ref{gamma5} of Appendix~\ref{loopintegrals}. 
All the Dirac algebra is performed in $D$-dimensions
with the help of the {\it HIP}~\cite{Hsieh92}\index{HIP} 
{\it Mathematica 2.0.} package.\index{Mathematica}
We work in the Feynman gauge.\index{Feynman!gauge}

\section{Virtual corrections}

The radiative corrections to the process
\beq
  Z(q) \rightarrow b(p_1) + \bar{b}(p_2) + g(p_3)~,
\end{equation}

%%%%%%%%%%%%%%%
\mafigura{8.5cm}{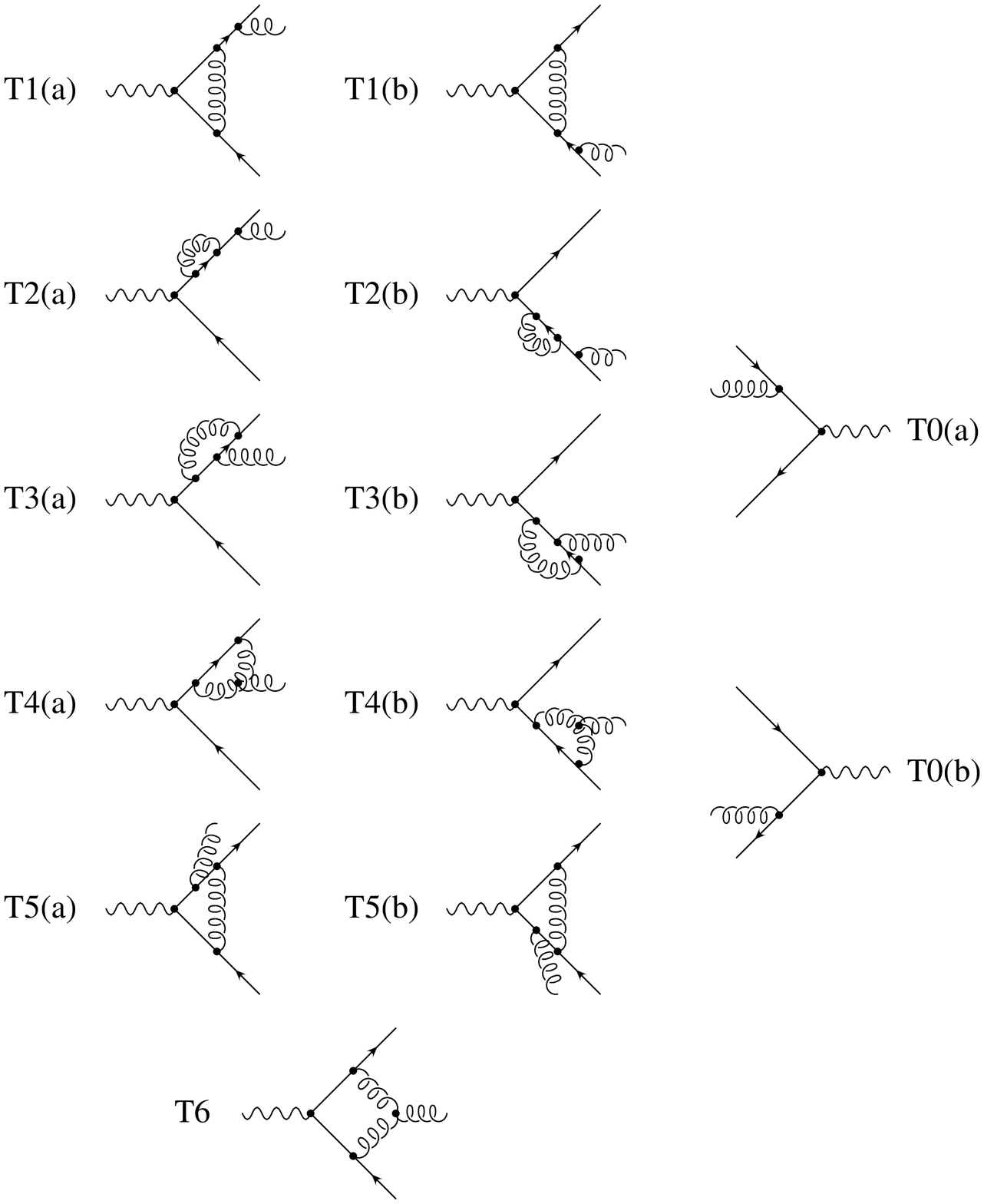}
{Radiative corrections to the process $Z\rightarrow b\bar{b}g$.
Selfenergies in external legs have not been shown.
Interference with the lowest order Feynman diagrams, $T0(a)$ and
$T0(b)$, gives rise to the $O(\as^2)$ correction.\protect\index{T0} }
{loop}
%%%%%%%%%%%%%%%

\noindent
are shown in figure~\ref{loop}. 
They contribute to the 
three-jet decay rate at $O(\as^2)$ through their 
interference with the lowest order Feynman diagrams $T0(a)$ and 
$T0(b)$.\index{T0} We have not depicted the diagrams 
with selfenergy insertions in the quark and gluon 
external legs since their contribution is just the 
wave function renormalization constant times 
the square of the lowest order matrix elements.

\index{T1}\index{T2}\index{T3}\index{T4}\index{T5}\index{T6}
Only the one loop Feynman diagrams $T1$, $T2$, $T3$ and $T4$ hold
UV divergences, the one-loop box integrals of 
diagrams $T5$ and $T6$ are UV finite
since they contain four propagators. 
Diagrams $T3$ and $T4$ are responsible for the renormalization 
of the strong coupling constant and diagram $T2$ renormalize 
the quark mass. 
The UV divergences of diagram $T1$ 
are cancelled by the UV piece of the quark wave function
renormalization constant.
We perform the UV renormalization in a mixed scheme where
the strong coupling constant is renormalized in the
$\overline{MS}$ scheme\index{Renormalization!scheme!$\overline{MS}$} 
while the rest is renormalized in the 
on-shell scheme.\index{Renormalization!scheme!on-shell}
After UV renormalization  $T1$, $T2$ and $T3$ become
completely finite, $T5$ remains with only simple IR poles 
whereas $T4$ and $T6$ contain up to double IR divergences.
Colour can be used to preliminary classify the 
one-loop diagrams. The interference of diagrams $T1$ and $T2$
with the lowest order diagrams $T0$ carries a colour factor $C_F^2$,
where $C_F=4/3$.\index{CF@$C_F$}
Diagrams $T3$ and $T5$ generate a 
$C_F(C_F-N_C/2)$ colour factor, with $N_C=3$ the number of colours,
whereas $T4$ and $T6$ produce $C_F N_C$. 

We found the most efficient way to calculate the loop diagrams is 
to directly perform all the traces over the Dirac matrices
on the interference matrix elements
to reduce them to a Lorentz scalar, before 
integrating over the virtual gluon loop momentum $k$.
After this, we end up with a series of loop integrals 
with scalar products of the internal momentum $k$ and 
the external momenta in the numerator.
All these vector and tensor loop integrals can be reduced
to simple scalar one loop integrals by following the
Passarino-Veltman reduction\index{Passarino-Veltman reduction}
procedure, see section \ref{pasa}
of Appendix \ref{loopintegrals}.
At the end, our problem is simplified to the 
calculation of only five one loop three propagator integrals,
two infrared divergent scalar box integrals and some well known
scalar one- and two-point functions, 
see Appendix~\ref{loopintegrals}. Moreover, 
the infrared divergent piece of the two four-point 
functions can be written in terms of two of the three-point
one-loop integrals.

\section{Emission of two real gluons}

The calculation of the transition probability for the process
\beq
Z(q) \rightarrow b(p_1) + \bar{b}(p_2) + g (p_3) + g (p_4),
\end{equation}
from the eight diagrams shown in figure \ref{bbgg} 
contains in principle 36 terms. However, many of them are related
by interchange of momentum labels and, at the end, only 13
transition probabilities need to be calculated.
We follow the notation of \cite{Ellis81}. 
The interference of graph $B_i$ with
$B_j$ is written as $B_{ij}$. The relevant 13 transitions
probabilities, which we display in fig \ref{bbggbuble},
can be classified into three different subsets depending on the 
colour factor. In table~\ref{permuta} we give the momentum label 
interchanges necessary to generate all the transitions probabilities 
from the thirteen which we choose to calculate.

%%%%%%%%%%%%%%%
\mafigura{6.5cm}{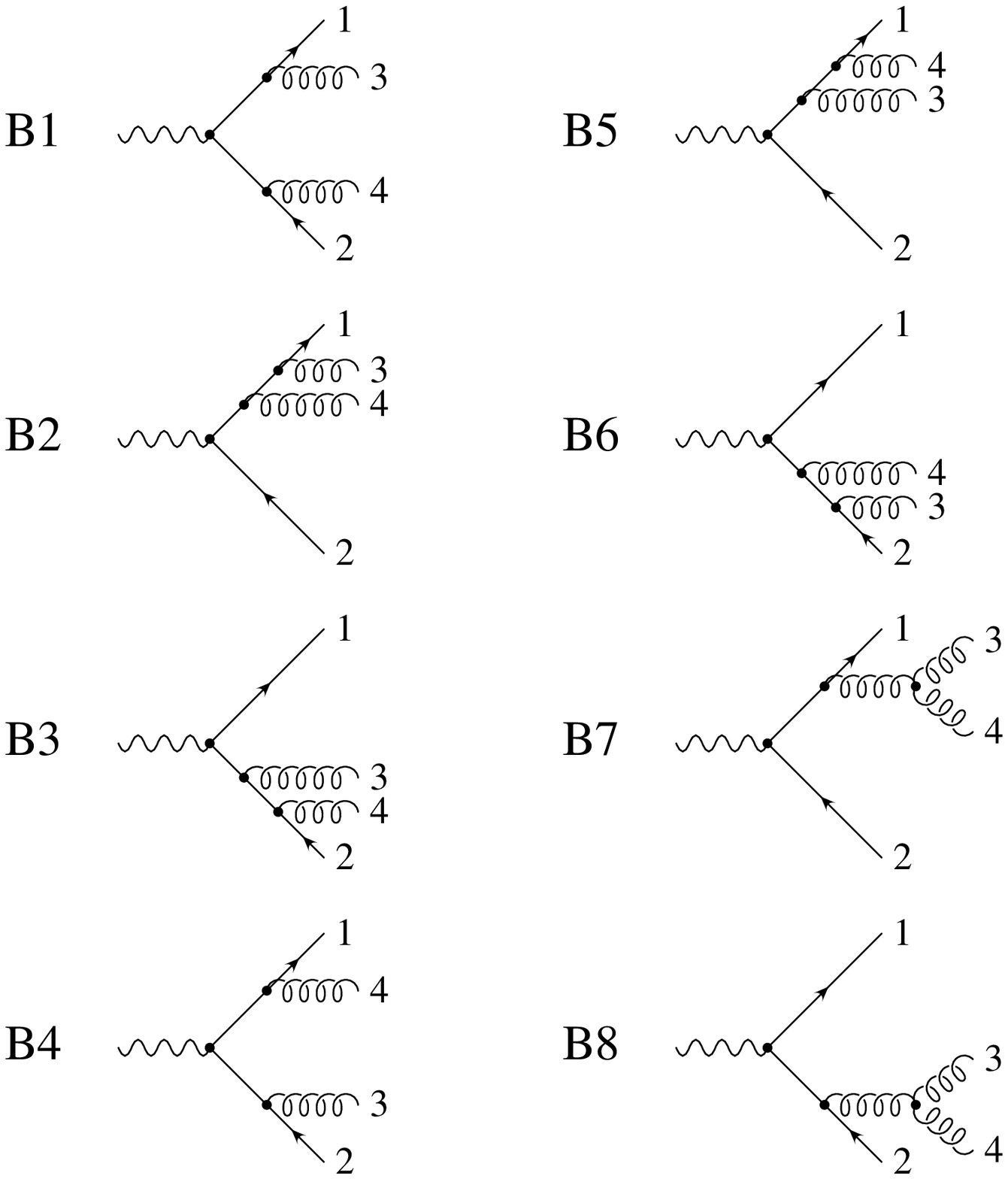}
{Feynman diagrams contributing to the process 
$Z\rightarrow b\bar{b}gg$.}
{bbgg}
%%%%%%%%%%%%%%%

\begin{table}[hbtp]
\begin{center}
\caption{The interchange table relating the graphs for 
$Z \rightarrow b \bar{b} g g$. \label{permuta}}
\begin{tabular}{|c|c|c|c|} \hline
label       & Class A & Class B & Class C \\
permutation & $C_F^2$ & $C_F(C_F-\frac{1}{2}N_C)$ & $C_F N_C$ \\
\hline
      & B11 B21 B22 B32 & B41 B42 B62 B52  & B71 B72 B82 B77 B87 \\
$(1 \leftrightarrow 2)$ 
      & B44 B64 B66 B65 & B41 B61 B62 B63 & B84 B86 B76 B88 B87 \\
$(3 \leftrightarrow 4)$ 
      & B44 B54 B55 B65 & B41 B51 B53 B52 & B74 B75 B85 B77 B87 \\
$(1 \leftrightarrow 2)$ $(3 \leftrightarrow 4)$ 
      & B11 B31 B33 B32 & B41 B43 B53 B63 & B81 B83 B73 B88 B87 \\
\hline
\end{tabular}
\end{center}
\end{table}

Therefore, it is  sufficient to consider the following combinations
of transitions amplitudes\index{Class A}\index{Class B}\index{Class C}
\beq
\bes
\mrm{Class A} &= \frac{1}{2} B11 + 2 B21 + B22 + B32, \\
\mrm{Class B} &= \frac{1}{2} B41 + 2 B42 + B62 + B52, \\
\mrm{Class C} &= 2 (B71 + B72+ B82) + \frac{1}{2} (B77+B87),
\ees
\end{equation}
plus the interchanges $(1 \leftrightarrow 2)$, $(3 \leftrightarrow 4)$
and $(1 \leftrightarrow 2)$ $(3 \leftrightarrow 4)$.

Since only gluons attached to external legs can generate
IR divergences in the three-jet region 
we can see immediately that
$B32$ and $B52$ are fully finite, $B21$, $B22$, $B42$ and $B62$ are
IR only in gluon labelled as 3 while $B11$ and $B41$ are IR
in both gluons, 3 and 4. 
All the matrix elements of Class A and B\index{Class A}\index{Class B}
are free of quark-gluon 
collinear divergences\index{IR singularities!collinear}
since we are working with massive quarks.
Therefore, we will find there just IR simple poles, $1/\epsilon$,
because only soft singularities remain.
On the other hand, we expect IR double poles
for diagrams of Class C\index{Class C}
because the gluon-gluon collinear divergences 
are still preserved at the
three gluon vertex.\index{Gluon!three gluon vertex}
This argument is not completely true for the transition probabilities
$B77$ and $B87$. Both diagrams individually contain IR double poles
but when we take into account all the momenta interchanges, i.e.,
in the square of the sum of diagrams $B7$ and $B8$, double poles cancel.
As we will see, these diagrams are related, 
by Cutkowsky rules\index{Cutkowsky rules},
to the diagram with a selfenergy insertion
in an external gluon leg. 
That is the reason why only simple IR poles can appear.

%%%%%%%%%%%%%%%
\mafigura{12cm}{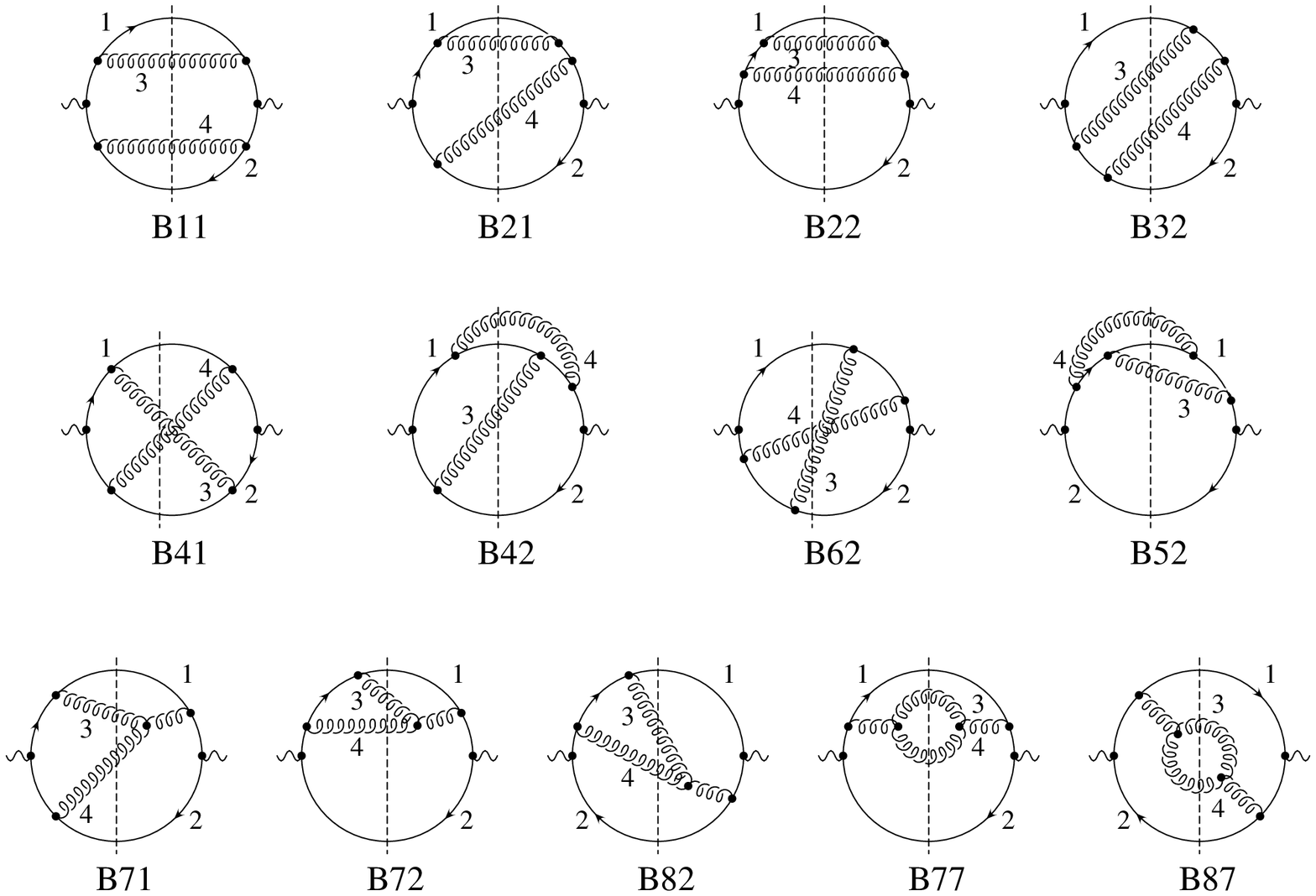}
{Transitions probabilities for the process $Z\rightarrow b\bar{b}gg$
On-shell particles are indicated by the dashed line and 
the numbers refer to the momentum labels. All the 
other transition probabilities can be obtained by permutation
of momentum labels.}
{bbggbuble}
%%%%%%%%%%%%%%%

We must sum only over the two physical polarizations of the
produced gluons. This is most easily accomplished by summing over
the polarizations with
\beq
\sum_{pol} \varepsilon^{\mu *} \varepsilon^{\nu} 
= - g^{\mu \nu}~,
\end{equation}
but including $B7$- and $B8$-like Feynman diagrams
with ``external'' ghosts\index{Ghosts} to take into account of the
fact that the gluon current is not conserved.

\section{Emission of four quarks}

Lastly, we calculate the matrix elements for the 
decay width of the $Z$-boson into four quarks. 
Two processes have to be considered
\beq
\bes
Z(q) & \rightarrow b(p_1) + \bar{b}(p_2) + q(p_3) + \bar{q}(p_4), \\
Z(q) & \rightarrow b(p_1) + \bar{b}(p_2) + b(p_3) + \bar{b}(p_4),
\ees
\end{equation}
where $q$ stands for a light quark.

%%%%%%%%%%%%%%%
\mafigura{7cm}{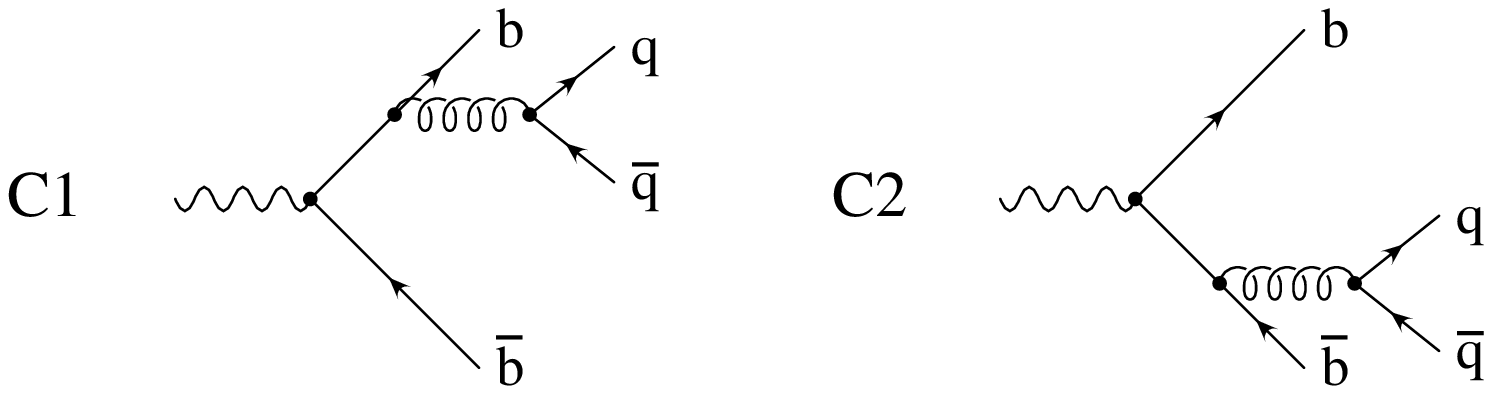}
{Feynman diagrams contributing to the process 
$Z \rightarrow b \bar{b} q \bar{q}$ where $q$ stands for a 
light quark.}
{bbqq}
%%%%%%%%%%%%%%%

For the first process we restrict to the case where
the pair of bottom-antibottom quarks
is emitted from the primary vertex, figure \ref{bbqq}.
Similar Feynman diagrams, where the heavy quark pair 
is radiated off a light $q\bar{q}$ system, are also
possible.
Despite the fact that $b$ quarks are present in 
this four fermion final state, the natural 
prescription is to assign these events to the light channels.
Obviously, they must be subtracted experimentally.
This should be possible since their signature is 
characterized by a large invariant mass of the light quark 
pair and a small invariant mass of the bottom system.
Since four fermion final states, $b\bar{b}q\bar{q}$,
originate from the interference between $q\bar{q}$
and $b\bar{b}$ induced amplitudes to assign them 
to a partial decay rate of a particular flavour is 
evidently not possible in a straightforward manner.

   The massless calculation~\cite{Ellis81,Kunszt89,Bethke92}
to which we want to compare our results was performed by summing
over all the allowed flavours. No ambiguity appears in this 
case. The massless QCD prediction is proportional to the 
the sum over all the squared vector and axial-vector
couplings 
\beq
\sum_{N_F} (g_{Vi}^2 + g_{Ai}^2),
\end{equation}
although this factor cancels in the three-jet decay rate ratios.
Our choice is the appropriate to get 1 for the
massive over the massless three-jet fraction ratio in
the limit of massless bottom quarks.
Conclusively, there is no way to solve this ambiguity and 
hence in the case of four fermion final states
the theoretical analysis should be tailored to the 
specific experimental cuts.

   The transition probabilities $C11$, $C22$ and $C12$
can generate infrared divergences, soft and quark-quark
collinear singularities, in the three-jet region due 
to the light quarks. In principle these divergences can manifest 
as double poles.~\footnote{For exact massless quarks. It is also
possible to regularize these infrared divergences
with a small quark mass. In this case, infrared divergences 
are softened into mass singularities and lead to 
large logarithms in the quark mass, $\log(m_q/\mu)$.
Infrared gluon divergences can be regulated at lowest order 
by giving a small mass, $\lambda$, to the gluons.\index{Gluon!mass}
At next-to-leading order we would violate gauge invariance
at the three gluon vertex.\protect\index{Gluon!three gluon vertex}} 
Nevertheless, as in the 
case of the gluon-gluon collinear divergences of the 
transition probabilities $B77$, $B88$ and $B87$ double 
poles cancel in the sum because these transition amplitudes
are related, by Cutkowsky rules\index{Cutkowsky rules},
to the diagrams with 
a gluon selfenergy\index{Gluon!selfenergy}
insertion in the Born amplitudes $T0(a)$
and $T0(b)$, where only simple infrared poles can appear.
All the other transition probabilities we will treat 
in this section are infrared finite.

%%%%%%%%%%%%%%%
\mafigura{7cm}{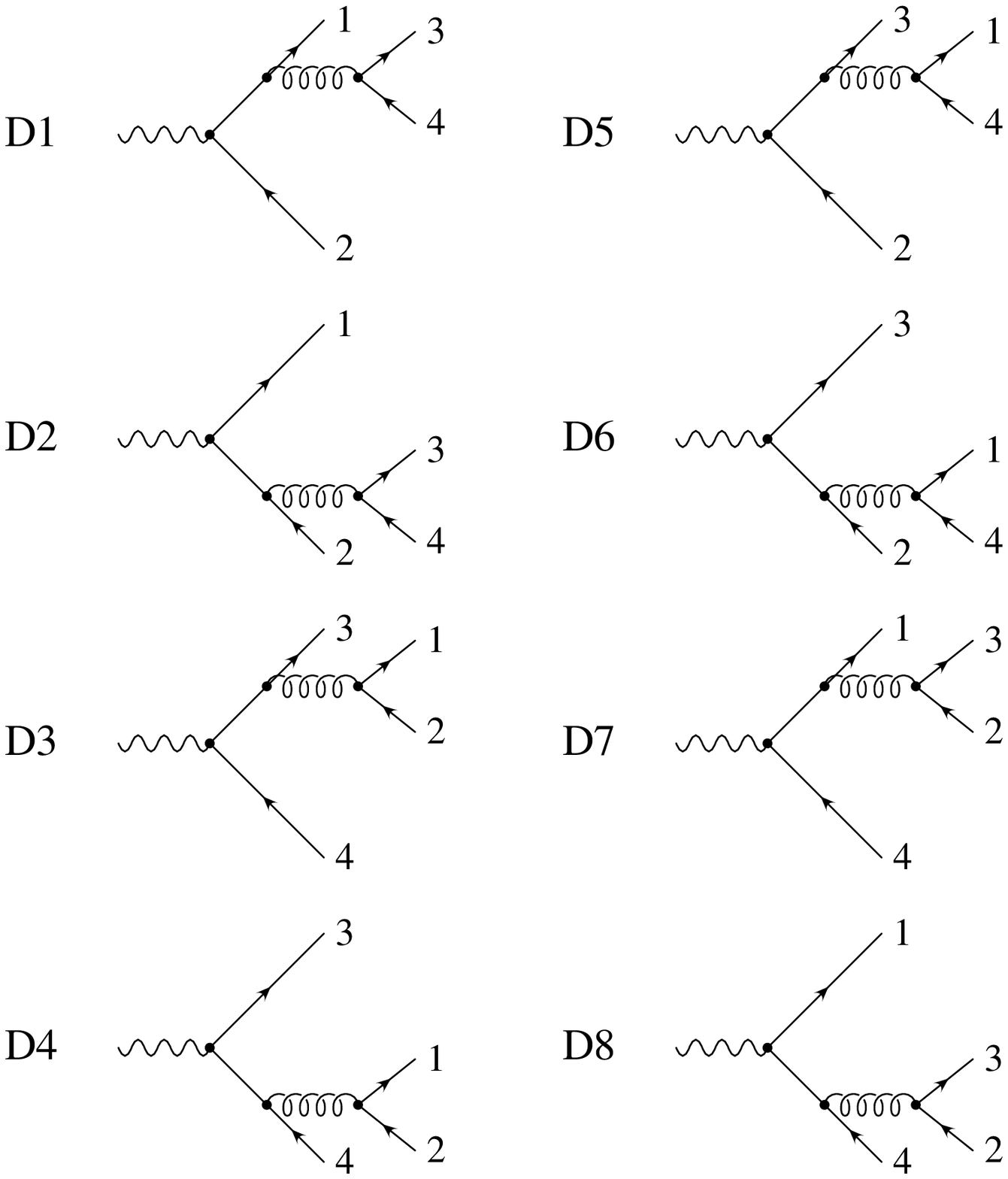}
{Feynman diagrams contributing to the decay width of the 
$Z$-boson into four massive bottom quarks.}
{bbbb}
%%%%%%%%%%%%%%%

We consider now the emission of four bottom quarks.
As in the case of the emission of two real gluons,
from the eight diagrams shown in figure \ref{bbbb} 
we should compute only twelve transition probabilities
because many of the, in principle, possible  
36 terms are related by interchange of momentum labels. 

\begin{table}[hbtp]
\begin{center}
\caption{The interchange table relating the graphs for 
$Z \rightarrow b \bar{b} b \bar{b}$. 
\label{permuta2}}
\begin{tabular}{|c|c|c|c|} \hline
label       & Class D   & Class E & Class F \\
permutation & $C_F T_R$ & $C_F(C_F-\frac{1}{2}N_C)$ & $C_F$ \\
\hline
      & D11 D12 D22 & D18 D25 D15 D28 D17 D26 & D13 D23 D24 \\
$(1 \leftrightarrow 3)$ 
      & D55 D56 D66 & D45 D16 D15 D46 D35 D62 & D57 D67 D68\\
$(2 \leftrightarrow 4)$ 
      & D77 D78 D88 & D27 D38 D37 D28 D17 D48 & D57 D58 D68\\
$(1 \leftrightarrow 3)$ $(2 \leftrightarrow 4)$ 
      & D33 D34 D44 & D36 D47 D37 D46 D35 D48 & D13 D14 D24\\
\hline
\end{tabular}
\end{center}
\end{table}

%%%%%%%%%%%%%%%
\mafigura{11cm}{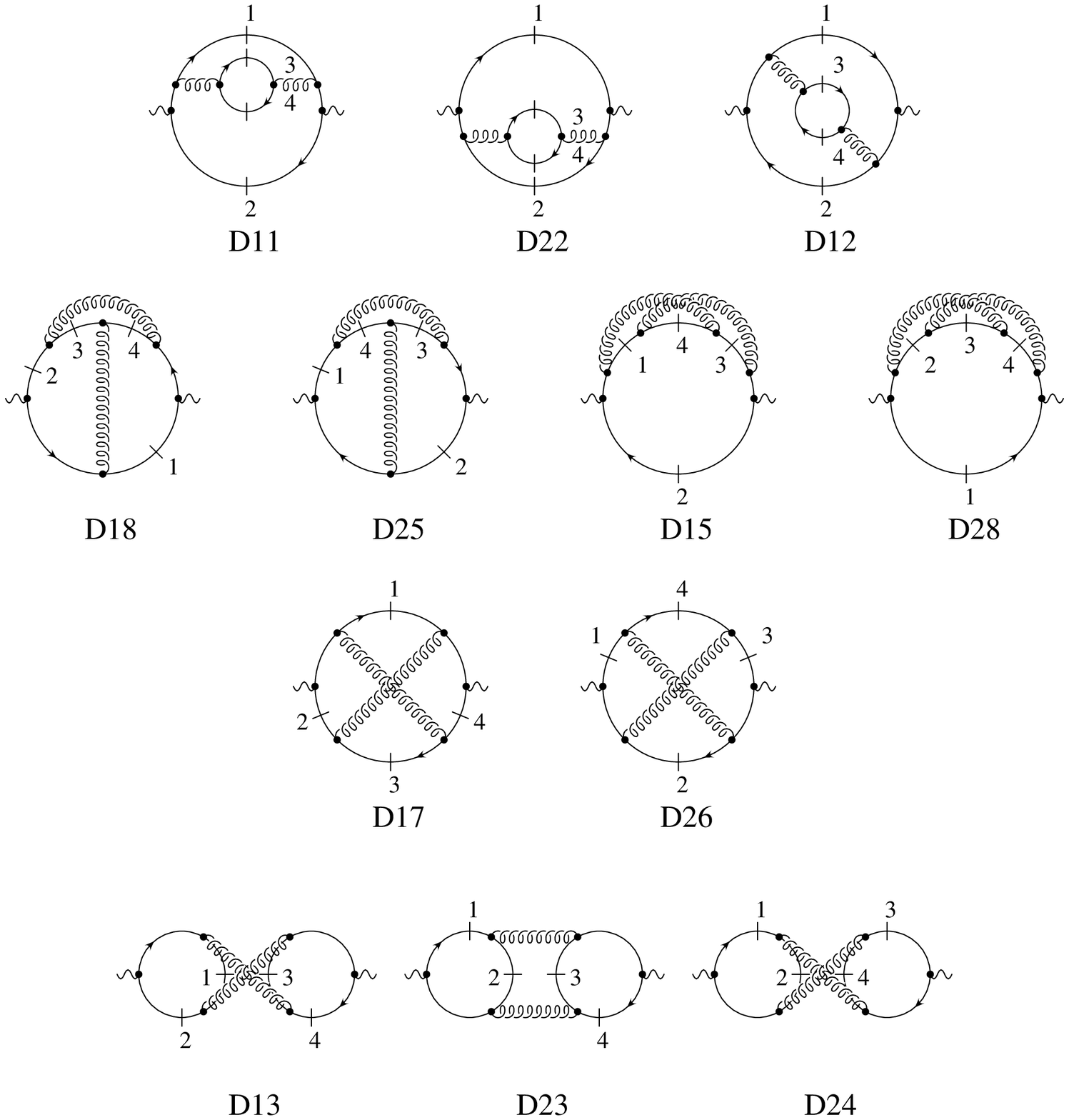}
{Transition probabilities for the process 
$Z \rightarrow b \bar{b} b \bar{b}$. 
On-shell particles are indicated
by a short cutting line and the numbers refer to the 
momentum labels. All the other transition probabilities can 
be obtained by exchange of momentum labels.} 
{babel}
%%%%%%%%%%%%%%%

It is sufficient to consider the following combinations
of transitions amplitudes\index{Class D}\index{Class E}\index{Class F}
\beq
\bes
\mrm{Class D} &= D11 + 2 D12 + D22~, \\
\mrm{Class E} &= - (D15 + D28 + D17 + D26 + 2(D18 + D25))~, \\
\mrm{Class F} &= D13 + 2 D23 + D24~,
\ees
\end{equation}
plus the interchanges $(1 \leftrightarrow 3)$, $(2 \leftrightarrow 4)$
and $(1 \leftrightarrow 3)$ $(2 \leftrightarrow 4)$.
Due to Fermi statistics\index{Fermi statistics}
there is a relative minus sign for diagrams $D5$ 
to $D8$ that is reflected in the transition probabilities
of Class E\index{Class E} 
and ensures that each helicity amplitude vanish when 
both fermions (antifermions) have identical quantum numbers.
Furthermore, we will call
\beq
\mrm{Class G} = \mid C1 + C2 \mid^2
\end{equation}
that evidently carries the same colour factor as the Class D 
transition probabilities, $C_F T_R$, where $T_R=1/2$.\index{TR@$T_R$}

The transition amplitudes of Class F\index{Class F}
are called in the literature~\cite{Chetyrkin94}, 
singlet contributions\index{Singlet contributions}
because they contain two different fermion loops and hence 
can be splited into two parts by cutting gluon lines only.
The first contribution to the vectorial part arises 
at $O(\as^3)$ as a consequence of the non-abelian 
generalization of Furry's theorem.\index{Furry's theorem} 
Singlet contributions to the axial part appear already 
at $O(\as^2)$. 

Lastly, let's comment on the Class F-like case where a bottom quark
is running in one of the loops and a light quarks is in the other.
This would produce a term proportional to the product 
of the vector-axial couplings of both quarks, $g_{Ab}g_{Aq}$,
and represents the interference of diagrams of figure \ref{bbqq}
with those we have not considered.
Nevertheless, since the vector-axial coupling of up and down quarks 
are different in sign, $g_{Au} = -g_{Ad} = 1$, their contribution 
will cancel when summing over all the light quarks and only
the diagram with bottom quarks running on both loops will survive.

This section concludes our classification of the 
transition probabilities we must compute in order to 
achieve the NLO QCD prediction for the three-jet decay 
rate of the $Z$-boson into massive quarks.
As we have commented, IR divergences are the main problem that 
appears when we try to compute them.
In the next chapter we will show how to extract from these transition
amplitudes the divergent pieces and how to keep away the finite 
parts. We will integrate analytically the divergent contributions over 
phase space and we will show how infrared divergences are cancelled.

\chapter{Infrared cancellations}
%\chapter{Cancellation of Infrared divergences}
\label{IRcancellations}

In the previous chapter we have presented the NLO 
virtual and real corrections to the decay width of the
$Z$-boson into three jets. We were able to reduce 
the contribution of the real corrections to the
calculation of a few transition probabilities
and the contribution of the virtual corrections 
to the calculation of a few scalar one-loop
n-point functions, see Appendix~\ref{app2}.
After UV renormalization\index{UV divergences}
of the loop diagrams we still 
encounter a plethora of IR divergences which should 
cancel with the soft and collinear singularities 
\index{IR singularities!soft}
\index{IR singularities!collinear}
of the tree level diagrams when we integrate over
the three-jet region of phase space. 
The theorems of Bloch-Nordsiek~\cite{IR1}
and Kinoshita-Lee-Nauenberg~\cite{IR2,IR3} guaranty such 
cancellation.
\index{Bloch-Nordsiek theorem}
\index{Kinoshita-Lee-Nauenberg theorem} 

We have classified our transition probabilities 
following their colour factors. It is clear that 
the cancellation of IR divergences can only occur 
inside groups of diagrams with the same colour factor.
The problem of the IR cancellation can be simplified
if we find a criteria to split our transition 
probabilities into different groups such that 
the cancellation of IR singularities can be shown
independently. The key is to depict them as bubble
diagrams\index{Bubble diagram}
as we did in figure \ref{bbggbuble} and
figure~\ref{babel} and perform all the 
possible cuts that could lead to three-jet final 
states, see figure~\ref{cancel}.

It is not difficult to convince ourselves that 
transition probabilities of Class A\index{Class A}
can be related only to the insertion of the quark selfenergy in the
external legs of the lowest order diagram $T0$.
The transition probabilities of Class B\index{Class B}
have the same IR structure as the one loop diagram $T5$.
Besides, $B71$, $B72$ and $B82$ lie in the 
same group as the one loop diagrams $T4$ and $T6$.
And finally, $B77$, $B87$, $C11$, $C12$, $D11$ and $D22$ 
are related with the gluon selfenergy insertion diagram  
at $T0$.\index{Gluon!selfenergy}
\index{T0}\index{T1}\index{T2}\index{T3}\index{T4}\index{T5}\index{T6}

In the following sections we are going to show analytically 
how the cancellation of the IR divergences occurs following 
the previous classification.

%\section{Groups 1 and 2}
\section{Soft divergences}\index{IR singularities!soft}

The phase space of $n+1$ particles can be written as the product of  
the $n$-body phase space times the integral over the 
energy and solid angle of the extra particle \cite{Was93}.
In arbitrary $D=4-2\epsilon$ dimensions we have
\beq
dPS(n+1) = \frac{1}{2(2\pi)^{D-1}} E^{D-3} \: dE \: d\Omega \: dPS(n)~. 
\end{equation}
Suppose $E_3$ is the energy of one soft gluon:
$E_3 < w$ where $w$, with $w$ very small, 
is an upper cut on the soft gluon energy. Let's consider
\beq
\int_0^w E_3^{D-3} dE_3 \: d\Omega \frac{1}{(2 p_1 \cdot p_3)^2}~, 
\end{equation}
that appears for instance in the four parton transition amplitudes
of Class A\index{Class A},
where $p_3$ is the momentum of the gluon and $p_1$ is the momentum of
the quark. After integration over the trivial angles we find
\beq
\frac{2\pi^{\frac{D}{2}-1}}{4 \Gamma \left(\frac{D}{2}-1\right)}
\int_0^w dE_3 E_3^{D-5} 
\int_{-1}^1 dx \frac{(1-x^2)^{\frac{D-4}{2}}}{(E_1-\mrm{\bf p}_1 x)^2}~, 
\end{equation}
with $\mrm{\bf p}_1$ the modulus of the threemomentum.
In the eikonal region we can suppose $E_1$ almost independent
from $E_3$
\beq
E_1 \simeq \frac{\sqrt{s}}{2} (1-y_{24})~, 
\end{equation}
with $y_{24} = 2 (p_2\cdot p_4)/m_Z^2$.
The integral over $E_3$ can be easily done and gives
a simple infrared pole in $\epsilon$. Since the $x$ dependent part
is completely finite we can expand in $\epsilon$ before
the integral over this variable is performed.
The final result we get is as follows
\begin{multline}
\frac{1}{2(2\pi)^{D-1}}
\int_0^w E_3^{D-3} dE_3 d\Omega \frac{1}{(2 p_1 \cdot p_3)^2} = \\
\frac{1}{16\pi^2} \frac{(4\pi)^{\epsilon}}{\Gamma(1-\epsilon)} 
\frac{w^{D-4}}{D-4}
\frac{1}{E_1^2-\mrm{\bf p}_1^2} \left[ 1 - \epsilon \left(
\frac{E_1}{\mrm{\bf p}_1} 
\log \frac{E_1-\mrm{\bf p}_1}{E_1+\mrm{\bf p}_1} + 2\log 2 \right) \right]~.
\label{energy1}
\end{multline}

Let's consider now
\beq
\int_0^w E_3^{D-3} dE_3 d\Omega 
\frac{1}{(2 p_1 \cdot p_3)(2 p_2 \cdot p_3)}~, 
\end{equation}
that appears in the four parton transition amplitudes of 
Class B\index{Class B},
with $p_2$ the momentum of the antiquark.
In this case we can perform a 
Feynman parametrization\index{Feynman!parametrization}
over the two momenta scalar products
\beq
\frac{1}{(p_1 \cdot p_3)(p_2 \cdot p_3)} = 
\int_0^1 dy \frac{1}{[p_3 \cdot (p_1+(p_2-p_1)y)]^2}~.
\end{equation}
We get therefore the same integral structure as in \Eq{energy1},
integral over $y$ left, but with the fourmomentum 
\beq
p_A = p_1+(p_2-p_1)y~.
\end{equation}
instead of $p_1$.
For the divergent part we get
\begin{multline}
\frac{1}{2(2\pi)^{D-1}}
\int_0^w E_3^{D-3} dE_3 d\Omega 
\frac{1}{(2 p_1 \cdot p_3)(2 p_2 \cdot p_3)} = \\
- \frac{1}{16\pi^2} \frac{(4\pi)^{\epsilon}}{\Gamma(1-\epsilon)} 
\frac{w^{D-4}}{D-4} \left[
\frac{2}{(y_{12}+2r_b) \beta_{12}} \log c_{12} + O(\epsilon) \right]~, 
\label{energy2}
\end{multline}
where
\beq
r_b = m_b^2/m_Z^2~,  \qquad 
\beta_{12} = \sqrt{1-\frac{4r_b}{y_{12}+2r_b}}~,  \qquad 
c_{12} = \frac{1-\beta_{12}}{1+\beta_{12}}~. 
\end{equation}
It is not easy to get a full analytical expression for the finite part.
Nevertheless, the integral over the $y$ parameter can be done numerically,
using for instance a simple GAUSS integration. Notice this last
integral has the same divergent structure as the scalar one loop
$C05$ function defined in \Eq{C05}.

The matrix element of diagram $B2$ reads
\begin{multline}
B2 = i \frac{g}{4c_W} g_s^2 (T^a T^b) \bar{u}(p_1) \\ 
\left\{ \gamma_{\nu} \frac{1}{\pb_{13}-m_b} \gamma_{\sigma} 
\frac{1}{\pb_{134}-m_b} \gamma_{\mu}(g_V+g_A \gamma_5)
\right\} v(p_2)  
\varepsilon_a^{*\nu}(p_3) \varepsilon_b^{*\sigma}(p_4)
\varepsilon_Z^{\mu}(q)~,
\end{multline}
where $T^a=\lambda^a/2$ are the colour Gell-Mann matrices,
$\varepsilon_a^{*\nu}$, $\varepsilon_b^{*\sigma}$
and $\varepsilon_Z^{\mu}$ stand for the polarization vector
of the two gluons and the $Z$-boson 
and $g_V$ ($g_A$) are the vector (axial-vector) neutral 
current couplings of the quarks in the Standard Model. 
At tree-level and for the bottom quark we have 
\beq
g_V = -1 + \frac{4}{3} s_W^2, \qquad g_A = 1~.
\end{equation}
We denote by $c_W$ and $s_W$ the cosine and the sine of the 
weak mixing angle.

    With the help of the equation of motion we can write
\beq
\bar{u}(p_1) \gamma^{\nu} (\pb_{13}+m_b) =
\bar{u}(p_1) (\gamma^{\nu} \pb_3 + 2 p_1^{\nu})~.
\end{equation}
In the limit of gluon labelled as 3 soft, $p_3 \rightarrow 0$,
it reduces to $2 p_1^{\nu} \bar{u}(p_1)$ and 
$B2$ behaves as the Born amplitude $T0(a)$
\beq
B2 \simeq - g_s (T^a) \frac{2 p_1^{\nu}}{y_{13}}
T0(a) \varepsilon_a^{*\nu}(p_3)~.
\label{B2eikonal}
\end{equation}
Hence, in this limit we find for the transition amplitude 
$B22$ the following result
\beq
B22 \simeq - g_s^2 C_F \frac{4 r_b}{y_{13}^2} \mid T0(a) \mid^2~.
\end{equation}
Integrating over phase space with the help of \Eq{energy1}
we find
\beq
dPS(4) B22 \simeq C_F \frac{\as}{4\pi} 
\frac{(4\pi)^{\epsilon}}{\Gamma(1-\epsilon)}
\frac{2 w^{-2\epsilon}}{\epsilon} dPS(3) [1+O(\epsilon)] \mid T0(a) \mid^2~. 
\end{equation}

Same argument can be applied to the transitions amplitudes
$B11$ and $B21$ for which we get
\beq
\bes
B11 &\sim \frac{1}{\epsilon} ( \mid T0(a) \mid^2 + \mid T0(b) \mid^2 )~,\\
B21 &\sim \frac{1}{\epsilon} T0(a)^* T0(b)~.
\label{B11}
\ees
\end{equation}
For the moment we don't need to consider $B32$
because it is infrared finite.
Since the transition amplitude $B11$ 
can be soft in both gluons, labelled 3 and 4,
we have splited off the contribution of each one
by performing a partial fractioning
\beq
\frac{1}{y_{13}^3 y_{24}^2} = \frac{1}{y_{13}^2+y_{24}^2} 
\left( \frac{1}{y_{13}^2} + \frac{1}{y_{24}^2} \right)~.
\end{equation}
The first term should be integrated in the 1-3 system, second 
term in the equivalent 2-4 system. Relabelling in both cases
the remaining hard gluon as 3 we arrive at the result quoted
in \Eq{B11}. 

Finally, taking into account all the momenta permutations, 
we find for the eikonal contribution of the diagrams of
Class A\index{Class A} to the decay width of the Z-boson into 
massive bottom quarks the following result 
\beq
\Gamma_w(\mrm{Class A}) = 
\frac{8}{2!} C_F \frac{\as}{4\pi} 
\frac{(4\pi)^{\epsilon}}{\Gamma(1-\epsilon)}
\frac{w^{-2\epsilon}}{\epsilon} \frac{1}{2m_Z}
\int dPS(3) \mid T0(a) + T0(b) \mid^2~,
\label{ClassA}
\end{equation}
where $2!$ is the statistical factor and
\beq
\bes
& \frac{16 c_W^2}{g^2} \frac{1}{g_s^2 C_F} \mid T0(a) + T0(b) \mid^2  = \\ 
& g_V^2 \left[ 8(D-2+4r_b)\frac{h_p}{y_{13}^2 y_{23}^2}
+ 2(D-2)^2 \left( \frac{y_{13}}{y_{23}}+\frac{y_{23}}{y_{13}} \right)
+ 4(D-2)(D-4) \right] \\
+ &  g_A^2 (D-2) \left[ 8(1-4r_b)\frac{h_p}{y_{13}^2 y_{23}^2}
+ 2(D-2+4r_b) \left( \frac{y_{13}}{y_{23}}+\frac{y_{23}}{y_{13}} \right)
+ 4(D-4+4r_b) \right]~, 
\ees
\end{equation}
with
\beq
h_p = y_{13} y_{23}(1-y_{13}-y_{23}) - r_b (y_{13}+y_{23})^2~,
\end{equation}
is the squared lowest order transition amplitude
in $D$-dimensions.

The wave function renormalization constant of the quark
propagator contains two pieces
\beq
\bes
Z_2 = 1 - & C_F \frac{\as}{4\pi} \tau_0 
\left[ \Delta_{UV} - \log \frac{m_b^2}{\mu^2} + 2 \right] \\
        - & C_F \frac{\as}{4\pi} \left\{ 
(3-\tau_0) \left[ \Delta_{IR} - \log \frac{m_b^2}{\mu^2} \right]
+ 2(2-\tau_0) \right\}~,
\ees
\end{equation}
where $\Delta = 1/\epsilon - \gamma_E + \log 4\pi$ and
$\tau_0$ is the gauge parameter, $\tau_0=1$ since 
we work in the Feynman gauge.\index{Feynman!gauge}
The first one is the usual UV piece and cancels the UV divergences 
of diagram $T1$. The second piece comes from the residue 
of the $\overline{MS}$ renormalized propagator in the pole.
When we include it in the self-energy diagrams of external 
quarks we get a contribution  
\beq
- 2 \frac{\as}{4\pi} C_F 2 \Delta_{IR} 
\frac{1}{2m_Z} \int dPS(3) \mid T0(a) + T0(b) \mid^2~,
\end{equation}
which exactly cancels the IR divergences of \eqref{ClassA}.

As we made with $B2$ in \eqref{B2eikonal} 
the matrix element of diagram $B4$ reads
\beq
B4 \simeq  g_s (T^a) \left(C_F-\frac{N_C}{2}\right)
\frac{2 p_2^{\nu}}{y_{23}}
T0(a) \varepsilon_a^{*\nu}(p_3)~,
\end{equation}
for gluon labelled as 3 soft.
The colour factor comes from the following 
properties of the $SU(3)$ Gell-Mann matrices
\beq
\bes
& [T_a,T_b] = i \: f_{abc} T^c~, \\
& f_{abc} T^b T^c = i \frac{N_C}{2} T_a~.
\ees
\end{equation}

Therefore, the transition amplitude $B42$ behaves as 
\beq
B42 \simeq g_s^2 \left(C_F-\frac{N_C}{2}\right)
\frac{4 (p_1\cdot p_2)}{y_{13} y_{23}}
\mid T0(a) \mid^2~,
\end{equation}
in the eikonal region. Including phase space,
from \Eq{energy2} we get for its divergent piece
\begin{multline}
dPS(4) B42 \simeq \left(C_F-\frac{N_C}{2}\right) \frac{\as}{4\pi} 
\frac{(4\pi)^{\epsilon}}{\Gamma(1-\epsilon)} \\
\frac{2 w^{-2\epsilon}}{\epsilon} dPS(3)  
\left[ \frac{y_{12}}{(y_{12}+2 r_b) \beta_{12}} \log c_{12}
+O(\epsilon) \right] \mid T0(a) \mid^2~. 
\end{multline}
As before we find for the other transition amplitudes 
of Class B\index{Class B} the following behaviour
\beq
\bes
B41 &\sim \frac{1}{\epsilon} [ T0(a)^* T0(b) + T0(b)^* T0(a) ]~, \\
B62 &\sim \frac{1}{\epsilon}  T0(a)^* T0(b)~,
\ees
\end{equation}
where for $B41$ we made the following partial fractioning
\beq
\frac{1}{y_{13} y_{23} y_{14} y_{24}} = 
\frac{1}{y_{13} y_{23} + y_{14} y_{24}} 
\left( \frac{1}{y_{13} y_{23}} + \frac{1}{y_{14} y_{24}} \right)~.
\end{equation}
The final answer for this class of diagrams reads
\begin{multline}
\Gamma_w(\mrm{Class B}) = 
\frac{8}{2!} \left(C_F-\frac{N_C}{2}\right) \frac{\as}{4\pi} 
\frac{(4\pi)^{\epsilon}}{\Gamma(1-\epsilon)} \\
\frac{w^{-2\epsilon}}{\epsilon} 
\frac{1}{2m_Z} \int dPS(3) 
\frac{y_{12}}{(y_{12}+2 r_b) \beta_{12}} \log c_{12}
\mid T0(a) + T0(b) \mid^2~.
\label{ClassB}
\end{multline}

Let's consider now the one-loop box diagram $T5$
\beq
\bes
T5(a) & = -g_s^3 \bar{u}(p_1) (T^b T^a T^b)
\int \frac{d^D k}{(2\pi)^D}  \\
& \frac{\gamma^{\sigma} [(\kb+\pb_1)+m_b] \gamma^{\nu}
[(\kb+\pb_{13})+m_b] \gamma^{\mu}(g_V+g_A \gamma_5)
[(\kb-\pb_2)+m_b] \gamma^{\sigma'}}
{k^2 [(k+p_1)^2-m_b^2] [(k+p_{13})^2-m_b^2] [(k-p_2)^2-m_b^2]} \\ 
& \left[g_{\sigma \sigma'} - (1-\tau_0) \frac{k_{\sigma}k_{\sigma'}}{k^2}
\right] v(p_2) \varepsilon_a^{*\nu}(p_3) \varepsilon_Z^{\mu}(q)~.
\ees
\end{equation}
Again we apply the equation of motion of the quarks
\beq
\bes
\bar{u}(p_1) \gamma^{\sigma} [(\kb+\pb_1)+m_b] 
&= \bar{u}(p_1) ( \gamma^{\sigma} \kb + 2 p_1^{\sigma} )~, \\
[(\kb-\pb_2)+m_b] \gamma^{\sigma'} v(p_2)
&= (\kb \gamma^{\sigma'} - 2 p_2^{\sigma'} ) v(p_2)~, 
\ees
\end{equation} 
at both sides of the previous equation.
Dropping all the $\kb$ factors in the numerator
and expanding the third propagator for small loop 
momentum we arrive to
\beq
T5(a) \simeq - \frac{\as}{4\pi} (4 p_1\cdot p_2) C05(y_{12}) T0(a)~,
\label{T5a}
\end{equation}
where $C05$ is defined in~\Eq{allC0}. It is straitforward
to see from the solution \eqref{C05} to the one-loop integral 
$C05$ how the interference of the one loop amplitude of
\Eq{T5a} with the lowest order amplitude $T0$
cancels exactly the IR divergences of 
the diagrams grouped in Class B\index{Class B}, \Eq{ClassB}.

%\section{Group 3}
\section{Collinear divergences}\index{IR singularities!collinear}

To show how the cancellation of the infrared divergences 
for diagrams of Class C\index{Class C} occurs is quite more difficult.
Since we have gluon-gluon collinear divergences
it is not enough to make a cut in the energy of one of the gluons,
as we made in the previous section, to extract the divergent piece.
In this case the appropriate variable over which we have to
impose a cut is $y_{34}=2(p_3\cdot p_4)/m_Z^2$ as it includes
both kind of divergences, soft and collinear, in the limit of
$y_{34}\rightarrow 0$. Looking at the four partons phase space
in the so-called ``system 3-4'' defined in section \ref{secsystem34}
of appendix \ref{app2} we can notice several things.
First, in the limit $y_{34}\rightarrow 0$
the function $h_p$ that defines the limits of the phase
space reduces to the three parton phase space
$h_p$ function~\eqref{bornPS}.
Second, the momentum $p_{34}=p_3+p_4$ behaves as the 
momentum of a pseudo on-shell gluon because
$p_{34}^2\rightarrow 0$ in this limit.
Therefore, it is possible to factorize 
from the four-body phase space a three-body phase space 
with an effective gluon of momentum $p_{34}$. Only 
integration over the extra variables should be performed to 
show the cancellation of the infrared divergences.

\subsection{Diagrams containing gluon-gluon collinear divergences}

Let's consider the following phase space integral
\beq
PS(4) \frac{1}{y_{13}y_{34}}~,
\end{equation}
in the called system 3-4~\eqref{PSsystem34}, where as usual $p_1$
denotes the momentum of the quark and $p_3$ and $p_4$ are the 
two gluon momenta. This integral can be soft for gluon labelled as 
3 and collinear because of the scalar product of the two gluon
momenta $y_{34}$. In the eikonal region, $y_{34}<w$ with $w$ 
very small, we can decompose the four body-phase space as the 
product of a three-body phase space in terms of variables 
$y_{134}$ and $y_{234}$ times the integral over $y_{34}$ and 
the two angular variables $v$ and $\theta'$
\begin{multline}
PS(4) \frac{1}{y_{13}y_{34}} = PS(3)(y_{134},y_{234}) \\
\frac{S}{16 \pi^2}  \frac{(4\pi)^\epsilon}{\Gamma(1-\epsilon)}
\int_0^w dy_{34}
\int_0^1 dv (v(1-v))^{-\epsilon}
\frac{1}{N_{\theta'}} \int_0^{\pi} d\theta' \sin^{-2\epsilon} \theta'
\frac{y_{34}^{-1-\epsilon}}{y_{13}}~.
\end{multline}
$S=1/2!$, the statistical factor.
In the 3-4 system the two-momenta invariant $y_{13}$ can be written 
in terms of the integration variables as
\beq
y_{13} = \frac{1}{2} \left( y_{134} - \sqrt{y_{134}-4 r_b y_{34}} (1-2v)
\right)~.
\label{y13}
\end{equation}
The $y_{13}$ factor is independent of the $\theta'$ angle
so we can get rid of this first integral. Second observation, 
$y_{13}$ contain a piece independent of the $v$ angle and 
another one that is odd under the interchange
$v \leftrightarrow(1-v)$. With the help of this symmetry
\beq
\frac{1}{2} \left( \frac{1}{A+f(v)} + \frac{1}{A-f(v)} \right)
= \frac{A}{A^2-f(v)^2}~,
\end{equation}
we can avoid the use of the square root of \Eq{y13} and write 
our integral as 
\bea
& &PS(4) \frac{1}{y_{13}y_{34}} = PS(3)(y_{134},y_{234}) 
\frac{S}{16 \pi^2}  \frac{(4\pi)^\epsilon}{\Gamma(1-\epsilon)} \\
& & \int_0^w y_{34}^{-1-\epsilon} dy_{34}
\int_0^1 dv (v(1-v))^{-\epsilon}
\frac{2 y_{134}}{y_{134}^2-(y_{134}^2-4 r_b y_{34})(1-2v)^2}~.
\nonumber
\end{eqnarray}

After integration over $y_{34}$ we get the first infrared pole
\beq
\frac{-1}{\epsilon} \frac{w^{-\epsilon}}{2 y_{134}}
\int_0^1 dv (v(1-v))^{-1-\epsilon}
{}_2F_1[1,-\epsilon,1-\epsilon,-\frac{r_b w (1-2v)^2}{y_{134}v(1-v)}]~.
\end{equation}
Some simple mathematical manipulations~\cite{Kramer87,AS72,BA53,PT84}
allow us to express the 
hypergeometric function\index{Hypergeometric function}
in terms of a simple dilogarithm\index{Dilogarithm function}
\beq
\bes
{}_2F_1[1,-\epsilon,1-\epsilon,-\frac{a}{b}] &= 
\left(\frac{b}{a+b}\right)^{-\epsilon}
{}_2F_1[-\epsilon,-\epsilon,1-\epsilon,\frac{a}{a+b}] \\ &=
\left(\frac{b}{a+b}\right)^{-\epsilon}
\left[ 1+ \epsilon^2 Li_2\left(\frac{a}{a+b}\right) 
+ O(\epsilon^3) \right]~,
\ees
\end{equation}
to finally obtain 
\begin{multline}
\frac{-1}{\epsilon} \frac{w^{-\epsilon}}{2 y_{124}}
\int_0^1 dv (v(1-v))^{-1-2\epsilon}
\left( v(1-v) + A (1-2v)^2 \right)^{\epsilon} \\
\left[ 1 + \epsilon^2 
Li_2 \left( \frac{A (1-2v)^2}{v(1-v)+A (1-2v)^2} \right) +
O(\epsilon^3) \right]~,
\end{multline}
with $A = r_b w/y_{134}^2$.
The contribution of the dilogarithm function reduces to its value
at the border of the integration region, i.e., just additional 
$Li_2(1) = \pi^2/6$ factors to the finite part.
Further contributions would be next order in $\epsilon$.
Therefore, it is enough to analyze the first two factors.
This integrand is symmetric under the interchange 
$v \leftrightarrow (1-v)$. We perform the integral only 
in half of the integration region and apply the 
following change of variables
\beq
u = 4 v(1-v)~,
\end{equation}
to get
\beq
\frac{-1}{\epsilon} 
\frac{w^{-\epsilon} A^{\epsilon}}{y_{134}} 2^{4\epsilon}
\int_0^1 du u^{-1-2\epsilon} (1-u)^{-1/2}
\left[ 1- \left( 1-\frac{1}{4A} \right) u \right]^{\epsilon}~,
\end{equation}
which gives again an hypergeometric function
\beq
\frac{-1}{\epsilon} 
\frac{1}{y_{134}} 
\left( \frac{r_b}{y_{134}^2} \right)^{\epsilon} 2^{4\epsilon}
\sqrt{\pi} \frac{\Gamma(-2\epsilon)}{\Gamma(\frac{\D 1}{\D 2}-2\epsilon)}
{}_2F_1[-\epsilon, -2\epsilon,\frac{1}{2}-2\epsilon,
1-\frac{1}{4A}]~.
\end{equation}
The hypergeometric function is already $1+O(\epsilon^2)$
but now it is not possible to write it in terms of a simple 
dilogarithm function.
If we are interested just in the pole structure we can stop here.
Observe that neither the double pole or the simple one depend on
the eikonal cut $w$. To get the complete finite part further 
mathematical manipulations must be performed on the
hypergeometric function. 

Be
\beq
{}_2F_1[-\epsilon, -2\epsilon,\frac{1}{2}-2\epsilon,
1-\frac{1}{4A}] = \nonumber
\end{equation}
\begin{multline} 
(4A)^{2\epsilon} \Gamma(\frac{\D 1}{\D 2}-2\epsilon) \left\{ 
 \frac{\Gamma(\epsilon)}
{\Gamma(-\epsilon)\Gamma(\frac{\D 1}{\D 2})}
{}_2F_1[-2\epsilon, \frac{1}{2}-\epsilon, 1-\epsilon, 4A] \right. \\
+ \left. \frac{\Gamma(-\epsilon)}
{\Gamma(-2\epsilon) \Gamma(\frac{\D 1}{\D 2}-\epsilon)}
{}_2F_1[-\epsilon, \frac{1}{2}, 1+\epsilon, 4A] \right\}~.
\end{multline}
For $w$ enough small, $w \ll y_{134}^2/r_b$, 
we can use the Gauss series \cite{AS72} to expand the
hypergeometric functions around $A \rightarrow 0$.
The final result we obtain is as follows
\beq
\bes
dPS(4) \frac{1}{y_{13}y_{34}} = dPS(3)(y_{134},y_{234}) & \\
\frac{S}{16 \pi^2}  \frac{(4\pi)^\epsilon}{\Gamma(1-\epsilon)}
\frac{1}{y_{134}} \left\{
\frac{1}{2\epsilon^2} 
+ \frac{1}{2\epsilon} \log \frac{r_b}{y_{134}^2} \right. 
+ & \frac{1}{4} \log^2 \frac{r_b}{y_{134}^2} - \frac{1}{2} \log^2 A
- \frac{\pi^2}{4}  \\
- & 2 A \log A - A^2 (2+3 \log A) + O(A^3) \biggr\}~.
\ees
\end{equation}
Notice we get the same infrared poles as in the one-loop 
three point function $C03$, \Eq{C03},
appearing in the one-loop amplitudes $T4$ and $T6$.

%%%%%%%%%%%%%%%
\mafigura{10cm}{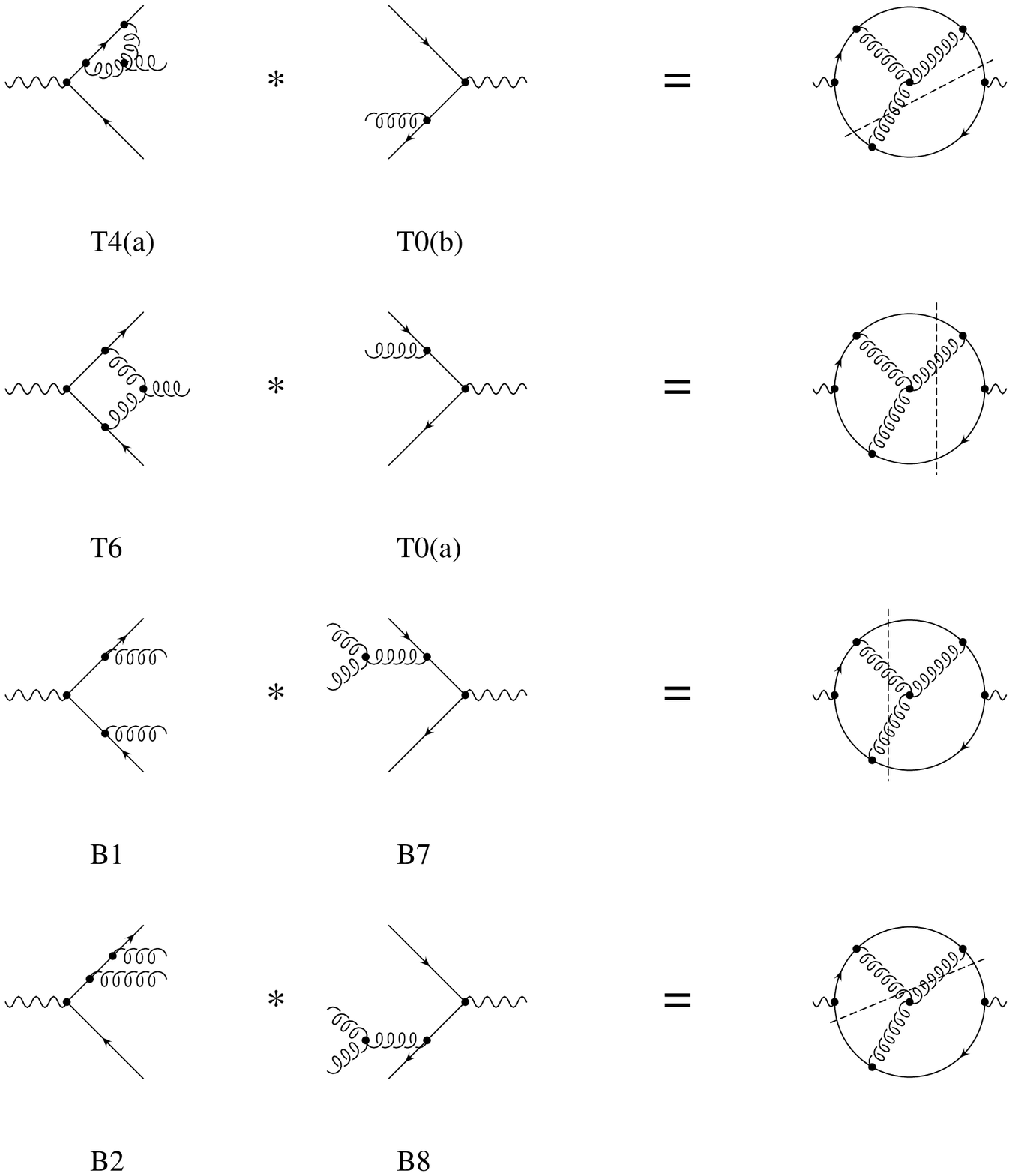}
{Cuts over a bubble diagram with 
three gluon vertex\protect\index{Gluon!three gluon vertex}
and transition probabilities with related IR structures
in the three-jet region. The other possible cuts 
lead to two-jet final states.\protect\index{Bubble diagram}}
{cancel}
%%%%%%%%%%%%%%%

\subsection{Gluon selfenergy-like diagrams}\index{Gluon!selfenergy}

We discuss now the infrared behaviour of
the transition amplitudes $B77$, $B87$, $C11$, $C12$ 
and those grouped in Class D.
Diagrams of Class D, that correspond to the emission 
of four bottom quarks, are infrared finite since 
the infrared singularities are softened into 
logarithms of the bottom quark mass and can be 
calculated in $D=4$ dimensions and integrated 
numerically without any trouble.

The infrared divergences of the transition probabilities
$C11$ and $C12$ that correspond to the emission 
of a pair of light quarks can be regulated by introducing
a small light quark mass. Nevertheless, we chose
Dimensional Regularization.\index{Dimensional Regularization}
Since kinematically massless quarks and 
gluons are equivalent we can integrate the $B77$, $B87$, $C11$ and
$C12$ transition probabilities in the same system 3-4.

For the transition amplitudes $B77$ and $B87$
we have to solve the following basic phase space integral
\beq
PS(4) \frac{1}{y_{34}}~.
\end{equation}
As we said we work in the system 3-4, \eqref{PSsystem34}.
All the integrals factorize giving rise to a simple result
in the eikonal region
\beq
dPS(4) \frac{1}{y_{34}} = 
\frac{S}{16 \pi^2}  \frac{(4\pi)^\epsilon}{\Gamma(1-\epsilon)} 
\left[ -\frac{1}{\epsilon} + \log w - 2 \right]
dPS(3)(y_{134},y_{234})~.
\end{equation}

As we have mentioned, the transition probabilities $B77$ and
$B87$ contain in principle double infrared poles.
Nevertheless, the sum of them, i.e., the square of the sum of 
amplitudes $B7$ and $B8$, is proportional to 
\beq
\left( \frac{y_{13}}{y_{134}} - \frac{y_{23}}{y_{234}} \right)^2 
\frac{1}{y_{34}^2}~.
\end{equation}
Expanding for small values of $y_{34}$ we find
\beq
4 v (1-v) \cos^2 \theta'
 \left( \frac{1-2r_b-y_{134}-y_{234}}{y_{134}y_{234}}
- \frac{r_b}{y_{134}^2} - \frac{r_b}{y_{234}^2}
\right) \frac{1}{y_{34}} + O(y_{34}^{-1/2})~.
\end{equation}
Therefore, only simple poles can appear.
Same arguments apply for 
the transition probabilities $C11$ and $C12$, that
we grouped under Class G.\index{Class G}

This completes the calculation for the collinear divergent 
diagrams grouped in Class C\index{Class C}
and Class G\index{Class G}.
Intermediate steeps are rather involved but the final result
can be written in a simple manner in terms of the Born amplitude
\beq
\bes
\Gamma_w(\mrm{Class C}) & = 
\frac{1}{2!} N_C \frac{\as}{4\pi} 
\frac{(4\pi)^{\epsilon}}{\Gamma(1-\epsilon)}  
\frac{1}{2m_Z} \int dPS(3)(y_{13},y_{23}) \\
\Bigg\{ & 4 \left[\frac{w^{-\epsilon}}{\epsilon} + 2 
+ I^w(y_{23}) + I^w(y_{13}) \right] \\
+ & \left[ 
\frac{10}{3} \frac{w^{-\epsilon}}{\epsilon} + \frac{62}{9} \right] \Bigg\}
\mid T0(a) + T0(b) \mid^2~,
\label{ClassC}
\ees
\end{equation}
where
\beq
I^w(x) = 
\frac{1}{2\epsilon^2} 
+ \frac{1}{2\epsilon} \log \frac{r_b}{x^2} 
+ \frac{1}{4} \log^2 \frac{r_b}{x^2} - \frac{1}{2} \log^2 \frac{r_b w}{x^2}
- \frac{\pi^2}{4} + O\left(\frac{r_b w}{x^2}\right)~,
\end{equation}
and
\beq
\bes
\Gamma_w(\mrm{Class G}) & = 
T_R (N_F-1) \frac{\as}{4\pi} 
\frac{(4\pi)^{\epsilon}}{\Gamma(1-\epsilon)}  
\frac{1}{2m_Z} \int dPS(3)(y_{13},y_{23}) \\
& \left[ 
-\frac{4}{3} \frac{w^{-\epsilon}}{\epsilon} - \frac{20}{9} \right] 
\mid T0(a) + T0(b) \mid^2~,
\label{ClassG}
\ees
\end{equation}
with $T_R=1/2$ and $(N_F-1)$ the number of light quarks.

Since, as we mentioned, $p_{34}$ behaves in the IR region
as the momentum of a pseudo on-shell massless particle,
$p_{34}^2 \rightarrow 0$, we have relabelled the integration
variables $(y_{134},y_{234})$ as $(y_{13},y_{23})$.
First contribution of \Eq{ClassC} comes from the transition 
probabilities $B71$, $B72$ and $B82$ and cancels 
the IR divergences of the one-loop diagrams $T4$ and $T6$.
To show it explicitly needs a considerable amount of algebra. 
We get terms proportional to the simple infrared pole and 
terms proportional to the $C03$ one-loop function that 
exactly cancels the $I^w$ function divergent piece.
Nevertheless, we can justify such cancellation using 
simple graphical arguments as we have shown in figure~\ref{cancel}.    
In fact, this was the main argument we used to classify 
all the transition probabilities in different groups such
that IR divergences cancel independently.
Second contribution comes from the transition amplitudes 
$B77$ and $B87$ and together with the 
Class G contribution is cancelled against
\beq
- 2 \frac{\as}{4\pi} 
\left[ \frac{5}{6}N_C -\frac{2}{3} T_R(N_F-1) \right]
\Delta_{IR} 
\frac{1}{2m_Z} \int dPS(3) \mid T0(a) + T0(b) \mid^2~,
\end{equation}
with $\Delta_{IR}=1/\epsilon-\gamma_E + \log 4\pi$, coming
from the gluon wave function renormalization constant.

\chapter{Numerical results and conclusions}

We have seen in the previous chapter how infrared singularities 
cancel between the three parton one-loop and the four parton
tree-level diagrams contributing at NLO to the $Z$-boson
three-jet decay rate. In fact, due to the
Bloch-Nordsiek~\cite{Bloch37}\index{Bloch-Nordsiek theorem} 
and Kinoshita-Lee-Nauenberg~\cite{Kinoshita62,Lee64} 
theorems,\index{Kinoshita-Lee-Nauenberg theorem}
we knew from the beginning they should do.
The cancellation of the IR divergences is of course the 
first test of our calculation.
Nevertheless, the real challenge is in the calculation 
of the finite contributions.
 
Ultraviolet singularities\index{UV divergences}
arise in loop integrals once the integration over the 
internal momentum is performed and in the 
form of a singular Gamma function.
Moreover, we are allowed to expand in $\epsilon$
before integrating over the 
Feynman parameters.\index{Feynman!parameters}
Main problem of IR divergences is that
they appear always at the borders. 
At the border of the one unit side box defined by the Feynman
parameters\index{Feynman!parameters}
in loop integrals, and at the border of the 
phase space for the tree-level transition amplitudes.
Therefore, we are forced to perform in arbitrary $D$-dimensions
the full calculation in the infrared region. 
This makes the computation of their finite contributions
extremely difficult.
Even for the finite contributions
strong cancellations occur among 
different groups of diagrams making very difficult the 
numerical approach.

We have taken as guide line 
the massless result~\cite{Bethke92,Kunszt89,Ellis81}
although the IR structure in the massive case
is completely different from the massless one.
With massive quarks we loose all the quark-gluon 
collinear divergences. The amplitudes behave better in the 
IR region. The disadvantage however is the mass itself 
because we have to perform quite more complicated
loop and phase space integrals. 
Furthermore, the gluon-gluon collinear divergences,
leading to IR double poles and whose finite contributions
are harder to solve, are still present.

Our procedure was as follows. For the 
one-loop transition probabilities 
we were able to reduce the loop integrals 
to a few scalar one-loop n-point functions whose 
finite parts are under control, see Appendix \ref{loopintegrals}.
We drop all the terms proportional to $1/\epsilon^2$ and 
$1/\epsilon$ since, as we have seen in chapter~\ref{IRcancellations},
they cancel against the four parton real corrections
to the three-jet decay rate. 
Since the boundary of the three-body phase space,
including cuts, is a complicated function 
we perform the remaining phase space integration
over the one-loop finite contributions with the 
help of {\it VEGAS}~\cite{Lepage78,Lepage80}\index{VEGAS}, an 
adaptative Monte Carlo\index{Monte Carlo} FORTRAN algorithm for 
multidimensional integration.  

We have seen that the four parton transition amplitudes have
a simple and well known behaviour in the soft and collinear regions,
basically proportional to the lowest order three parton transition 
probabilities, making the integration over these regions
of the four-body phase space feasible analytically.
Our strategy was to exclude from the numerical integration
domain the singular regions. We impose a cut, we called $w$,
with $w$ very small, in the energy of the soft gluons. For 
the diagrams with collinear singularities we cut 
on the two momenta invariant variable $y_{34}=2(p_3\cdot p_4)/s$.
In the excluded phase space regions we replace 
the transitions amplitudes by their limiting values and 
perform analytically the integration over 
this phase space singular region in 
arbitrary $D=4-2\epsilon$ dimensions 
(we only need to analytically integrate
three of the five independent variables defining
the four-body phase space).
The analytic integration gives the IR poles in 
$1/\epsilon^2$ and $1/\epsilon$ that again 
we drop\footnote{The IR poles appear multiplied by 
the Born amplitude calculated in $D$-dimensions 
and by some $\epsilon$-dependent factors like 
$(4\pi)^{\epsilon}/\Gamma(1-\epsilon)$ that in 
principle would generate finite contributions 
for $\epsilon \rightarrow 0$.
Nevertheless, since both the one-loop and the four parton
transition probabilities share the same factors
they will cancel as well as the IR poles and therefore we are 
allowed to drop these extra contributions.} 
and some finite contributions that we still have to integrate 
numerically over an effective three-body phase space.
Finally, the full transition probabilities, calculated 
in $D=4$ dimensions, are integrated numerically 
in the three-jet region of the four-body phase 
space above the $w$ cut.

Both results, the semi-analytical one and the full-numerical one
depend on the cut $w$. First requirement should be of course the 
sum of them is independent of $w$.
We have to keep in mind this is an approximate method
since we have substituted in the singular region the
transition amplitudes by their limiting values.
The smaller value of $w$ the better approximation we get.
On the other hand,
since the integrand near the boundaries of the singular 
region increases steeply, typically like $\log w$,
the $w$ cut can not be very small otherwise 
we immediately loose accuracy in the numerical integration.
In practice we should find a compromise over these two
conflicting requirements to achieve the best efficiency.
We have studied the $w$-dependence of the final result
and we found $w/m_Z \sim 10^{-5}$ provides a good approximation.

This method is called in the literature~\cite{Kunszt96CO}
the {\it phase space slicing method}\index{Slicing method}
as we exclude from the numerical integration
domain a {\it slice} of the phase space.
It was applied for instance
by~\cite{Giele92,Giele93,Baer89,Gutbrod84,Aversa90}.
Another method of analytic cancellation 
of IR singularities, called the 
{\it subtraction method}, \index{Subtraction method}
has been developed
by~\cite{Ellis81,Kunszt89,Kunszt89b,Kunszt96,Kunszt92,Mangano92,Catani96}.
The idea is to subtract and add back a quantity that 
must be a proper approximation to the differential real 
cross section such that it has the same pointwise singular
behaviour (in $D$-dimensions) and is analytically integrable 
(in $D$-dimensions) over the one parton subspace leading to
the soft and collinear divergences.

\section{Tests}

As we mentioned the introduction of massive quarks 
complicate extremely the calculation of the three-jet
decay rate in spite of the fact that 
some of the singularities are 
softened into logarithms of the quark mass.
Therefore, to check our results 
against already tested simpler calculations
is a crucial point. 

The knowledge of the transition amplitudes is a well established
result in the massless case~\cite{Ellis81,Fabricius81,Kramer89}
as well as the three- and four-jet fractions in the 
JADE, E and DURHAM jet clustering 
algorithms~\cite{Bethke92,Kunszt89}.
The four-jet decay rate has been calculated for 
massive quarks for instance
in~\cite{Ballestrero92,Ballestrero94,Kleiss90}.

First test was to check our four parton final state 
transition amplitudes in the four-jet region.
Since the four-jet phase space region is free of 
IR singularities in both the massless and the massive case
we are allowed to neglect the 
quark mass. The comparison with the 
massless result~\cite{Ellis81,Bethke92,Kunszt89}
was successful as well as with the massive
calculations~\cite{Ballestrero92,Ballestrero94,Kleiss90}.
Nevertheless, this is not a conclusive proof.
First, this checks only the four parton final state 
transition amplitudes. Second, the calculation can 
be performed in four dimensions. And finally, since 
the four-jet region is completely IR finite,
it does not matter the way we write the transition 
amplitudes, the Monte Carlo\index{Monte Carlo}
phase space integration converges quite quickly. 
The main problem is not how to calculate the 
transition probabilities but the way we write them.
The ERT~cite{Ellis81}\index{ERT} transition amplitudes
can not be directly used when we try 
to calculate the three-jet decay rate.

\subsection{Massless three-jet decay rate at next-to-leading order}

Next steep was to test our procedure of analytic cancellation of 
IR divergences over the ERT~\cite{Ellis81} transition amplitudes
in the three-jet region and to recalculate the 
$A^{(1)}(y_c)$ function that defines the NLO massless
correction to the three-jet decay rate. 
Furthermore, this test allowed us to check our 
phase space, algorithm cuts and numerical integration 
routines in the massless limit. We remind 
the three-jet decay rate can be written as 
\beq
\Gamma^{b}_{3j} = 
m_Z \frac{g^2}{c_W^2 64 \pi} \frac{\as}{\pi}
[g_V^2 H_V(y_c,r_b) + g_A^2 H_A(y_c,r_b)],
\end{equation}

\begin{table}
\begin{center}
\caption{Results of the five parameter fits to the function
$A^{(1)}(y_c)=\sum_{n=0}^4 k_n$ log${}^n y_c$ in the range $0.01<y_c<0.20$
\label{A1fit}}
\begin{tabular}{lrrrrr}
\hline
Algorithm   & $k_0 \qquad$ & $k_1 \qquad$ & $k_2 \qquad$ 
            & $k_3 \qquad$ & $k_4 \qquad$ \\
\hline
EM     & $17.25(12)$  & $20.37(10)$  & $-7.320(47)$  & $-12.621(21)$  
& $-1.9200(45)$ \\
JADE   & $17.872(26)$ & $26.599(21)$ & $1.265(10)$   & $-9.5644(48)$  
& $-1.8547(10)$ \\ 
E      & $12.41(13)$  & $20.22(10)$  & $-1.912(49)$  & $-11.173(23)$  
& $-1.7920(47)$ \\
DURHAM & $-.021(13)$  & $-.533(11)$ & $-3.8826(55)$ & $-3.3720(25)$ 
& $-.46956(49)$ \\
\hline
\end{tabular}
\end{center}
\end{table}

%%%%%%%%%%%%%%%
\mafigura{9.6cm}{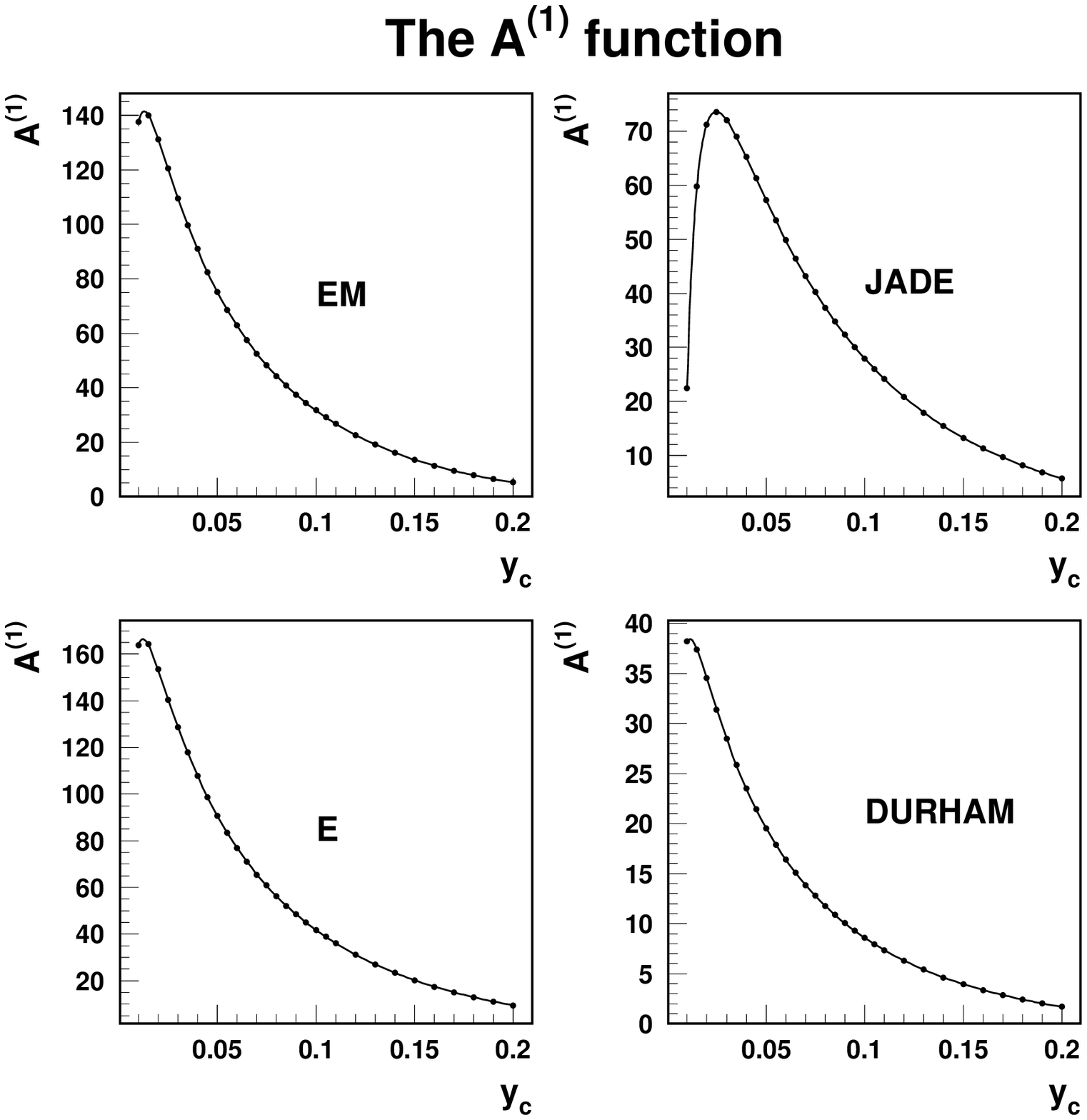}
{The function $A^{(1)}$ which determines the massless 
NLO three-jet decay rate for the EM, JADE, E and DURHAM
jet clustering algorithms.}
{ERTps}
%%%%%%%%%%%%%%%

\noindent
where $r_b=m_b^2/m_Z^2$, $y_c$ is the resolution parameter, 
and $g_V$ and $g_A$ are the vector and the axial-vector
neutral current quark couplings.
At tree-level and for the bottom quark 
$g_V = -1 + 4 s_W^2/3$ and $g_A = 1$ in the 
Standard Model.\index{Standard Model}
In the massless limit we wrote
\beq
H_{V(A)} = A^{(0)}(y_c) + \frac{\as}{\pi} A^{(1)}(y_c),
\end{equation}
where we have taken into account that for massless quarks vector
and axial contributions are identical\footnote{We do not consider the
small $O(\as^2)$ triangle anomaly \cite{hagiwara}.
$A^{(1)}(y_c) = B(y_c)/4$ with $B(y_c)$ defined
in~\cite{Bethke92,Kunszt89}.}

The EM jet clustering algorithm was introduced for the 
first time in~\cite{RO95} because in this algorithm 
at LO the two- and three-jet decay rates can be calculated
analytically for massive quarks. The function 
$A^{(1)}(y_c)$ is therefore unknown in this case.
We have calculated it for the first time.
Furthermore, we have not only tested but tried 
to improve the accuracy in the knowledge of the 
$A^{(1)}(y_c)$ function for the JADE and DURHAM 
algorithms. Quark mass corrections are very small
in these two jet clustering algorithms, specially
in DURHAM, and a better determination of the 
$A^{(1)}(y_c)$ function contributes to a better 
understanding of the quark mass effects.

To simplify the use of our results and to compare them 
with the~\cite{Bethke92,Kunszt89} result we
have performed a simple five parameter $\log y_c$ power series
fit to the $A^{(1)}(y_c)$ function
for the different jet clustering algorithms
\beq
A^{(1)}(y_c)=\sum_{n=0}^4 k_n \log^n y_c~.
\end{equation}
The results of our fits for the range $0.01 < y_c <0.20$
are presented in table~\ref{A1fit}. We plot in figure~\ref{ERTps}
the $A^{(1)}(y_c)$ function for the four algorithms we 
have considered.

\subsection{The massless limit test}

Finally, we integrate in the three-jet region 
the massive quark transition amplitudes
we have calculated and we try to see if extrapolating our result to 
the massless limit we can reach the $A^{(1)}(y_c)$ function
of the previous section.
This is the main test of our calculation.

In figures \ref{Jtest} and \ref{Etest} we present our
result for the vectorial contribution to the
$O(\as^2)$ three-jet decay rate of the Z-boson into
bottom quarks for the JADE and E algorithms.
çWe have performed the calculation for
different values of the bottom quark mass from
$1$ to $5(GeV)$ for fixed $y_c$.
We remind we factorized the leading dependence on the quark mass 
in the functions $H_{V(A)}$ as follows
\beq
H_{V(A)} = A^{(0)}(y_c) + r_b B_{V(A)}^{(0)}(y_c,r_b) 
+ \frac{\as}{\pi} \left( A^{(1)}(y_c) + r_b B_{V(A)}^{(1)}(y_c,r_b) 
\right). 
\end{equation}

%%%%%%%%%%%%%%%
\mafigura{10cm}{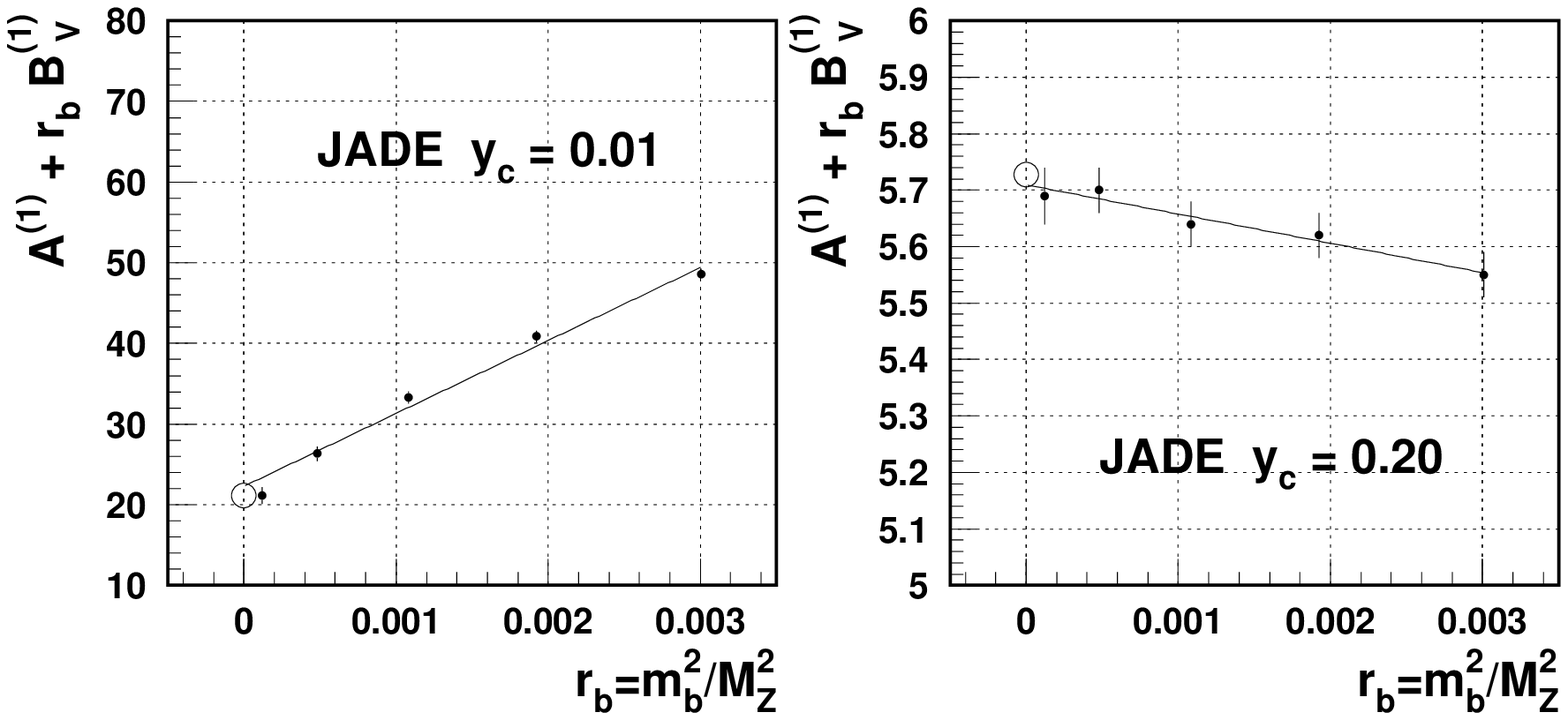}
{NLO contribution to the three-jet decay rate of $Z\rightarrow b\bar{b}$
for bottom quark masses from $1$ to $5(GeV)$ and fixed $y_c$
in the Jade algorithm. Big circle is the massless case.}
{Jtest}
%%%%%%%%%%%%%%%
\mafigura{10cm}{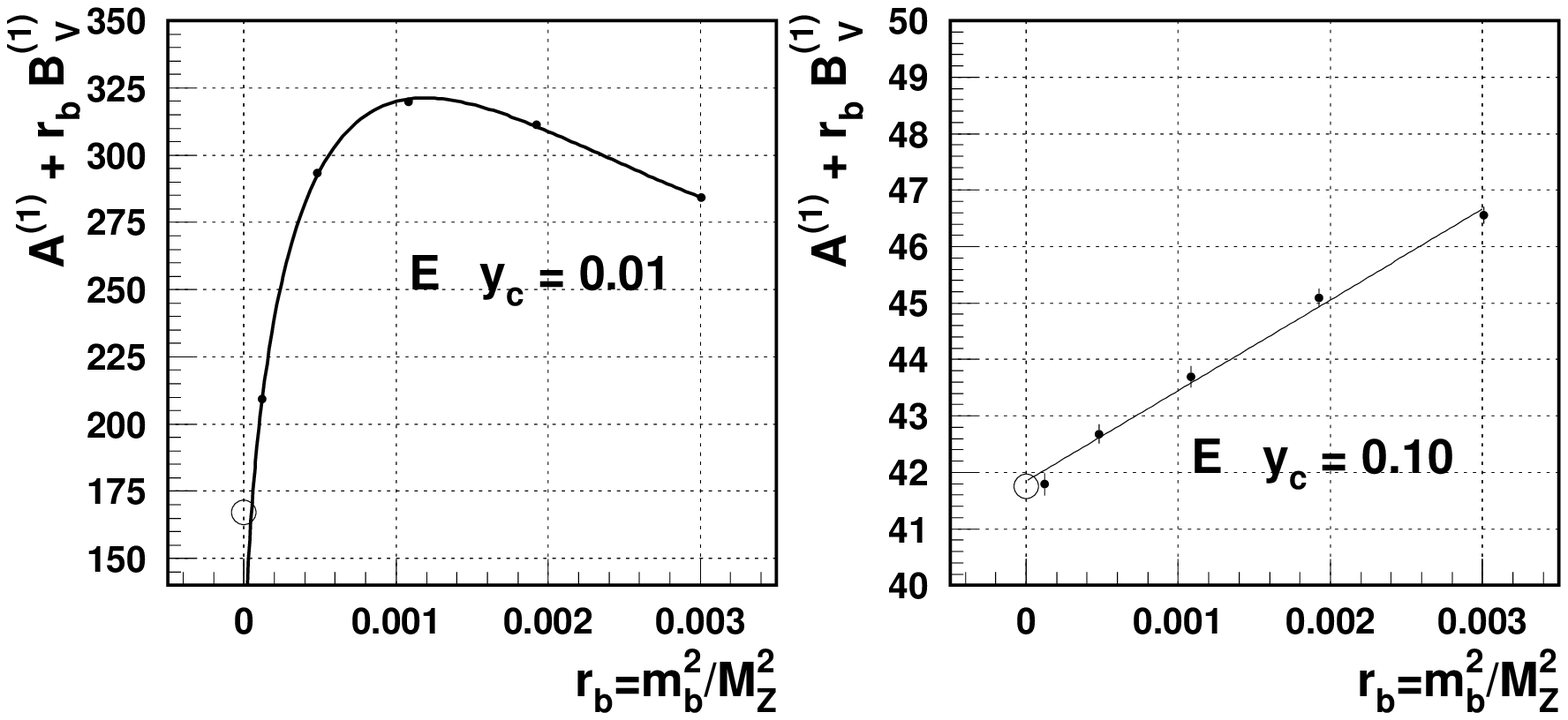}
{NLO contribution to the three-jet decay rate of $Z\rightarrow b\bar{b}$
for bottom quark masses from $1$ to $5(GeV)$  and fixed $y_c$
in the E algorithm. Big circle is the massless case.}
{Etest}
%%%%%%%%%%%%%%%

In the JADE algorithm we can see that for big values of
$y_c$ the NLO corrections due to the quark mass are
very small and below the massless result.
Notice they increase quite a lot for small values of $y_c$
and give a positive correction that will produce a change 
in the slope of the LO prediction for 
$R_3^{bd}$.\index{R3bd@$R_3^{bd}$}
In any case we recover the massless limit and a linear
parametrization in the quark mass squared would
provide a good description. 
The DURHAM and EM algorithm would 
exhibit a very similar behaviour 
to the JADE algorithm.

The E algorithm behaves also linearly in the quark mass squared
although only for big values of $y_c$.
Corrections in the E algorithm are always very
strong. The reason is the following, the resolution parameter
for the E algorithm explicitly incorporates the quark mass,
$y_{ij}=(p_i+p_j)/s$, i.e., for the same value of $y_c$
we are closer to the two-jet IR region and the difference
from the other algorithms is precisely the quark mass.
This phenomenon is already manifested
at the LO. For massive quarks the behaviour of the E algorithm
is completely different from the other algorithms.
It is difficult to believe in the E algorithm as a
good prescription for physical applications 
since mass corrections as so big. However for the same reason,
it seems to be the best one for testing massive calculations.

Furthermore, since we have encoded 
the ERT formulas, we were able to perform some partial tests,
not only the final three-jet decay rate.
The IR structure of the massless transition amplitudes 
of~\cite{Ellis81} is completely different from the massive case.
This means, we can not test our calculation amplitude by
amplitude. Nevertheless, in the previous chapters we were able to 
split our transition probabilities into different groups 
such that IR singularities cancel independently.
Checking can be performed group by group.
The difference lies in the singularities that are softened into 
$\log r_b$ for massive quarks. They must cancel too.
In the massless case we have 
to cancel these singularities analytically. In the massive 
case we do it numerically. We have not calculated 
analytically the terms proportional to $\log r_b$. A 
future improvement of our calculation could be to do that.

%\section{Fits}
\section{Massive three-jet decay rate at next-to-leading order}

To conclude, we present our final calculation 
for the $B^{(1)}_{V(A)}(y_c,r_b)$ functions defining
the NLO mass correction to the three-jet ratio
$R_3^{bd}$\index{R3bd@$R_3^{bd}$}
in the four jet clustering algorithms EM, JADE, E and DURHAM,
see figures~\ref{Mfunc}, \ref{Jfunc}, \ref{Efunc} and \ref{Dfunc}.
We have performed the calculation for $m_b=3(GeV)$
and $m_b=5(GeV)$ pole masses.
Error bars come from the Monte Carlo\index{Monte Carlo}
phase space numerical integration.
As before, to simplify the use of our results,
we have performed simple five parameter $\log y_c$ power series
fits to the ratio of the function $B^{(1)}_{V,A}(y_c,r_b)$ and
the massless NLO function $A^{(1)}(y_c)$.
We use the following parametrization: 
\beq
B^{(1)}_{V,A}(y_c,r_b)/A^{(1)}(y_c) = 
\sum_{n=0}^4 k^n_{V,A} \log^n y_c~.
\end{equation}
The results of our fits for the range $0.01 < y_c <0.20$
($0.025 < y_c <0.20$ for JADE and E algorithms)
are presented in table~\ref{A2fit}.
For the moment we perform independent fits to the 
$m_b=3(GeV)$ and $m_b=5(GeV)$ results.

\begin{table}
\begin{center}
\caption{Results of the five parameter fits to the function
$B^{(1)}_{V,A}(y_c)/A^{(1)}(y_c)=\sum_{n=0}^4 k^n_{V,A}$ log${}^n y_c$
in the range $0.01<y_c<0.20$ for EM and DURHAM algorithms and 
$0.025<y_c<0.20$ for JADE and E for bottom quark pole masses
of $3(GeV)$ and $5(GeV)$.
\label{A2fit}}
\begin{tabular}{lrrrrr}
\hline
Algorithm   & $k_0 \qquad$ & $k_1 \qquad$ & $k_2 \qquad$ 
            & $k_3 \qquad$ & $k_4 \qquad$ \\
\hline
            & & & & &  \\
EM          & & & & &  \\
\hline
$k_V (3 GeV)$  & $6.9(48)$   & $90.2(30)$   & $84.15(93)$  & $27.79(27)$  
& $2.957(47)$ \\
$k_A        $  & $-54.5(16)$ & $-43.32(94)$ & $-6.85(28)$  & $3.024(84)$  
& $0.664(14)$ \\ 
$k_V (5 GeV)$  & $-10.7(52)$ & $65.4(31)$   & $71.98(93)$  & $25.39(27)$  
& $2.798(47)$ \\
$k_A        $  & $-38.4(16)$ & $-24.09(95)$ & $1.49(28)$   & $4.731(83)$  
& $0.800(14)$ \\
            & & & & &  \\
JADE        & & & & &  \\
\hline
$k_V (3 GeV)$  & $396.3(39)$ & $676.8(25)$  & $410.59(92)$ & $109.93(32)$  
& $10.941(75)$ \\
$k_A        $  & $85.5(12)$  & $170.40(75)$ & $111.77(28)$ & $33.256(97)$  
& $3.747(23)$ \\
$k_V (5 GeV)$  & $224.2(38)$ & $403.2(25)$  & $249.43(92)$ & $68.35(32)$  
& $6.989(75)$ \\
$k_A        $  & $94.0(12)$  & $181.87(75)$ & $117.55(28)$ & $34.689(97)$  
& $3.893(23)$ \\
            & & & & &  \\
E           & & & & &  \\
\hline
$k_V (3 GeV)$  & $589.6(52)$ & $1080.1(32)$ & $759.5(12)$  & $234.24(40)$  
& $27.599(92)$ \\
$k_A        $  & $-584.7(17)$& $-1059.1(10)$& $-673.70(38)$& $-183.63(13)$  
& $-17.305(30)$ \\
$k_V (5 GeV)$  & $621.8(52)$ & $1112.7(32)$ & $767.8(12)$  & $233.85(40)$  
& $27.343(92)$ \\
$k_A        $  & $-535.4(17)$& $-975.3(10)$ & $-622.80(38)$& $-170.11(13)$  
& $-16.004(30)$ \\
             & & & & &  \\
DURHAM       & & & & &  \\
\hline
$k_V (3 GeV)$  & $105.0(38)$ & $183.4(25)$  & $107.27(77)$ & $27.17(23)$  
& $2.498(38)$ \\
$k_A        $  & $69.8(11)$  & $110.34(76)$ & $61.53(23)$  & $15.364(69)$  
& $1.421(11)$ \\
$k_V (5 GeV)$  & $132.5(38)$ & $212.5(25)$  & $120.40(77)$ & $29.99(23)$  
& $2.739(38)$ \\
$k_A        $  & $74.9(11)$  & $116.88(76)$ & $65.51(23)$  & $16.435(69)$  
& $1.525(11)$ \\
            & & & & &  \\
            & & & & &  \\
\hline
\end{tabular}
\end{center}
\end{table}

DURHAM seems to be the most promising jet algorithm
to keep under control the higher order corrections  
since not only the NLO massless function $A^{(1)}(y_c)$
is the smallest but also because the ratio 
$B^{(1)}_{V,A}(y_c,r_b)/A^{(1)}(y_c)$  
holding the mass information is the smallest too.
We get $B^{(1)}_{V,A}(y_c,r_b)/A^{(1)}(y_c) \simeq -10$.
On the other hand, mass corrections are so big in the 
E algorithm, for the reasons we mentioned in the 
previous section, and
LO and NLO predictions for $R_3^{bd}$\index{R3bd@$R_3^{bd}$} 
are so different that
perturbation theory probably breaks down 
in this case.
The normalized mass correction in the JADE and 
EM algorithms are around two and three times the mass
correction in the DURHAM algorithm,
we obtain $B^{(1)}_{V,A}(y_c,r_b)/A^{(1)}(y_c) \simeq -20, -30$
respectively, although still under reasonable values.

In all the three algorithms, EM, JADE and DURHAM, the 
$B^{(1)}_{V,A}(y_c,r_b)/A^{(1)}(y_c)$ ratio is 
negative since massive quarks are expected to 
radiate less gluons than massless quarks.
Nevertheless, for enough small value of the resolution 
parameter $y_c$, depending on the algorithm,
the mass correction becomes positive
and steeply increasing. This means 
there is one point where the NLO mass corrections
are zero. 

This phenomenon happens because for enough small values 
of $y_c$ the four parton transition probabilities 
contribute mainly to the four-jet fraction and therefore 
in the three-jet region the virtual corrections 
dominate. This translates into a change in 
the slope of the LO prediction for 
$R_3^{bd}$,\index{R3bd@$R_3^{bd}$} i.e., the NLO 
prediction for $R_3^{bd}$ going down for small 
values of $y_c$ becomes increasing and closer to unity.
For the range of the resolution parameter $y_c$ we 
are considering, $0.01< y_c < 0.20$, we can see in 
figures~\ref{Mfunc} and \ref{Jfunc} this already occurs
in the EM and JADE algorithm for $y_c \simeq 0.02$.
For the DURHAM algorithm we have to wait for a resolution 
parameter smaller than $y_c < 0.01$. 
In fact, although this phenomenon is a sign that 
perturbation theory breaks down and therefore our 
perturbative prediction is meaningless, we can 
see in figure~\ref{fuster} that for the preliminary 
DELPHI\index{DELPHI} results in the DURHAM algorithm
this tendency really happens.

\subsection{Estimate of higher order contributions}

Up to now we made all the calculations with the 
assumption $\mu^2=s=m^2_Z$. This does not restrict the 
validity of our result since the $\mu$ dependence is 
completely determined by the Renormalization Group.
The scale dependence is introduced in $R_3^{bd}$
by the replacements
\beq
\bes
\as(m_Z) & \rightarrow \as(\mu), \\ 
r_b & \rightarrow \bar{r}_b(\mu) 
\left\{ 1 + 2 \frac{\as(\mu)}{\pi} \left[\frac{4}{3} - 
\log \frac{\bar{m}_b^2(\mu)}{\mu^2} \right] \right\}. 
\ees
\end{equation}
Assuming we have factorized in the functions 
$B_{V,A}^{(0)}(y_c,r_b)$ and $B_{V,A}^{(1)}(y_c,r_b)$ 
the leading dependence on the quark mass and 
the remnant mass dependence is 
soft~\footnote{For the moment we will assume the 
$B_{V,A}^{(i)}(y_c)$ function comes from the calculation 
with $m_b=3(GeV)$.}, i.e., 
$B_{V,A}^{(i)}(y_c,r_b) \simeq B_{V,A}^{(i)}(y_c)$ 
the three-jet fraction ratio $R_3^{bd}$\index{R3bd@$R_3^{bd}$}
 reads
\bea
R_3^{bd} &=& \Bigg[ 1 + \bar{r}_b(\mu) \Bigg\{ \nonumber \\
& & c_V
\frac{B_V^{(0)}(y_c)}{A^{(0)}(y_c)}
\left[ 1 + \frac{\as(\mu)}{\pi}
\left( \frac{B_V^{(1)}(y_c)}{B_V^{(0)}(y_c)}
+2\left(\frac{4}{3}-\log \frac{\bar{m}^2_b(\mu)}{\mu^2} \right)
- \frac{A^{(1)}(y_c)}{A^{(0)}(y_c)}
\right) \right] \nonumber \\
&+& c_A
\frac{B_A^{(0)}(y_c)}{A^{(0)}(y_c)}
\left[ 1 + \frac{\as(\mu)}{\pi}
\left( \frac{B_A^{(1)}(y_c)}{B_A^{(0)}(y_c)}
+2\left(\frac{4}{3}-\log \frac{\bar{m}^2_b(\mu)}{\mu^2} \right)
- \frac{A^{(1)}(y_c)}{A^{(0)}(y_c)}
\right) \right] \Bigg\} \Bigg] \nonumber \\
& \times & \left[ 1 + 6 \bar{r}_b(\mu)
\left\{ c_A \left( 1+ 2 \frac{\as(\mu)}{\pi} 
\left( \frac{4}{3}-\log \frac{s}{\mu^2} \right) \right)
-c_V 2 \frac{\as(\mu)}{\pi}  \right\} \right]~.
\label{r3mu}
\end{eqnarray}

At the lowest order we define
\beq
\bar{r}_b^0 = \frac{R_3^{bd}-1}{
c_V \frac{\D B_V^{(0)}(y_c)}{\D A^{(0)}(y_c)}
+ c_A \left(\frac{\D B_A^{(0)}(y_c)}{\D A^{(0)}(y_c)}+6\right) }~,
\end{equation}
with $\bar{m}_b^0 = m_Z \sqrt{\bar{r}_b^0}$. We can 
solve iteratively \Eq{r3mu} to get 
\beq
\bes
\bar{r}_b(m_Z) & = \left. \bar{r}_b^0 
\left( \frac{\as(m_Z)}{\as(\mu)} \right)^{\frac{\D 4 \gamma_0}{\D \beta_0}}
\right/
\Bigg[ 1 + \frac{\as(\mu)}{\pi} \frac{\bar{r}_b^0}{R_3^{bd}-1} 
\Bigg\{ \\
& c_V \frac{B_V^{(0)}(y_c)}{A^{(0)}(y_c)}
\left( \frac{B_V^{(1)}(y_c)}{B_V^{(0)}(y_c)}
+2\left(\frac{4}{3}-2\log \frac{\bar{m}^0_b}{\mu} \right)
- \frac{A^{(1)}(y_c)}{A^{(0)}(y_c)} \right)  \\
+ & c_A \frac{B_A^{(0)}(y_c)}{A^{(0)}(y_c)}
\left( \frac{B_A^{(1)}(y_c)}{B_A^{(0)}(y_c)}
+2\left(\frac{4}{3}-2\log \frac{\bar{m}^0_b}{\mu} \right)
- \frac{A^{(1)}(y_c)}{A^{(0)}(y_c)} \right) \\
+ & 12 c_A \left(\frac{4}{3}-\log \frac{s}{\mu^2} \right) - 12 c_V 
\Bigg\} \Bigg],
\label{mbmzmu}
\ees
\end{equation}
where $\beta_0 = 11 - 2/3 N_F$ and $\gamma_0 = 2$,
with the number of flavours $N_F = 5$.

Using our fit for a pole mass of $m_b=3(GeV)$
and following~\Eq{r3mu} we have plotted in
figures~\ref{Mfunc}, \ref{Jfunc}, \ref{Efunc} and \ref{Dfunc}
our prediction for $R_3^{bd}$\index{R3bd@$R_3^{bd}$}
for a running bottom quark mass at the $m_Z$ scale of
$\bar{m}_b(m_Z) = 3 (GeV)$. We include also our
prediction for $\mu=m_Z/2$ and $\mu=2\: m_Z$ for  
$\bar{m}_b(m_Z) = 3 (GeV)$ fixed that could be taken as 
our estimate of the higher order uncertainty.

%%%%%%%%%%%%%%%
\mafigura{12cm}{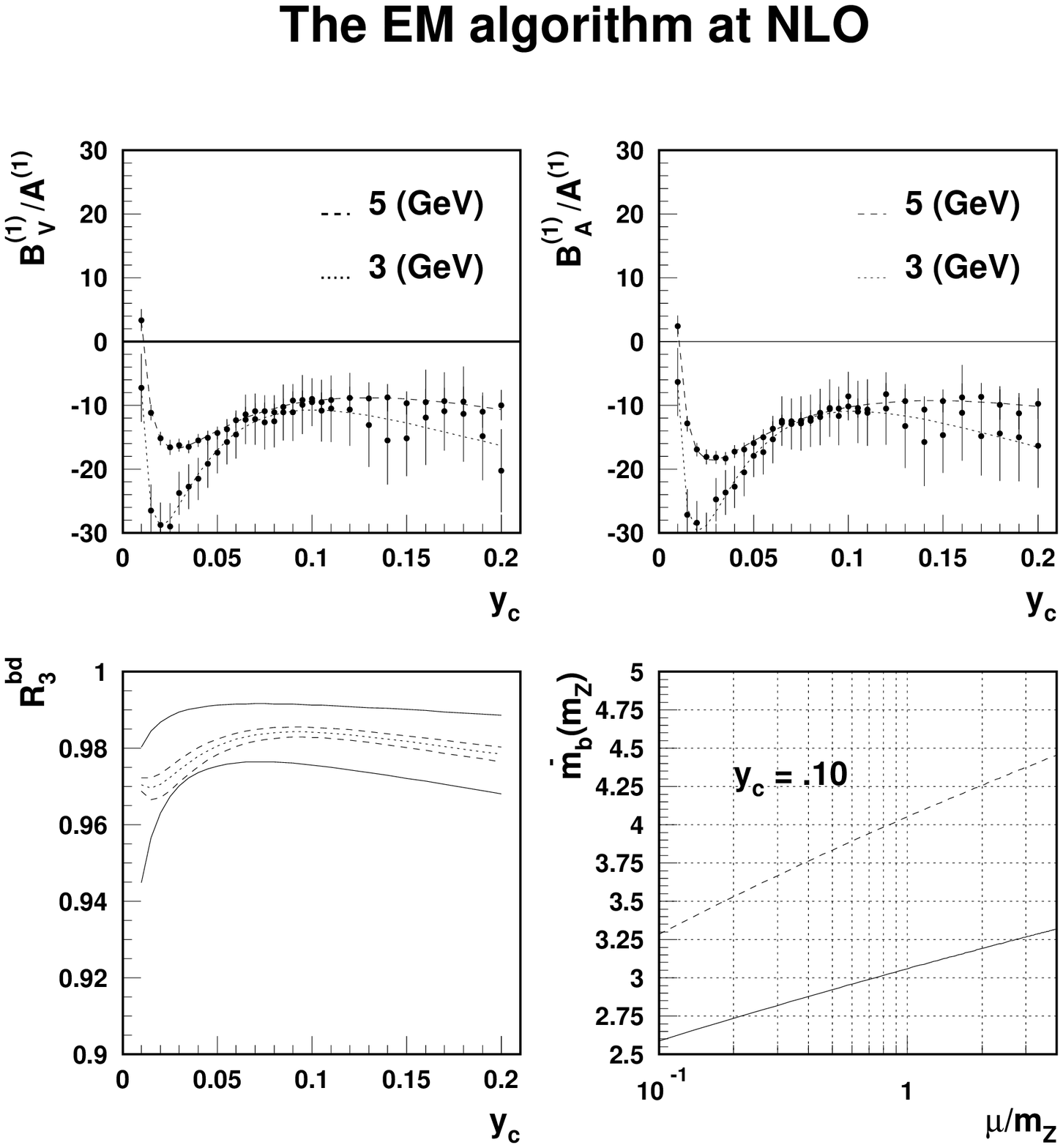}
{EM algorithm: functions $B^{(1)}_V(y_c)$ and $B^{(1)}_A(y_c)$
normalized to he massless NLO function $A^{(1)}(y_c)$
in the range $0.01 < y_c < 0.20$,
three-jet fraction ratio $R_3^{bd}$ (solid lines are 
LO for $m_b=3(GeV)$ and $m_b=5(GeV)$, short dashed line is the NLO 
prediction for a running mass $\bar{m}_b(m_Z)=3(GeV)$
and long dashed lines for $\mu=m_Z/2$ and $\mu=2\: m_Z$) 
and NLO scale dependence for the bottom running mass extracted from 
a fixed value of the ratio $R_3^{bd}$ (dashed line is LO). }
{Mfunc}
%%%%%%%%%%%%%%%
%%%%%%%%%%%%%%%
\mafigura{12cm}{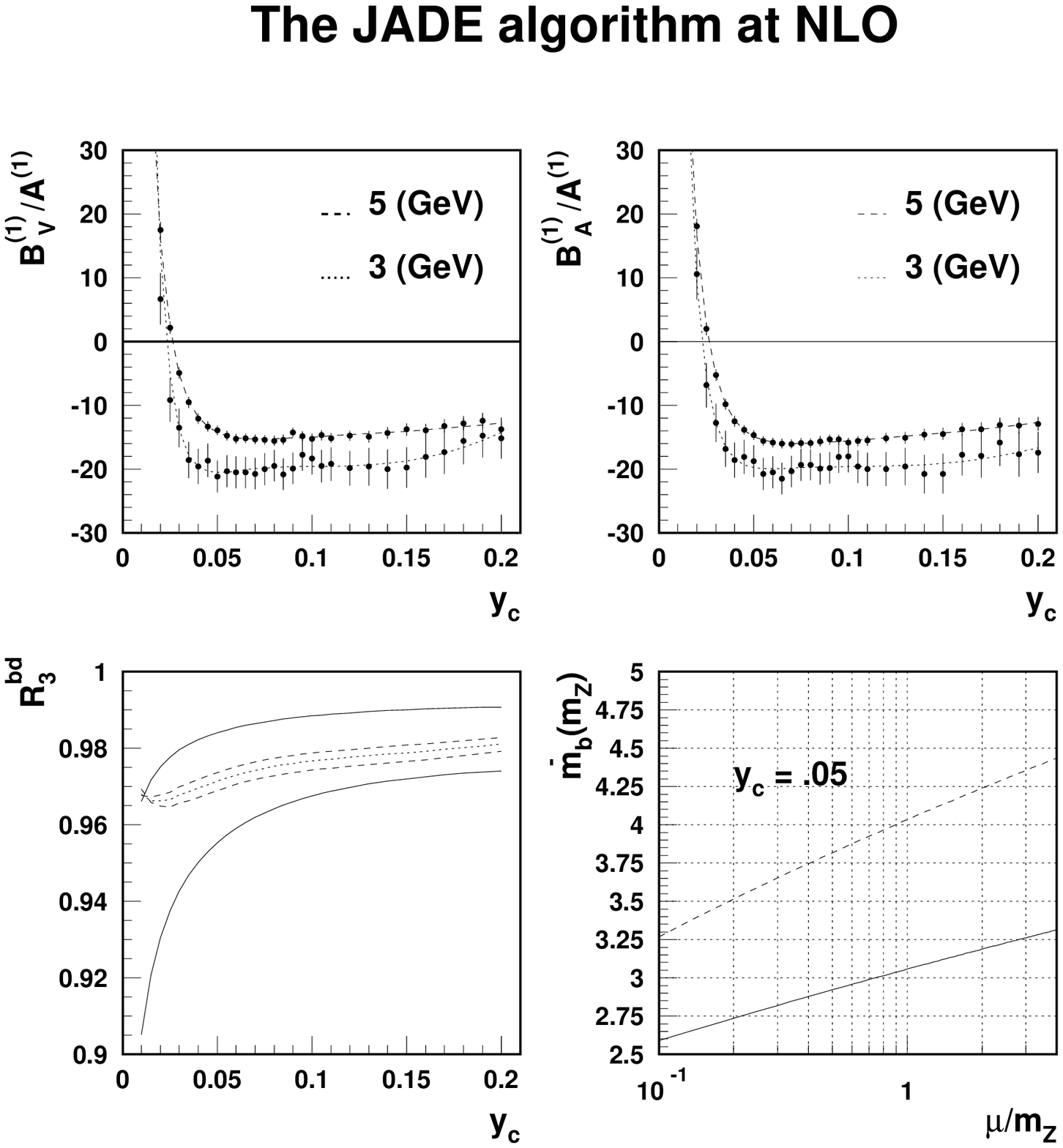}
{JADE algorithm: functions $B^{(1)}_V(y_c)$ and $B^{(1)}_A(y_c)$
normalized to he massless NLO function $A^{(1)}(y_c)$
in the range $0.01 < y_c < 0.20$,
three-jet fraction ratio $R_3^{bd}$ (solid lines are 
LO for $m_b=3(GeV)$ and $m_b=5(GeV)$, short dashed line is the NLO 
prediction for a running mass $\bar{m}_b(m_Z)=3(GeV)$
and long dashed lines for $\mu=m_Z/2$ and $\mu=2\: m_Z$) 
and NLO scale dependence for the bottom running mass extracted from 
a fixed value of the ratio $R_3^{bd}$ (dashed line is LO). }
{Jfunc}
%%%%%%%%%%%%%%%
%%%%%%%%%%%%%%%
\mafigura{12cm}{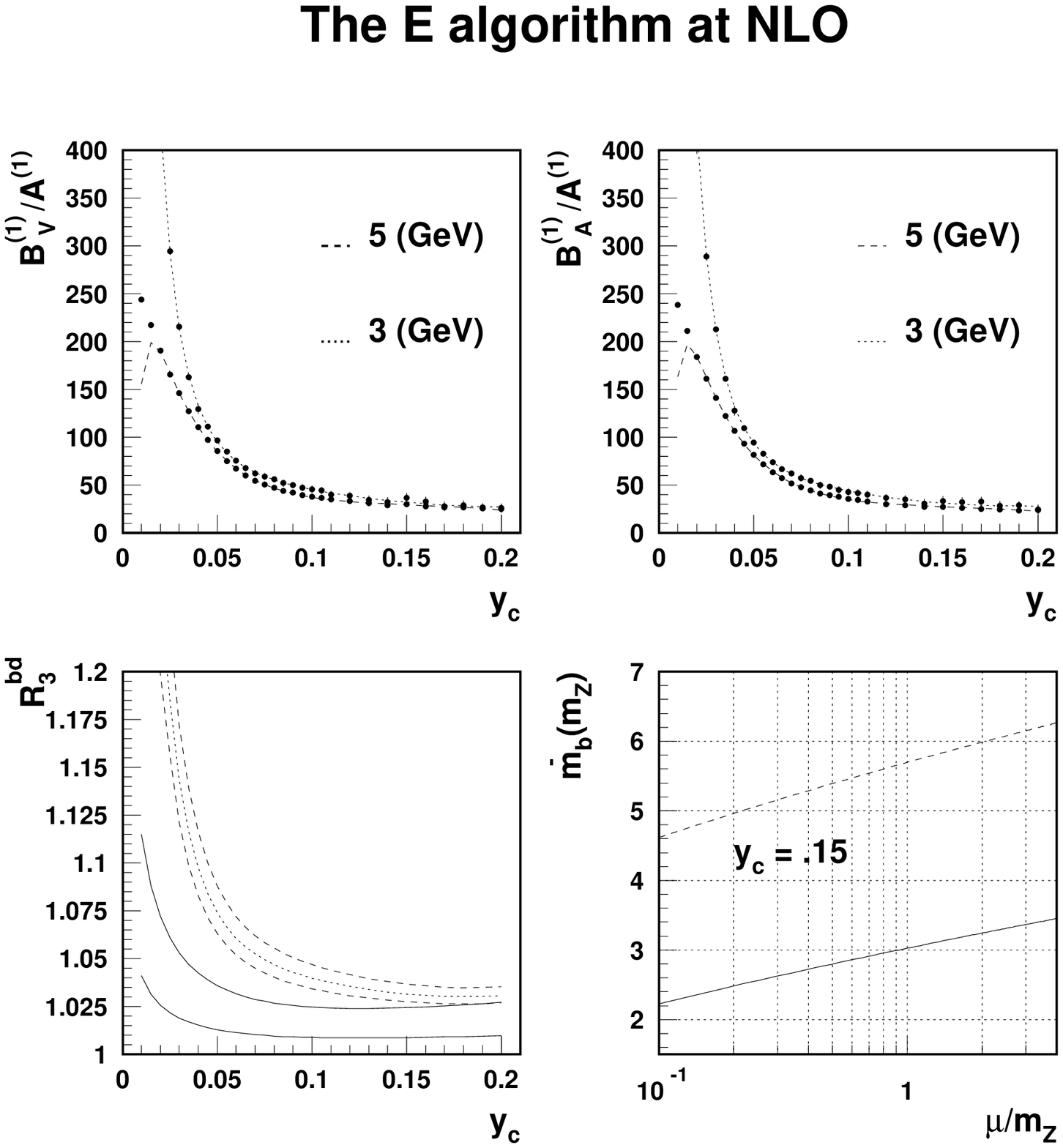}
{E algorithm: functions $B^{(1)}_V(y_c)$ and $B^{(1)}_A(y_c)$
normalized to he massless NLO function $A^{(1)}(y_c)$
in the range $0.01 < y_c < 0.20$,
three-jet fraction ratio $R_3^{bd}$ (solid lines are 
LO for $m_b=3(GeV)$ and $m_b=5(GeV)$, short dashed line is the NLO 
prediction for a running mass $\bar{m}_b(m_Z)=3(GeV)$
and long dashed lines for $\mu=m_Z/2$ and $\mu=2\: m_Z$) 
and NLO scale dependence for the bottom running mass extracted from 
a fixed value of the ratio $R_3^{bd}$ (dashed line is LO). }
{Efunc}
%%%%%%%%%%%%%%%
%%%%%%%%%%%%%%%
\mafigura{12cm}{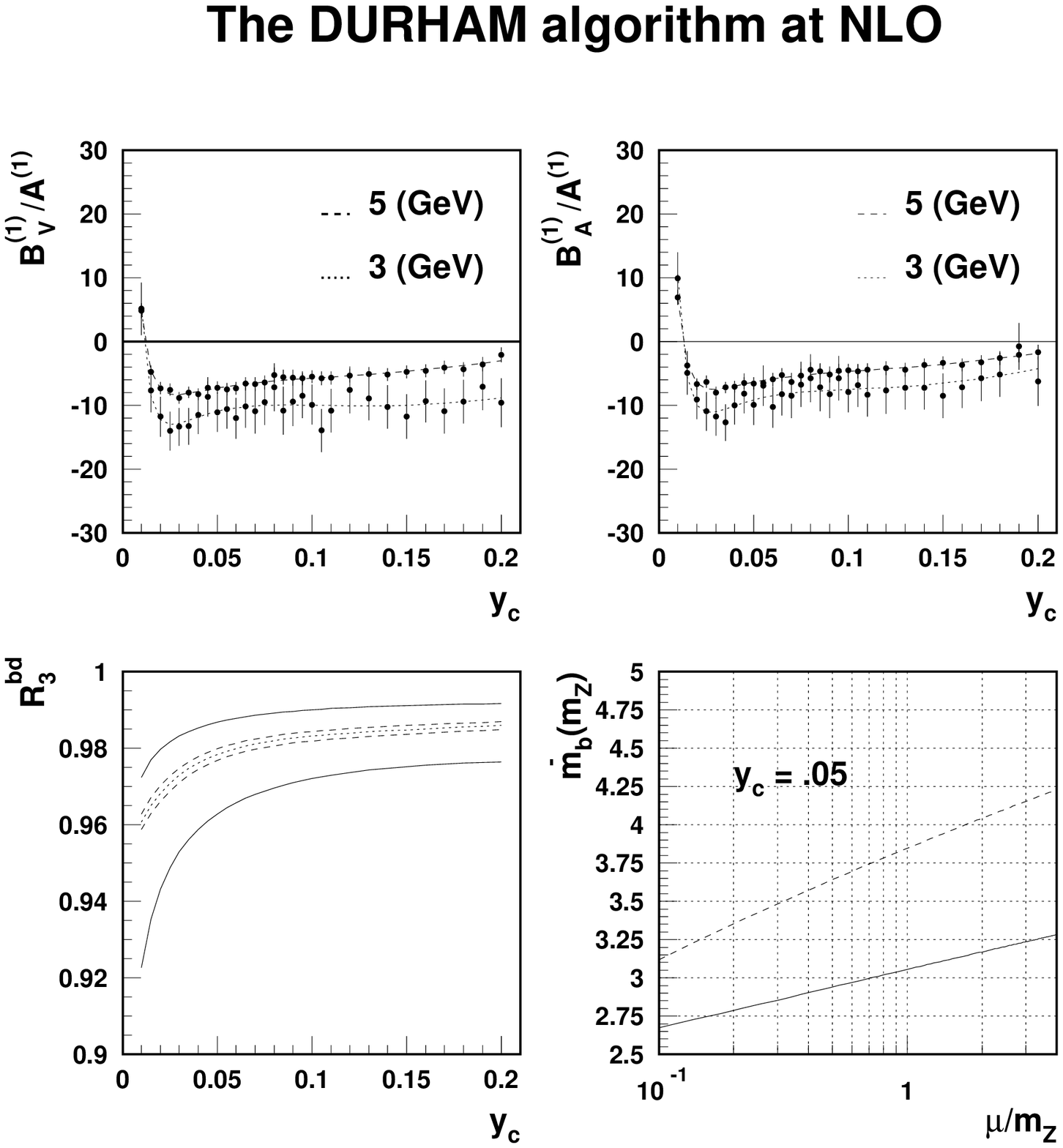}
{DURHAM algorithm: functions $B^{(1)}_V(y_c)$ and $B^{(1)}_A(y_c)$
normalized to he massless NLO function $A^{(1)}(y_c)$
in the range $0.01 < y_c < 0.20$,
three-jet fraction ratio $R_3^{bd}$ (solid lines are 
LO for $m_b=3(GeV)$ and $m_b=5(GeV)$, short dashed line is the NLO 
prediction for a running mass $\bar{m}_b(m_Z)=3(GeV)$
and long dashed lines for $\mu=m_Z/2$ and $\mu=2\: m_Z$) 
and NLO scale dependence for the bottom running mass extracted from 
a fixed value of the ratio $R_3^{bd}$ (dashed line is LO). }
{Dfunc}
%%%%%%%%%%%%%%%

Conversely, for a fixed value of the resolution parameter 
$y_c$ and a fixed value of the ratio $R_3^{bd}$ 
we have explored the $\mu$ dependence of 
our prediction for the bottom quark mass at the $Z$-boson
mass scale, $\bar{m}_b(m_Z)$ , following \Eq{mbmzmu},
see figures~\ref{Mfunc}, \ref{Jfunc}, \ref{Efunc} and \ref{Dfunc}.
We include, to compare, the LO $\mu$-dependence. 

%%%%%%%%%%%%%%%
\mafigura{10cm}{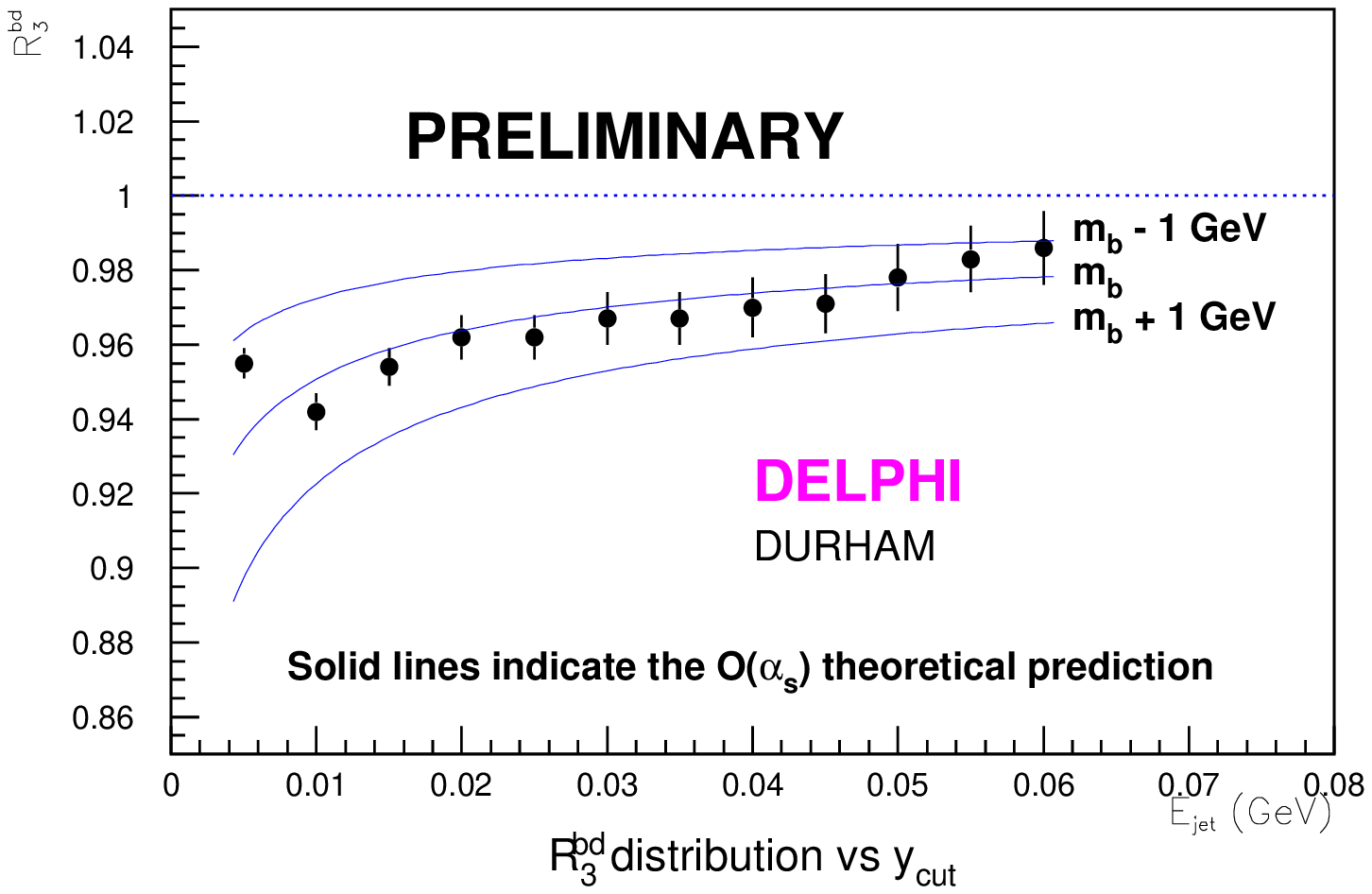}
{Preliminary experimental results for the ratio $R_3^{bd}$
from DELPHI Collaboration.\protect\index{DELPHI}}
{fuster}
%%%%%%%%%%%%%%%

Taking as the theoretical uncertainty for the 
running bottom quark mass at $m_Z$ half of the 
difference from the $\mu=m_Z/2$ and $\mu=2\: m_Z$
predictions we find as expected the smallest 
uncertainty correspond to the DURHAM algorithm,
$\Delta \bar{m}_b \simeq 110(MeV)$. JADE and EM share the 
same behaviour, $\Delta \bar{m}_b \simeq 130(MeV)$, slightly 
bigger than in the DURHAM case.
Finally, the E algorithm grows to $\Delta \bar{m}_b \simeq 220(MeV)$.
Case we enlarge the lower limit on $\mu$ we get for $\mu=m_Z/10$,
$\Delta \bar{m}_b \simeq 250$, $300$ and $500 (MeV)$ respectively.

%For a fixed $y_c$ and a fixed value of $R_3^{bd}$ in the 
%DURHAM scheme we explore the $\mu$ dependence of 
%our prediction for the bottom quark mass at the $Z$-boson
%mass scale, $m_b(m_Z)$ , following \Eq{mbmzmu},
%see figures~\ref{Mfunc}, \ref{Jfunc}, \ref{Efunc} and \ref{Dfunc}.
%Although, we make the plot for a very generous wide range 
%of the $\mu$-scale, from $\mu = m_Z/10$ to $\mu = 4\: m_Z$,
%the NLO prediction is quite stable against the 
%$\mu$ dependence.
%The strong $\mu$-dependence of the LO prediction, about $1(GeV)$ 
%of variation, drastically decreases to at most $100(MeV)$
%when the NLO is included. 
%If expected experimental error~\cite{FUQCD96}
%on the bottom quark mass is of the order of $300(MeV)$ 

\section{Discussion and conclusions}

In this thesis we have analyzed at NLO the quark mass dependence 
of the three-jet observable\index{R3bd@$R_3^{bd}$} 
\[
R^{bd}_3 = \frac{\Gamma^b_{3j}(y_c)/\Gamma^b}{\Gamma^d_{3j}(y_c)/\Gamma^d}~,
\]
for different jet clustering algorithms such as EM, JADE, E and DURHAM.
In particular, we have studied the possibility of extracting 
the bottom quark mass from $R^{bd}_3$ at LEP.

Since quarks are not free particles the 
study of its mass needs a precise theoretical framework
and, in fact, they should be treated more
like coupling constants.
The two most commonly used quark mass definitions are
the perturbative pole mass and the $\overline{MS}$ scheme
running mass. 
We reviewed in Chapter~\ref{UIMPchapter}
some of the most recent determinations of all
the quark masses, in particular, the bottom quark mass
extracted from the bottomia spectrum  
from QCD Sum Rules and Lattice calculations.
The bottom quark perturbative pole mass
appeared to be around $M_b = 4.6-4.7 (GeV)$ whereas
the running mass at the running mass scale read
$\bar{m}_b(\bar{m}_b) = (4.33 \pm 0.06) GeV$.
We analized its running until the $Z$-boson mass scale
and we found $\bar{m}_b(m_Z) = (3.00 \pm 0.12) GeV$.

Since for the bottom quark the difference between the
perturbative pole mass and the running mass at the $m_Z$ scale 
is quite significant it is crucial 
to specify, in any theoretical perturbative
prediction at $m_Z$, which mass should we use.
In practice, this means we have to go up to 
NLO, where a renormalization scheme has to be chosen,
if we want to extract a meaningful value 
for the bottom quark mass, since the LO does not 
allow to distinguish among the different quark 
mass definitions.

We have explored a new method: the study of 
three-jet observables at LEP, different from QCD Sum Rules and
Lattice calculations, for determining the bottom quark mass.
Among other advantages, it would allow to extract
the bottom quark mass far from threshold, in contrast to 
the other methods described above.

From our calculation, we have estimated, 
by exploring the $\mu$-dependence of our prediction at $m_Z$,
a theoretical bottom quark mass uncertainty of
at most $250(MeV)$ in the DURHAM algorithm, $300(MeV)$ for
JADE and EM and $500(MeV)$ for the E algorithm.
The preliminary experimental bottom quark mass 
uncertainty from the DELPHI\index{DELPHI} Collaboration~\cite{FUQCD96},
see figure~\ref{fuster},
in the DURHAM algorithm is expected to be $300(MeV)$.
Nevertheless, that is not the last word but the 
first step. The experimental error would 
be improved, specially once the four experiments at LEP are taken 
into account, and we are still discussing how to improve the 
theoretical error. Resummed expressions would probably give 
a better result.

Therefore, even if, for the moment, this method can not
be considered competitive with QCD Sum Rules and 
Lattice predictions, a total error of $300(MeV)$  
in the DURHAM algorithm, slightly bigger in JADE and EM algorithm, 
should be enough to check, for the first time, 
how the bottom quark mass evolves, 
from low scales $\mu\sim 5(GeV)$ to the $m_Z$ scale,
following the Renormalization Group prediction, 
in the same way the running of the strong coupling 
constant has been checked before.

Among other applications,
our calculation would provide a better understanding
of the systematic errors in the measurement of
$\as(m_Z)$ from $b\bar{b}$-production at LEP from the 
ratio of three to two jets~\cite{juano}
and therefore a better measurement of the strong 
coupling constant $\as^b(m_Z)$.

Furthermore, quark mass effects are expected to be 
quite more important in top production at 
NLC (Next Linear Collider)\index{NLC} and our calculation 
would be of a great interest.

In this thesis, we have concentrated in the study of 
the three-jet observable $R^{bd}_3$ ratio.
To do that, we had to calculate 
the NLO matrix elements for the decay of the $Z$-boson 
into massive quarks. Main problem of such calculation 
was, in addition to UV divergences, 
the appearance of IR singularities. 
We were able first, by simple graphical arguments, to classify 
our transition amplitudes into different groups such
that IR singularities cancel independently
and second to build matrix elements free of singularities
since we exclude from the phase space the IR region 
where integration is performed analytically.
Once the IR behaviour is under control, 
it would be possible to build, with our matrix elements,
a Monte Carlo\index{Monte Carlo} generator\index{Generator}
opening the door to the study
of other kind of three-jet observables 
(angular distributions, differential cross sections,
etc $\ldots$ ) where massive quarks are involved.
Furthermore, the two-jet decay rate can be extracted 
by subtracting from the well known total decay rate 
the three- and the, full infrared finite, four-jet decay rates.
Therefore, the study of two-jet observables is also possible
from our calculation.

Concluding, in a previous paper~\cite{RO95}, we raised the 
question of the possibility of measuring the $b$-quark 
mass at LEP by using three-jet observables.
In close collaboration with the DELPHI experimental 
group at Valencia~\cite{FUQCD96}, we have developed the necessary 
theoretical tools in order to extract a meaningful value for 
the $b$-quark mass from LEP data, in particular, the 
NLO calculation of the three-jet decay rate of the $Z$-boson
into massive quarks.
Clearly, more work has to be done but the
effort is worth since it will allow for an independent 
measurement of $m_b$, far from threshold, and it will provide 
a test of the QCD Renormalization Group predictions.

\appendix
\chapter{Loop integrals}

\label{loopintegrals}

     We present in this appendix all the scalar one loop integrals 
we encountered in the calculation of the diagrams of figure~\ref{loop}.
Since we are just interested in the interference transition
probabilities of these diagrams with the born amplitudes we 
will only need the real part of such n-point scalar one 
loop integrals. Although we don't specify it, the following results
refer, in almost all the cases, just to the real part of these
functions.

\section{One- and two-point functions}

\index{One-point function}\index{Two-point function}
One- and two-point one loop functions are defined in
$D$ dimensions as 
\beq \bes
& \frac{i}{16\pi^2} A(m) 
      = \mu^{4-D} \int \frac{d^D k}{(2\pi)^D} \frac{1}{k^2-m^2}~, \\
& \frac{i}{16\pi^2} B_0(p^2,m_1^2,m_2^2) 
      = \mu^{4-D} \int \frac{d^D k}{(2\pi)^D} 
        \frac{1}{[k^2-m_1^2] [(k-p)^2-m_2^2]}~.\\
\ees
\end{equation}
They are the source of UV divergences.\index{UV divergences}

For the scalar one-point integral in $D=4-2\epsilon$ dimensions we have
the following result
\beq
A(m) = m^2 (\Delta - \log \frac{m^2}{\mu^2} + 1)~,
\end{equation}
with
\beq
\Delta = \frac{1}{\epsilon} - \gamma + \log 4\pi~.
\end{equation}
the UV pole
while for the two point functions we get
\begin{alignat}{2}
& B_0(m_b^2,0,m_b^2) &=& \: 
\Delta - \log \frac{m_b^2}{\mu^2} + 2~, \nonumber \\
& B_0(0,m_b^2,m_b^2) &=& \:
\Delta - \log \frac{m_b^2}{\mu^2}~,     \nonumber \\
& B_0(0,0,m_b^2)   &=& \:
\Delta - \log \frac{m_b^2}{\mu^2} + 1~,   \\
& B_0(p_{13}^2,0,m_b^2) &=& \: \Delta - \log \frac{m_b^2}{\mu^2} 
    - \frac{y_{13}}{r_b+y_{13}} \log \frac{y_{13}}{r_b} + 2~,    \nonumber \\
& B_0(s,m_b^2,m_b^2) &=& \: \Delta - \log \frac{m_b^2}{\mu^2} 
    + \beta \log c + 2~, \nonumber \\
& B_0(p_{12}^2,m_b^2,m_b^2) &=& \: \Delta - \log \frac{m_b^2}{\mu^2} 
    + \beta_{12} \log c_{12} + 2~, \nonumber \\
& B_0(0,0,0) &=& \: 0~, \nonumber 
\end{alignat}
where
\begin{alignat}{2}
r_b   &= m_b^2/m_Z^2~,       & \qquad & 
y_{ij} = 2(p_i \cdot p_j)/m_Z^2 \nonumber \\
\beta &= \sqrt{1-4r_b}~,     & \qquad &
\beta_{12} = \sqrt{1-4\frac{r_b}{y_{12}+2r_b}}~, \\
c &= \frac{1-\beta}{1+\beta}~,   & \qquad &
c_{12} = \frac{1-\beta_{12}}{1+\beta_{12}}~. \nonumber 
\end{alignat}

We have set to zero the $B_0(0,0,0)$ integral.
Strictly speaking that is not completely true. 
On analyzing this integral in detail we encounter 
it receives contributions in terms of UV and IR 
divergences separately, i.e., we should write  
\beq
B_0(0,0,0) = \mu^{4-D} \int \frac{d^D k}{(2\pi)^D} \frac{1}{k^4} 
= \frac{1}{\epsilon_{UV}} - \frac{1}{\epsilon_{IR}}~.
\end{equation}
If we try to distinguish between UV and IR poles we have to keep them.
Since this kind of integrals just change the label of the
divergences but not the number of poles in $1/\epsilon$
we will identify $\epsilon_{IR}=\epsilon_{UV}$. 
We can try to follow the origin of the UV and IR poles
but we don't gain nothing. 
Ultraviolet singularities\index{UV divergences} 
are cancelled by the renormalization 
procedure that is very well known.

\section{Three-point functions}

The scalar three-point integral reads
\begin{multline}
\frac{i}{16\pi^2} C_0(A,B,C,m_1^2,m_2^2,m_3^2) = \\ 
\mu^{4-D} \int \frac{d^D k}{(2\pi)^D}
\frac{1}{[k^2-m_1^2][(k+p_1)^2-m_2^2][(k+p_2)^2-m_3^2]}
\end{multline}
where $A=p_1^2$, $B=(p_1-p_2)^2$ and $C=p_2^2$. After
Feynman parametrization\index{Feynman!parametrization}
and momentum integration we get 
\begin{multline}
C_0(A,B,C,m_1^2,m_2^2,m_3^2) = \\
-\left( \frac{4\pi \mu^2}{q^2} \right)^{\frac{4-D}{2}}
\Gamma\left(3-\frac{D}{2}\right)
\int_0^1 dx \int_0^x dy 
\frac{1}{(a x^2 + b y^2 + c x y + d x + e y + f)^{3-\frac{D}{2}}}~, \\
\label{C0feynman}
\end{multline}
where
\beq \bes
   a &= C~, \\
   b &= B~, \\
   c &= A - B - C~, \\
   d &= m_3^2 - m_1^2 - C~, \\
   e &= m_2^2 - m_3^2 + C - A~, \\
   f &= m_1^2 - i \eta~.  
\ees
\end{equation}

For $A$, $B$ and $C$ different from zero 
and if the scalar three-point function does not contain 
infrared divergences the integral can be performed
in $D=4$ dimensions and can be expressed
as a sum over twelve dilogarithms~\cite{tHooft79}
\beq
C_0 = \frac{1}{\sqrt{\lambda}} \sum_{i=1}^{3} \sum_{j=1}^{2} (-1)^i
\left\{ Li_2\left(\frac{x_i}{x_i-y_{ij}}\right)
     - Li_2\left(\frac{x_i-1}{x_i-y_{ij}}\right) \right\}~,
\label{Li12}
\end{equation}
with the Spence function or dilogarithm 
function\index{Spence function}\index{Dilogarithm function} 
defined as~\cite{PT84}
\beq
Li_2(z) = - \int_0^z dt \frac{\log (1-t)}{t}~,
\end{equation}
and
\beq \bes
x_1 &= -\frac{d+2a+(c+e)\alpha}{\sqrt{\lambda}}~,    \\ 
x_2 &= -\frac{d+e\alpha}{(1-\alpha)\sqrt{\lambda}}~, \\ 
x_3 &=  \frac{d+e\alpha}{\alpha \sqrt{\lambda}}~,    \\ 
y_{1j} &= \frac{-(c+e) \pm \sqrt{(c+e)^2-4b(a+d+f)}}{2b}~, \\
y_{2j} &= \frac{-(d+e) \pm \sqrt{(d+e)^2-4f(a+b+c)}}{2(a+b+c)}~, \\
y_{3j} &= \frac{-d \pm \sqrt{d^2-4af}}{2a}~, 
\ees
\end{equation}
where $\alpha$ is a real solution of the equation
\beq
b\alpha^2 + c\alpha + a = 0~,
\end{equation}
and $\sqrt{\lambda}=c+2\alpha b$.

We define the following one loop integrals\index{C01}
\beq \bes
\frac{i}{16\pi^2} C01(y_{13}) &= \mu^{4-D} \int \frac{d^D k}{(2\pi)^D}
           \frac{1}{k^2 [(k+p_{13})^2-m_b^2][(k-p_2)^2-m_b^2]}~, 
\\ & \\
\frac{i}{16\pi^2} C02(y_{13}) &= \mu^{4-D} \int \frac{d^D k}{(2\pi)^D}
           \frac{1}{k^2 [(k+p_{13})^2-m_b^2][(k+p_1)^2-m_b^2]}~, 
\\ & \\
\frac{i}{16\pi^2} C03(y_{13}) &= \mu^{4-D} \int \frac{d^D k}{(2\pi)^D}
           \frac{1}{k^2 (k-p_3)^2 [(k+p_1)^2-m_b^2]}~,
\\ & \\
\frac{i}{16\pi^2} C04(y_{12}) &= \mu^{4-D} \int \frac{d^D k}{(2\pi)^D}
           \frac{1}{[k^2-m_b^2][(k+q)^2-m_b^2][(k+p_{12})^2-m_b^2]}~,
\\ & \\
\frac{i}{16\pi^2} C05(y_{12}) &= \mu^{4-D} \int \frac{d^D k}{(2\pi)^D}
           \frac{1}{k^2[(k+p_1)^2-m_b^2][(k-p_2)^2-m_b^2]}~.    
\label{allC0}
\ees
\end{equation}

For integral $C01$ we can apply directly~\Eq{Li12} and take just its
real part. For $C02$ and $C04$ we can calculate them either 
directly from the Feynman parametrization~\eqref{C0feynman}
\index{Feynman!parametrization}
or from~\Eq{Li12} by taking the appropriate limit. At the end we get a 
quite simplified 
result~\footnote{It is understood only the real part.}
\index{C02}\index{C04}
\beq \bes
C02(y_{13}) &= \frac{1}{y_{13}} \left[ 
\frac{1}{2} \log^2 (A) + Li_2(A) - \frac{\pi^2}{6} \right]~, \\ & \\
C04(y_{12}) &= \frac{1}{1-(y_{12}+2r_b)} \frac{1}{2}
\left[ \log^2 c - \log^2 c_{12} \right]~,
\ees
\end{equation}
with $A = r_b/(r_b+y_{13})$.
The three-point functions $C03$ and $C05$ are IR divergent 
and therefore we can not apply the general result of~\Eq{Li12}.
In these two cases we obtain, after integration in 
$D=4-2\epsilon$ dimensions~\footnote{$C05$ is the same 
one-loop integral that appears in the lowest order 
one-loop diagram.}\index{C03}\index{C05}
\begin{multline}
C03(y_{13}) = \frac{(4\pi)^{\epsilon}}{\Gamma(1-\epsilon)} 
\frac{1}{y_{13}} \left[ \frac{1}{2\epsilon^2} 
+ \frac{1}{2\epsilon} \log \frac{r_b}{y_{13}^2} \right. \\
\left. + \frac{1}{4} \log^2 \frac{r_b}{y_{13}^2} - \frac{1}{2} \log^2 A
- Li_2(A) - \frac{7\pi^2}{12} \right]~,
\label{C03}
\end{multline}

\begin{multline}
C05(y_{12}) = \frac{(4\pi)^{\epsilon}}{\Gamma(1-\epsilon)}
\frac{1}{(y_{12}+2r_b)\beta_{12}} \left[ \left( 
\frac{1}{\epsilon} - \log r_b \right) \log c_{12} - 2 L(y_{12}) \right]~.
\label{C05}
\end{multline}
where
\beq
L(y_{12}) = Li_2(c_{12}) + \frac{\pi^2}{3} + \log (1-c_{12}) \log c_{12}
-\frac{1}{4} \log^2 c_{12}~.
\end{equation}

\section{Four-point functions}

\index{Four-point function}
In our one loop matrix elements we encounter the following 
four-point functions\index{D05}\index{D06}
\begin{multline}
\frac{i}{16\pi^2} D05(y_{13},y_{12}) = \\ 
   \mu^{4-D} \int \frac{d^D k}{(2\pi)^D}
   \frac{1}{k^2 [(k+p_1)^2-m_b^2][(k+p_{13})^2-m_b^2][(k-p_2)^2-m_b^2]}~, 
\end{multline}

\begin{multline}
\frac{i}{16\pi^2} D06(y_{13},y_{23}) = \\ 
   \mu^{4-D} \int \frac{d^D k}{(2\pi)^D}
   \frac{1}{k^2 (k+p_3)^2 [(k+p_{13})^2-m_b^2][(k-p_2)^2-m_b^2]}~. 
\end{multline}

Both four-point functions are IR divergent. The integral $D05$ has
simple poles in $1/\epsilon$ while $D06$ presents double poles.
To get their finite part is not an easy task but in order to extract the
divergent piece we just need to split its non on-shell propagator.

For $D05$ we can write
\beq
\frac{1}{(k+p_{13})^2-m_b^2} = \frac{1}{p_{13}^2-m_b^2} 
- \frac{k^2+2k \cdot p_{13}}{(p_{13}^2-m_b^2)[(k+p_{13})^2-m_b^2]}~.
\end{equation}
The IR divergence is isolated in
the first term of the righthand side of the previous equation and
give rise to $C05(y_{12})/y_{13}$.\index{C05}
The other term generates two IR finite one loop integrals. 

%As the second part is IR finite we can 
%calculate it with any regulator. We will use gluon mass. In this 
%way we can apply the general result of~\cite{Denner90}. Be 
%\begin{multline}
%\frac{i}{16\pi^2} D05^{\lambda}(y_{13},y_{12}) = \\ 
%   \int \frac{d^4 k}{(2\pi)^4}
%   \frac{1}
%   {[k^2-\lambda^2][(k+p_1)^2-m_b^2][(k+p_{13})^2-m_b^2][(k-p_2)^2-m_b^2]}~, 
%\end{multline}
%and 
%\beq
%\frac{i}{16\pi^2} C05^{\lambda}(y_{12}) =
%     \int \frac{d^4 k}{(2\pi)^4}
%     \frac{1}{[k^2-\lambda^2][(k+p_1)^2-m_b^2][(k-p_2)^2-m_b^2]}~,   
%\end{equation}
%the same $D05$ and $C05$ functions but regulated with a gluon mass.
%Therefore, for our four point function in $D$ dimension we have
%the following result
%\beq
%D05(y_{13},y_{12}) = \frac{1}{y_{13}} 
%\left( C05(y_{12}) - C05^{\lambda}(y_{12}) \right)
%+ D05^{\lambda}(y_{13},y_{12})~.
%\end{equation}

For $D06$ things are more complex. 
The splitting of the non on-shell propagator does not give
directly the divergent piece due to the presence of IR double poles.
First of all we have to notice $D06$ is invariant under the interchange
of particles 1 and 2, i.e., the result of the integral should be 
symmetric in the two momenta invariants $y_{13}$ and $y_{23}$.
We found the proper way for extracting the IR piece
is as follows\index{C03}
\beq 
D06(y_{13},y_{23}) = 
  \frac{1}{y_{13}} C03(y_{23}) 
+ \frac{1}{y_{23}} C03(y_{13}) + \mrm{finite terms}~.
\end{equation}

%\beq \bes
%D06(y_{13},y_{23}) = 
%& \frac{1}{y_{13}} 
%\left( C03(y_{23}) - C03^{\lambda}(y_{23}) \right) \\
%+ & \frac{1}{y_{23}} 
%\left( C03(y_{13}) - C03^{\lambda}(y_{13}) \right)
%+ D06^{\lambda}(y_{13},y_{23})~,
%\ees
%\end{equation}
%where 
%\begin{multline}
%\frac{i}{16\pi^2} D06^{\lambda}(y_{13},y_{23}) = \\ 
%   \int \frac{d^4 k}{(2\pi)^4}
%   \frac{1}{[k^2-\lambda^2][(k+p_3)^2-\lambda^2]
%            [(k+p_{13})^2-m^2][(k-p_2)^2-m^2]}~, 
%\end{multline}
%and
%\beq \bes
%\frac{i}{16\pi^2} C03(y_{13}) & = \int \frac{d^4 k}{(2\pi)^4}
%   \frac{1}{[k^2-\lambda^2][(k-p_3)^2-\lambda^2][(k+p_1)^2-m_b^2]}~,\\ 
%\frac{i}{16\pi^2} C03(y_{23}) & = \int \frac{d^4 k}{(2\pi)^4}
%   \frac{1}{[k^2-\lambda^2][(k+p_3)^2-\lambda^2][(k-p_2)^2-m_b^2]}~.
%\ees
%\end{equation}

We leave for a future publication the detailed calculation 
of the finite contributions of the box one-loop integrals
$D05$ and $D06$.

\section{Explicit calculation of $C03$}\index{C03}

For the $C03$ function a Feynman parametrization, \eqref{C0feynman},
\index{Feynman!parametrization}
with both Feynman parameters running from $0$ to $1$ is more convenient 
because the two integrals decouple each other 
\begin{multline}
C03 = -(4\pi)^{\epsilon} \Gamma(1+\epsilon) 
\int_0^1 x^{-1-2\epsilon} dx \\
\int_0^1 \left[(r_b+ y_{13}) y^2 - (2r_b+y_{13}) y + r_b - i \eta  
         \right]^{-1-\epsilon} dy~.
\end{multline}
The integral over the x-parameter is immediate and we left just with 
\beq
C03 = (4\pi)^{\epsilon} \frac{\Gamma(1+\epsilon)}{2\epsilon} 
(r_b+y_{13})^{-1-\epsilon} 
\int_0^1 (1-y)^{-1-\epsilon} (y_0-y)^{-1-\epsilon} dy~,
\end{equation}
where $y_0$ is the root of the y-parameter polynomial.
The last integral gives rise to an 
Hypergeometric function\index{Hypergeometric function}
\begin{multline}
C03 = \frac{(4\pi)^{\epsilon}}{\Gamma(1-\epsilon)}
\frac{\Gamma(1+\epsilon) \Gamma(-\epsilon)}{2\epsilon} \\ 
(r_b+y_{13})^{-1-\epsilon} y_0^{-1-\epsilon}
{}_2F_1 [1+\epsilon,1,1-\epsilon,\frac{1}{y_0}]~,
\end{multline}
that with the help of some mathematical properties~\cite{AS72,BA53}
can be written in terms of a dilogarithm function~\cite{PT84,DD84}
\index{Dilogarithm function}
\beq
\bes
{}_2F_1 [1+\epsilon,1,1-\epsilon,\frac{1}{y_0}] &=
\left(1-\frac{1}{y_0}\right)^{-1-2\epsilon}
{}_2F_1 [-2\epsilon,-e,1-\epsilon,\frac{1}{y_0}] \\ &=
\left(1-\frac{1}{y_0}\right)^{-1-2\epsilon}
\left[ 1+2\epsilon^2 Li_2\left(\frac{1}{y_0}\right) 
+ O(\epsilon^3) \right]~.
\ees
\end{equation}
In the region of physical interest $y_0=r_b/(r_b+y_{13})$ is 
smaller than one and we should perform an analytic continuation 
taking into account the $-i\eta$ prescription. The final result
is quoted in~\Eq{C03}.

\section{Passarino-Veltman reduction}

\label{pasa}
\index{Passarino-Veltman reduction}
\index{Vector one-loop integrals}
\index{Tensor one-loop integrals}
We illustrate here a simple example of how to reduce vectorial
and tensorial loop integrals~\cite{Passarino79} to the scalar
one loop functions defined in the previous sections.
Let's consider the following set of integrals
\beq
 \mu^{4-D} \int \frac{d^D k}{(2\pi)^D}
 \frac{k_{\mu};k_{\mu} k_{\nu}}{k^2[(k+p_1)^2-m_b^2][(k-p_2)^2-m_b^2]}
= \frac{i}{16\pi^2} C0_{\mu;\mu \nu}~.    
\end{equation}
By Lorentz invariance\index{Lorentz invariance}
\beq
\bes
C0^{\mu} \: \: &= p_1^{\mu} C_{11} + p_2^{\mu} C_{12}~, \\
C0^{\mu \nu} &= g^{\mu \nu} C_{20} +
p_1^{\mu} p_1^{\nu} C_{21} + p_2^{\mu} p_2^{\nu} C_{22} \\
& + (p_1^{\mu} p_2^{\nu} + p_1^{\nu} p_2^{\mu}) C_{23}~.
\ees
\end{equation}
Now, we have to perform all the possible contractions with 
external momenta to extract the coefficients $C_{ij}$.

Let's try with 
\begin{multline}
 \mu^{4-D}  \int \frac{d^D k}{(2\pi)^D}
 \frac{(k\cdot p_1)}{k^2[(k+p_1)^2-m_b^2][(k-p_2)^2-m_b^2]} = \\
 \frac{i}{16 \pi^2} \left[  p_1^2 C_{11} + (p_1\cdot p_2) C_{12}
\right]~.
\end{multline}
The scalar product in the numerator can be written, 
including the on-shell condition $p_1^2=m_b^2$, as 
\beq
2(k\cdot p_1) = (k+p_1)^2 - k^2 - p_1^2 = [(k+p_1)^2-m_b^2] - k^2~.
\end{equation}
Therefore, our initial vectorial three propagators integral reduces 
automatically to the sum of two scalar two-point functions
\begin{multline}
 \mu^{4-D}  \int \frac{d^D k}{(2\pi)^D}
 \frac{2(k\cdot p_1)}{k^2[(k+p_1)^2-m_b^2][(k-p_2)^2-m_b^2]} = \\    
 \mu^{4-D}  \int \frac{d^D k}{(2\pi)^D}
\left[ \frac{1}{k^2[(k-p_2)^2-m_b^2]}    
- \frac{1}{[(k+p_1)^2-m_b^2][(k-p_2)^2-m_b^2]} \right]  \nonumber
\end{multline}
\beq  
= \frac{i}{16 \pi^2} 
\left[ B_0(p_2^2,0,m_b^2) - B_0(p_{12}^2,m_b^2,m_b^2) \right]~.    
\end{equation}

Same argument can be applied to the scalar product with $p_2$. 
In this simple case the final result is as follows
\beq
C_{11} = - C_{12}= \frac{1}{y_{12}-2r_b} 
\left[B_0(p_{12}^2,m_b^2,m_b^2) - B_0(m_b^2,0,m_b^2) \right]~.
\end{equation}
To reduce the tensorial integrals to scalar integrals we have 
to follow the same mathematical procedure. 
Solutions appear to be quite more complex, specially when 
we have to treat with box integrals, i.e., 
one loop integrals with four propagators.
To solve all the Pasarino-Veltman reductions we needed 
we have used the {\it FeynCalc 1.0}~\cite{Mertig90,Denner93}  
algebraic package for {\it Mathematica 2.0.}
whereas all the Dirac algebra was performed with
{\it HIP}~\cite{Hsieh92}.
\index{FeynCalc}\index{Mathematica}\index{HIP}

\section{The problem of $\gamma_5$ in $D$-dimensions}

\label{gamma5}

\index{gamma5@$\gamma_5$}
Multi-loop calculations with dimensional regularization 
often encounter the question of how to treat $\gamma_5$ 
in $D$ dimensions.
Occasionally the problem can be circumvented by 
exploiting chiral symmetry.\index{Chiral symmetry}
In our case, in the limit 
of massless quarks axial and vector contributions 
to the three jets decay rate of the $Z$ boson should
be equal\footnote{Singlet contribution not included.}. 
In general, however, a consistent definition must 
be formulated. 

A rigourous choice based on the original definition 
of 't Hooft and Veltman~\cite{tHooft72}  
\beq
\gamma_5 = \frac{i}{4!} \epsilon_{\mu \nu \rho \sigma}
\gamma^{\mu} \gamma^{\nu} \gamma^{\rho} \gamma^{\sigma}~.
\end{equation}
can be found in~\cite{Breitenlohner77}.
As a consequence of the lost of anticommuntativity of 
$\gamma_5$ in this new definition, standard properties of the 
axial current as well as the 
Ward identities\index{Ward identities} are violated.
In particular, extra finite renormalization constant should 
be introduced to restore the correctly renormalized 
non-singlet axial current.\index{Non-singlet axial current}.

Nevertheless, it has been checked~\cite{Broadhurst93} that 
for diagrams with an even number of $\gamma_5$ connected 
to the external current the treatment based on a 
na\"{\i}ve anticommutating $\gamma_5$ leads to the same answer.
This is our case because for unpolarized final state quarks 
only terms proportional to the square of the 
vector and axial-vector neutral current couplings survive.
Therefore, in our approach we have restricted to work with 
anticommutating $\gamma_5$ avoiding extra complications.

\chapter{Phase space in $D=4-2\epsilon$ dimensions} 
\label{app2}

\index{Phase space}
The phase space for $n$-particles in the  final state
in $D$-dimensions~\cite{dimreg1,dimreg2,dimreg3}
$(D=4-2\epsilon)$
has the following general form
\beq
\bes
dPS(n)
&= (2\pi)^D \prod_{i=1,n} \frac{d^{D-1}p_i}{(2\pi)^{D-1}2E_i}
\delta^D \left( q-\sum_{i=1,n}p_i \right) \\
&=
(2\pi)^D \prod_{i=1,n} \frac{d^{D}p_i}{(2\pi)^{D-1}}
\delta(p_i^2-m_i^2)\Theta(E_i) \delta^D 
\left( q-\sum_{i=1,n}p_i \right)~.
\label{phased}
\ees
\end{equation}
Then doing several trivial integrations we have
the following phase-space factor for the process 
$Z\rightarrow b\bar{b}$\index{PS(2)}
\beq
PS(2)=\frac{1}{4\pi} \frac{\beta}{2}
\frac{\Gamma(1-\epsilon)}{\Gamma(2-2\epsilon)}
{\left( \frac{\beta^2 m_Z^2}{4\pi} \right) }^{-\epsilon}~,
\end{equation}
where 
$\beta=\sqrt{1-4 r_b}$   with $r_b=m_b^2/m_Z^2$,

For the case of the decay into three particles,
$Z\rightarrow b(p_1) \bar{b}(p_2) g(p_3)$, we get\index{PS(3)}
\beq
PS(3)=\frac{m_Z^2}{16(2\pi)^3}
\frac{1}{\Gamma(2-2\epsilon)}
{\left( \frac{m_Z^2}{4\pi} \right) }^{-2\epsilon} \int
\theta(h_p) h_p^{-\epsilon} dy_{13} dy_{23}~,
\end{equation}
where the function $h_p$ which gives the phase-space boundary
in terms of variables $y_{13}=2(p_1 \cdot p_3)/m_Z^2$ and 
$y_{23}=2(p_2 \cdot p_3)/m_Z^2$ has the form\index{hp@$h_p$}
\beq
h_p = y_{13} y_{23} (1-y_{13}-y_{23}) - r_b (y_{13}+y_{23})^2~.
\label{bornPS}
\end{equation}

\section{System 1-3}\index{System 1-3}

For those four parton transitions amplitudes of the process
$Z\rightarrow b(p_1) + \bar{b}(p_2) + g(p_3) + g(p_4)$
containing the denominator
$y_{13}=2 (p_1 \cdot p_3) / m_Z^2 $, i.e., with gluon labelled as 3 
soft, it is convenient to write the four-body phase space as a 
quasi three body decay
\beq
\bes
q \rightarrow & p_{13} + p_2 + p_4    \\
              & \hookrightarrow p_1 + p_3~.
\ees
\end{equation}

%%%%%%%%%%%%%%%
\mafigura{7cm}{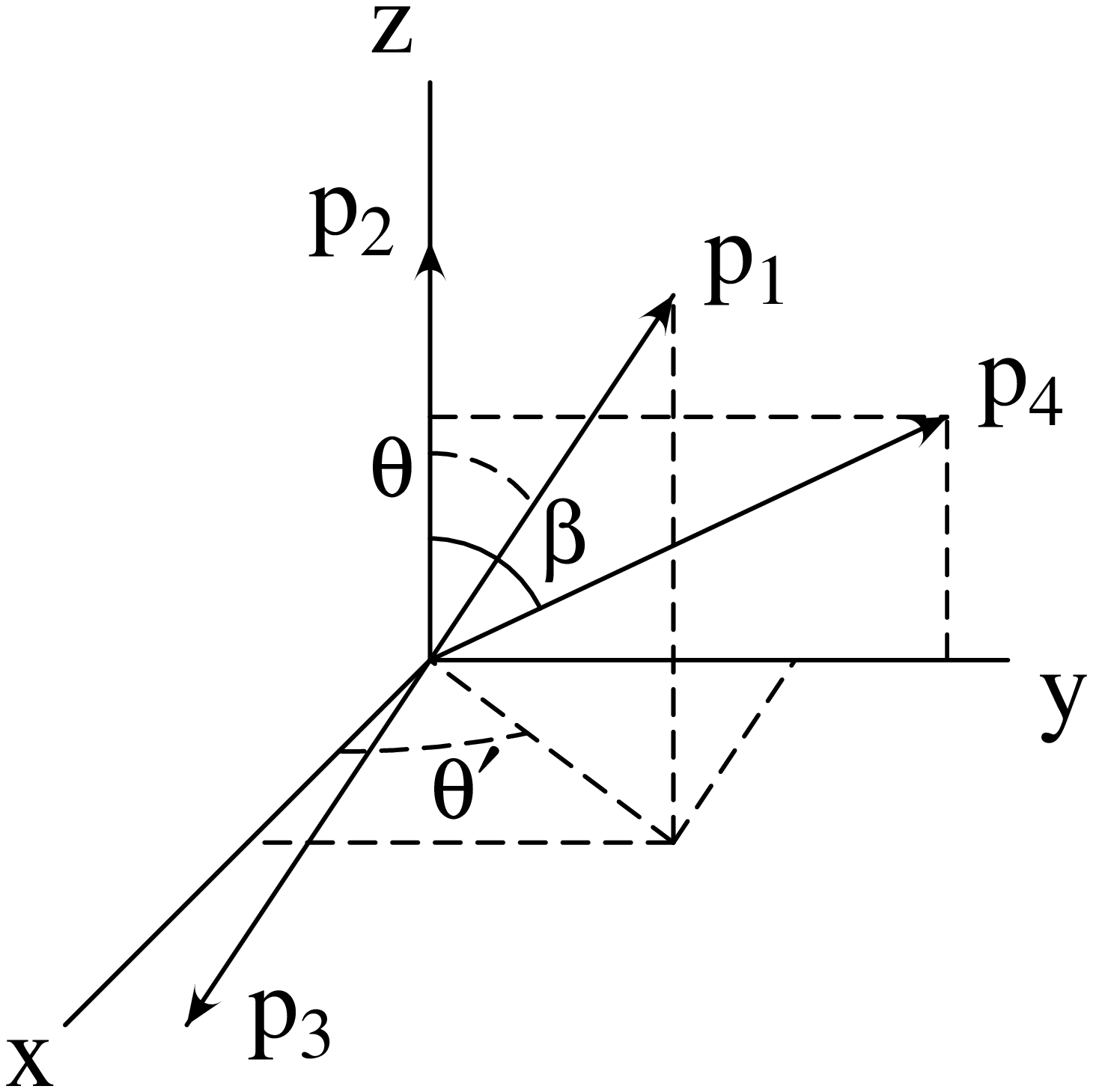}
{Threemomenta in the c.m. frame of particles 1 and 3.}
{system13}
%%%%%%%%%%%%%%%

We will refer to this as the ``1-3 system''~\cite{Ellis81}.
In the c.m. frame of particles 1 and 3 the four-momenta can
be written as 
\beq
\bes
p_1 &= (E_1, \ldots, \mrm{\bf p}_1 \sin \theta \cos \theta', 
\mrm{\bf p}_1 \cos \theta)~,  \\
p_2 &= (E_2, \ldots, 0, \mrm{\bf p}_2)~,\\
p_3 &= E_3 (1, \ldots, - \sin \theta \cos \theta', - \cos \theta)~, \\
p_4 &= E_4 (1,\ldots,\sin \beta, \cos \beta)~,
\ees
\end{equation}
where the dots in $p_1$ and $p_3$ indicate $D-3$ unspecified,
equal and opposite angles (in $D$ dimensions) and $D-3$ zeros
in $p_2$ and $p_4$. 
In terms of variables\index{y13@$y_{13}$} 
\beq
y_{13}  = \frac{2 (p_1 \cdot p_3)}{m_Z^2}~,    \qquad
y_{123} = \frac{2 (p_2 \cdot p_{13})}{m_Z^2}~, \qquad
y_{134} = \frac{2 (p_4 \cdot p_{13})}{m_Z^2}~, 
\end{equation}
with $p_{13}=p_1+p_3$, energies and threemomenta read ($s=m_Z^2=1$)
\begin{alignat}{2}
E_1 &= \frac{y_{13}+2r_b}{2 \sqrt{y_{13}+r_b}}~, & \qquad &
\mrm{\bf p}_1  = \frac{y_{13}}{2 \sqrt{y_{13}+r_b}}~, \nonumber \\
E_2 &= \frac{y_{123}}{2 \sqrt{y_{13}+r_b}}~,     & \qquad &
\mrm{\bf p}_2  = 
\frac{1}{2 \sqrt{y_{13}+r_b}} \sqrt{y_{123}^2-4r_b(y_{13}+r_b)}~, \\
E_3 &= \frac{y_{13}}{2 \sqrt{y_{13}+r_b}}~,      & \qquad &
E_4  = \frac{y_{134}}{2 \sqrt{y_{13}+r_b}}~. \nonumber
\end{alignat}

Setting $v=\frac{1}{2}(1-\cos \theta)$, we obtain for the $D$-dimensional
phase space in this system\index{PS(4)}
\beq
\bes
PS(4) &= \frac{m_Z^2}{16(2\pi)^3}
\frac{1}{\Gamma(2-2\epsilon)}
{\left( \frac{m_Z^2}{4\pi} \right) }^{-2\epsilon} \int dy_{134} dy_{234} \\
& m_Z^{2(1-\epsilon)}
\frac{S}{16 \pi^2}  \frac{(4\pi)^\epsilon}{\Gamma(1-\epsilon)}
\int dy_{13} \theta(h_p) h_p^{-\epsilon}  
\frac{y_{13}^{1-2\epsilon}}{(y_{13}+r_b)^{1-\epsilon}} \\
& \int_0^1 dv (v(1-v))^{-\epsilon}
\frac{1}{N_{\theta'}} \int_0^{\pi} d\theta' \sin^{-2\epsilon} \theta' ~, 
\ees
\end{equation}
where $S=1/2!$ is the statistical factor, 
$N_{\theta'}$\index{Ntheta@$N_{\theta'}$} is a
normalization factor determined such that
\beq
\int_0^{\pi} d\theta' \sin^{-2\epsilon} \theta' 
= N_{\theta'} = 2^{2\epsilon} \pi 
\frac{\D \Gamma(1-2\epsilon)}{\D \Gamma^2(1-\epsilon)}~,
\end{equation}
and the function\index{hp@$h_p$}
\beq
\bes
h_p &= y_{123}y_{134} (1-y_{123}-y_{134}) \\
& + r_b \left( -1+2(y_{123}+y_{134})-(y_{123}^2+4y_{123}y_{134}+2y_{134}^2) 
\right) \\
& + 4r_b^2(1-y_{123}-y_{134})-4r_b^3 \\
& + \left(-1+6r_b-8r_b^2 \right. \\
& \qquad \left. + 2(1-3r_b)(y_{123}+y_{134})-y_{123}^2
-3y_{123}y_{134}-y_{134}^2 \right) y_{13} \\
& + \left(2-5r_b-2y_{123}-2y_{134} \right) y_{13}^2 - y_{13}^3~,
\ees
\end{equation}
defines the limits of the phase space.

\section{System 3-4}\index{System 3-4}

\label{secsystem34}

For diagrams containing the denominator $y_{34} = 2 (p_3 \cdot p_4) / m_Z^2$,
that is, gluons labelled as 3 and 4 soft and/or collinear we will 
work in the c.m. frame of both gluons where
\beq
\bes
p_1 &= (E_1, \ldots, 0, \mrm{\bf p}_1)~,\\
p_2 &= (E_2, \ldots, \mrm{\bf p}_2 \sin \beta, \mrm{\bf p}_2 \cos \beta)~, \\
p_3 &= E_3 (1, \ldots, \sin \theta \cos \theta', \cos \theta)~,  \\
p_4 &= E_4 (1, \ldots, - \sin \theta \cos \theta', - \cos \theta)~. 
\ees
\end{equation}

%%%%%%%%%%%%%%%
\mafigura{7cm}{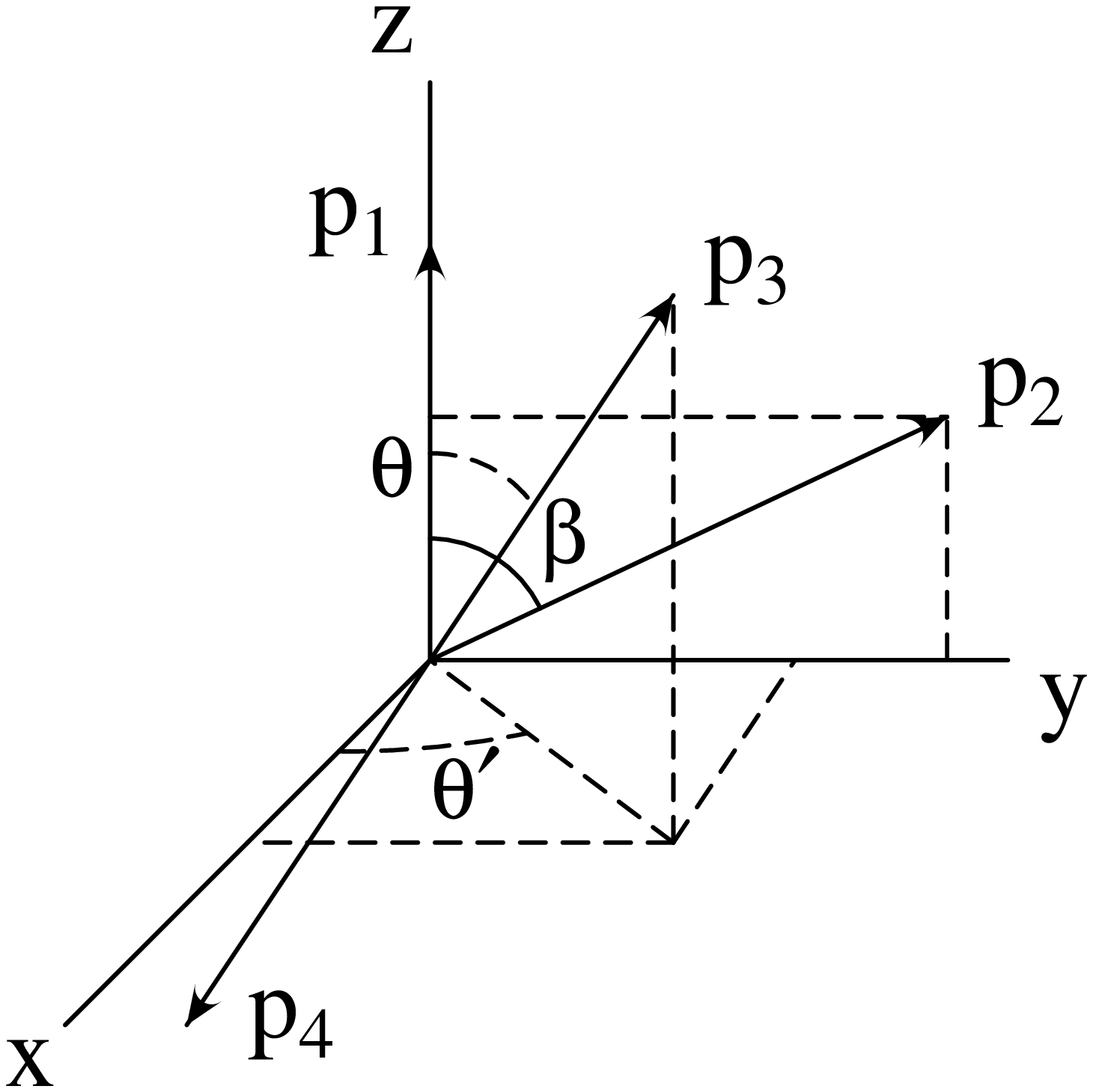}
{Threemomenta in the c.m. frame of particles 3 and 4.}
{system34}
%%%%%%%%%%%%%%%

In this case we will use the following set of 
variables\index{y34@$y_{34}$} 
\beq
y_{34}  = \frac{2 (p_3 \cdot p_4)}{m_Z^2}~,    \qquad
y_{134} = \frac{2 (p_1 \cdot p_{34})}{m_Z^2}~, \qquad
y_{234} = \frac{2 (p_2 \cdot p_{34})}{m_Z^2}~, 
\end{equation}
with $p_{34}=p_3+p_4$, in terms of which energies and threemomenta read
\begin{alignat}{2}
E_1 &= \frac{y_{134}}{2 \sqrt{y_{34}}}~, & \qquad &
\mrm{\bf p}_1 = 
\frac{1}{2 \sqrt{y_{34}}} \sqrt{y_{134}^2-4r_b y_{34}}~, \nonumber \\
E_2 &= \frac{y_{234}}{2 \sqrt{y_{34}}}~, & \qquad &
\mrm{\bf p}_2  = \frac{1}{2 \sqrt{y_{34}}} \sqrt{y_{234}^2-4r_b y_{34}}~, \\
E_3 &= \frac{\sqrt{y_{34}}}{2}~, & \qquad &
E_4  = \frac{\sqrt{y_{34}}}{2}~. \nonumber
\end{alignat}

We obtain for the $D$-dimensional phase space in this system\index{PS(4)}
\beq
\bes
PS(4) &= \frac{m_Z^2}{16(2\pi)^3}
\frac{1}{\Gamma(2-2\epsilon)}
{\left( \frac{m_Z^2}{4\pi} \right) }^{-2\epsilon} \int dy_{134} dy_{234} \\
& m_Z^{2(1-\epsilon)}
\frac{S}{16 \pi^2}  \frac{(4\pi)^\epsilon}{\Gamma(1-\epsilon)}
\int dy_{34} \theta(h_p) h_p^{-\epsilon} y_{34}^{-\epsilon} \\
& \int_0^1 dv (v(1-v))^{-\epsilon}
\frac{1}{N_{\theta'}} \int_0^{\pi} d\theta' \sin^{-2\epsilon} \theta' ~,
\label{PSsystem34}
\ees
\end{equation}
with\index{hp@$h_p$}
\beq
\bes
h_p &= y_{134}y_{234}(1-y_{134}-y_{234})-r_b(y_{134}+y_{234})^2 \\
& + \left( -1+4r_b+2(1-2r_b)(y_{134}+y_{234})
-y_{134}^2-3y_{134}y_{234}-y_{234}^2 \right) y_{34} \\
& + 2 (1-2r_b-y_{134}-y_{234}) y_{34}^2-y_{34}^3~.
\ees
\end{equation}
Observe $h_p$ reduce to \eqref{bornPS} in the case $y_{34}\rightarrow 0$.
Furthermore, in this limit $p_{34}=p_3+p_4$ behaves as the momentum of 
a pseudo on-shell gluon because $p_{34}^2\rightarrow 0$.

\bibliographystyle{amsplain}
\bibliography{tesis}
\printindex
%\documentstyle[times]{article}
%\newcommand{\D}{\displaystyle}
%\begin{document}

\thispagestyle{empty}

\begin{center}
{\sc Agradecimientos} 
\end{center}

\vspace{1cm}

Quisiera agradecer, en primer lugar, a mi director de tesis
por los consejos recibidos durante estos a\~nos, 
por haberse involucrado en este largo y pesado c\'alculo en 
algunos momentos clave de su desarrollo
y por haberme corregido incluso el mal ingl\'es de esta tesis.

Me gustar\'{\i}a agradecer al grupo experimental de DELPHI 
en Valencia el apoyo moral recibido, el haber acogido   
a {\it ``el te\'orico''} en sus viajes por la Helvetia 
y el soporte t\'ecnico e inform\'atico sin el cual esta tesis 
no habr\'{\i}a podido ser realizada.

Mi profundo agradecimiento a todos los 
miembros del Departamento de F\'{\i}sica Te\'orica, en particular
a aquellos que en alg\'un momento se interesaron por 
el estado del c\'alculo e intentaron ayudarme y muy 
especialmente a la Asociaci\'on de Becarios {\it ``Juan Vald\'es''} 
por las horas que desahogo que supuso la visita 
peri\'odica al bar de la Facultad.

Por \'ultimo, quisiera agradecer a mis amigos de siempre 
simplemente el haber continuado ah\'{\i} y a mis padres, hermanas, 
abuela y a Chelo por haberme aguantado tanto cuando lo 
\'unico que me preocupaba eran los quarks y los gluones.

%\end{document}

\end{document}